\documentclass[structabstract]{aa}  
\usepackage{graphicx}
\usepackage{savesym}
\usepackage{amsmath}
\usepackage{amssymb}
\savesymbol{iint}
\usepackage{txfonts}
\restoresymbol{TXF}{iint}
\usepackage{natbib}
\usepackage{lscape}
\usepackage{rotating}

\begin{document}
   \title{The AMBRE Project: Stellar parameterisation of the ESO:FEROS archived spectra}

   \subtitle{}

   \author{C.~C.~Worley
          \inst{1}
          \and
          P.~de~Laverny
	  \inst{1}
          \and
	  A.~Recio--Blanco
	  \inst{1}
          \and
	  V.~Hill
	  \inst{1}
          \and
	  A.~Bijaoui
	  \inst{1}
          \and
	  C.~Ordenovic
	  \inst{1}
	  %\and
	  %ESO Members
	  %\inst{2}
	  %\fnmsep\thanks{Thanks to ESO,OCA,CNES}
          }

   \institute{1. Laboratoire Lagrange (UMR7293), Universit\'e de Nice Sophia Antipolis, CNRS, Observatoire de la C\^ote d'Azur, BP 4229, F-06304 Nice Cedex 04, France\\
              \email{cworley@oca.eu} \\
      %2. European Southern Observatory
             }

   \date{Received 17 January, 2012; accepted 20 March, 2012}

% \abstract{}{}{}{}{} 
% 5 {} token are mandatory
 
  \abstract
   {The AMBRE Project is a collaboration between the European Southern Observatory (ESO) and the Observatoire de la C\^{o}te d'Azur (OCA) that has been established in order to carry out the determination of stellar atmospheric parameters for the archived spectra of four ESO spectrographs.}
   {The analysis of the FEROS archived spectra for their stellar parameters (effective temperatures, surface gravities, global metallicities, alpha element to iron ratios and radial velocities) has been completed in the first phase of the AMBRE Project. From the complete ESO:FEROS archive dataset that was received, a total of 21551 scientific spectra have been identified, covering the period 2005 to 2010. These spectra correspond to 6285 stars.}
   {The determination of the stellar parameters was carried out using the stellar parameterisation algorithm, MATISSE (MATrix Inversion for Spectral SynthEsis), which has been developed at OCA to be used in the analysis of large scale spectroscopic studies in galactic archaeology. An analysis pipeline has been constructed that integrates spectral normalisation, cleaning and radial velocity correction procedures in order that the FEROS spectra could be analysed automatically with MATISSE to obtain the stellar parameters. The synthetic grid against which the MATISSE analysis is carried out is currently constrained to parameters of FGKM stars only.}
   {Stellar atmospheric parameters, effective temperature, surface gravity, metallicity and alpha element abundances, were determined for 6508 (30.2\%) of the FEROS archived spectra ($\sim$3087 stars). Radial velocities were determined for 11963 (56\%) of the archived spectra. 2370 (11\%) spectra could not be analysed within the pipeline due to very low signal-to-noise ratios or missing spectral orders. 12673 spectra (58.8\%) were analysed in the pipeline but their parameters were discarded based on quality criteria and error analysis determined within the automated process. The majority of these rejected spectra were found to have broad spectral features, as probed both by the direct measurement of the features and cross-correlation function breadths, indicating that they may be hot and/or fast rotating stars, which are not considered within the adopted reference synthetic spectra grid. The current configuration of the synthetic spectra grid is devoted to slow-rotating FGKM stars. Hence non-standard spectra (binaries, chemically peculiar stars etc) that could not be identified may pollute the analysis.}
   {}

   \keywords{Methods/data analysis, Astronomical databases/miscellaneous, Stars/fundamental parameters, Techniques/spectroscopic}

   \maketitle
%
%________________________________________________________________

\section{Introduction}
Astronomy has entered an era of large scale astronomical surveys, the scientific goals of which have the potential to considerably expand our understanding of the formation and evolution of the Universe. In particular current and future large scale spectroscopic surveys of the Milky Way will allow astronomers to trace in incredible detail the chemical and kinematic history of our Galaxy.

These surveys are being undertaken over a range of resolutions. Low-resolution surveys (R = 2,000 to 8,000) are now widespread, for example RAVE \citep{Steinmetz2006} and SEGUE \citep{Yanny2009}, while high-resolution surveys are a more recent endeavour such as APOGEE \citep{Majewski2007}) and the Gaia-ESO Survey (P.I.s: Gerry Gilmore \& Sofia Randich). The science goals of these ground-based surveys will significantly contribute to the studies of galactic archaeology as well as provide complementary information to the upcoming astrometric survey, the European Space Agency (ESA) Gaia Mission. The ESA Gaia satellite will observe approximately a billion stars in the Milky Way for which distances will be determined to milliarcsecond accuracies. Its Radial Velocity Spectrometer (RVS) will observe spectra at medium-resolution (R$\simeq$7000 - 11,500) which will be used to obtain radial velocities for all the targets, as well as to determine stellar atmospheric parameters and chemical abundances for some ten's of millions of stars.

The multi-object instruments at high-resolution (R$\sim$20,000; MIKE on Magellan and FLAMES on the Very Large Telescope), medium-resolution (R$\sim$8,000; AAOmega on the Australian Astronomical Telescope (AAT)) and low-resolution (R$\le$2,200; Fibre Multi Object Spectrograph (FMOS) on the Subaru Telescope) are also being used to address science goals specific to the field of galactic archaeology. One key instrument that is currently being built and has been designed primarily for galactic archaeology research is the High Efficiency and Resolution Multi-Element Spectrograph \citep[HERMES][]{Barden2010} on the AAT. 

This ongoing accumulation of large spectral datasets from surveys and individual spectrographs has compelled the development of automated stellar parameterisation algorithms that can reliably and effectively analyse every spectrum. The stellar parameterisation algorithm, MATISSE, has been developed at the Observatoire de la C\^{o}te d'Azur (OCA) \citep{Recio-Blanco2006} to be included in the automated pipeline for the analysis and parameterisation of the Gaia-RVS stellar spectra. MATISSE has also been developed as a standalone java application for use in a wide variety of projects \citep[See for instance,][]{Gazzano2010,Kordopatis2011}. The AMBRE Project team at OCA, which oversees the development of MATISSE, is connected to the Gaia Data Processing Consortium (DPAC) under the Generalized Stellar Parametrizer-spectroscopy (GSP-spec) Top Level Work Package which is overseen by Coordination Unit 8 (CU8).

With such large datasets soon to be available a crucial aspect of the analysis is to be able to compare the results from the different surveys. To do, this a comprehensive set of standard objects is required that can be used to calibrate the datasets. Preparation for this is in the form of the development of spectral libraries. These are datasets of spectra with homogeneously determined characteristics, such as stellar parameters, which can be used as calibration stars for these surveys. In particular standard star lists are being developed for Gaia for both the radial velocity \citep{Crifo2010} and stellar parameter measurements \citep{Soubiran2010}.

The work carried out for the AMBRE Project in effect converts the spectra in the European Southern Observatory (ESO) archive into a comprehensive spectral library of homogeneously determined stellar parameters. There are three primary objectives for the AMBRE Project:
\begin{enumerate}
 \item To rigorously test MATISSE on large spectral datasets over a range of wavelengths and resolutions, including those for the Gaia RVS, 

 \item To provide ESO with a database of stellar temperatures, gravities, metallicities, alpha to iron ratios and radial velocities for the associated archived spectra that will then be made available to the international scientific community via the ESO Archive,

 \item To create a chemical map of the Galaxy from the combined ESO archived sample upon which stellar and galactic formation and evolution archaeological analysis can be carried out. 
\end{enumerate}

The first phase of the AMBRE Project was the analysis of the FEROS archived spectra. From the complete ESO:FEROS archive dataset that was delivered to OCA, a total of 21551 scientific spectra have been identified, covering the period 2005 to 2010. These spectra correspond to 6285 different stars based on a coordinate matching calculation with a radius of 10\textquotedblright.

The structure of this paper is as follows: Section~\ref{sec:AMBRE} introduces the AMBRE Project; Section~\ref{sec:MATISSE} introduces the MATISSE algorithm and the synthetic spectra grid; Section~\ref{sec:FEROSpipeline} describes the analysis pipeline that has been built around MATISSE for the analysis of the archived spectra; Section~\ref{sec:int_error} discusses the internal errors that have been calculated for the pipeline; Section~\ref{sec:ext_error} discusses the external errors analysis based on a reference sample of stars; Section~\ref{sec:esotab_rej} presents the application of key rejection criteria to the spectral dataset; Section~\ref{sec:AFstellarparams} presents the stellar parameter results for FEROS and finally Section~\ref{sec:conclusion} concludes the paper.

\section{The AMBRE Project}\label{sec:AMBRE}
Under a contract with ESO the archived spectra of four ESO spectrographs are being analysed using the automated parameterisation programme MATISSE in a project overseen by the AMBRE Project team. Table~\ref{tab:spectrographs} lists the main characteristics of the four spectrographs: the Fiberfed Extended Range Optical Spectrograph (FEROS) \citep{Kaufer1999}; the High Accuracy Radial velocity Planet Searcher (HARPS) \citep{Mayor2003}; the Ultraviolet and Visual Echelle Spectrograph \citep{Dekker2000}; and the Fibre Large Array Multi Element Spectrograph/GIRAFFE (FLAMES/GIRAFFE) \citep{Pasquini2002}. The atmospheric stellar parameters of effective temperature ($T_{\textrm{eff}}$), surface gravity ($\log g$), global metallicity ([M/H]), $\alpha$ element to iron ratio ([$\alpha$/Fe]) and radial velocity (V$_{rad}$) will be derived for each of the archived stellar spectra. These will be delivered to ESO for inclusion in the ESO database and then made available to the astronomical community via the ESO archive. It is intended that the availability of stellar parameters for each spectra will encourage further use of the archived spectra. Previously unconsidered samples can be found through searches on the parameters, for examples extracting all the spectra in a particular metallicity range, or with very similar temperatures and gravities. 

\begin{table}
\caption{Details of the four ESO spectrographs and their publicly available archived spectra sample that are part of the AMBRE Project.}\label{tab:spectrographs}
%\vspace{0.2cm}
\begin{tabular}{cccc}
\hline\hline
Spectrograph & Resolution & $\lambda$ (nm) & No. spectra \\
\hline
%\vspace{0.2cm}
{\scriptsize FEROS} & 48,000 & 350--920 & 21,551 \\
{\scriptsize HARPS} & 115,000 & 378--690 & 126,694 \\
{\scriptsize UVES} & 20,000--110,000 & 300--1100 & 78,593 \\
{\scriptsize FLAMES/GIRAFFE} & 5,600--46,000 & 370--900 & $>$100,000 \\
\hline
% & & Total Sample & 198,000 \\
\end{tabular}
\end{table}  

The analysis of the archived spectra of these four spectrographs presents a unique opportunity to test the performance of MATISSE on large datasets of real spectra. In particular key instrument configurations of FLAMES/GIRAFFE cover the Gaia RVS wavelength domain and resolutions. Rigorous testing of MATISSE is necessary in order to optimise its performance in the Gaia-RVS analysis pipeline that is being compiled at the Centre National d\textquoteright Etudes Spatiales (CNES). As such the AMBRE Project has been formally designated as a sub-work package under GSP-spec. The stars analysed by AMBRE will also be available for use as standard or calibration stars for the Gaia-ESO survey, and as secondary standards for the Gaia Mission.

The first phase of the AMBRE Project, the analysis of the FEROS archived spectra, is now complete. FEROS is a state-of-the-art bench-mounted high resolution spectrograph that was built by a consortium\footnote{http://www.lsw.uni-heidelberg.de/projects/instrumentation/Feros/} of four astronomical institutes: Landessternwarte Heidelberg, Astronomical Observatory Copenhagen, Institut d'Astrophysique de Paris and Observatoire de Paris/Meudon. In 1998 FEROS was installed and commissioned on the ESO 1.52~m telescope at La Silla, Chile. In 2002 FEROS was moved to the MPG/ESO 2.2~m telescope where it permanently resides\footnote{http://www.eso.org/sci/facilities/lasilla/instruments/feros/index.html}. The high resolution and full coverage of the optical domain means that FEROS is suitable for a wide range of astronomical projects from radial velocities and variability studies to chemical abundances in stellar populations.

In 2009-10 the FEROS spectra from October 2005 to December 2009 were reanalysed by the ESO archivists using an improved reduction pipeline. These reduced spectra were the sample that was delivered to OCA for analysis in AMBRE. 

\begin{figure}[!h]
\centering
\begin{minipage}{80mm}
\centering
\includegraphics[width=80mm]{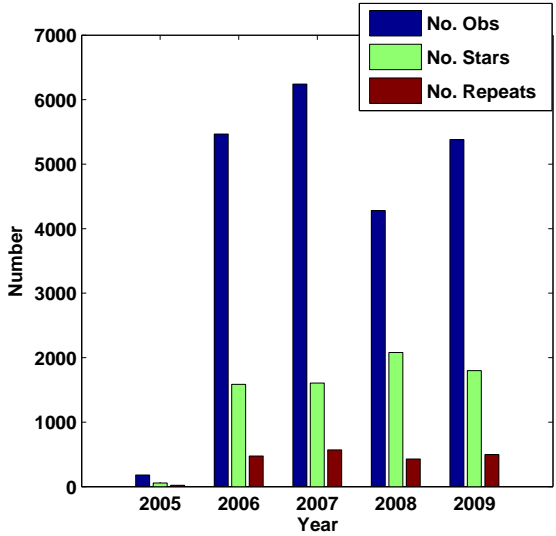}
\caption{Number of FEROS observations per year for analysis in AMBRE. The 2005 observations begin in October. The number of distinct objects observed and the number with repeated observations are as indicated in the key.}\label{fig:feros_yrobs}
\end{minipage}
\end{figure}

Figure~\ref{fig:feros_yrobs} shows a histogram of the number of FEROS observations per year that were received by OCA, the number for 2005 representing only 3 months of observations. The number counts for the number of distinct objects that have been observed by FEROS in this timeframe and the number with repeated observations are also shown.

The analysis procedure and results for these FEROS spectra are presented in this paper. This analysis has been the testbed for producing many of the tools that will also be used in the analysis of the UVES, HARPS and FLAMES/GIRAFFE archived spectra. These tools have been integrated into an analysis pipeline that feeds the processed spectra into MATISSE for derivation of the spectral parameters. 

\section{MATISSE \& the synthetic spectra grid}\label{sec:MATISSE}

MATISSE (MATrix Inversion for Spectral SynthEsis) is an automated stellar parameterisation algorithm based on a local multi-linear regression method. It derives stellar parameters ($\theta$ = $T_{\textrm{eff}}$, $\log \ g$, [M/H], individual chemical abundances) by the projection of an input observed spectrum on a vector function $B_{\theta}(\lambda)$,

   \begin{equation}
    \hspace{0cm}  \hat{\theta}_{i} = \sum\limits_{\lambda} \ B_{\theta}(\lambda) \cdot O_{i}(\lambda)
   \end{equation}

\noindent with $\hat{\theta}$ being the derived value. The $B_{\theta}(\lambda)$ vector function is an optimal linear combination of theoretical spectra, $S_{i}(\lambda)$, calculated from a synthetic spectra grid. 

   \begin{equation}
    \hspace{0cm} B_{\theta_{i}}(\lambda) = \sum\limits_{j} \ \alpha_{ij} \ S_{j}(\lambda)
   \end{equation}

Key features in the observed spectrum due to a particular $\theta$ are reflected in the corresponding $B_{\theta}(\lambda)$ vector indicating the particular regions which are sensitive to $\theta$ \citep{Recio-Blanco2006, Bijaoui2008}).

MATISSE also generates synthetic spectra interpolated to each set of output stellar parameters. For each set of parameters, the synthetic spectrum on the grid, $S_{0}(\lambda)$, with the closest stellar parameters is identified along with the associated $B_{\theta}(\lambda)$ function. The $B_{\theta}(\lambda)$ functions located locally about this point are used to estimate the variations in the flux between the synthetic spectra at the corresponding grid points and the final interpolated synthetic spectrum corresponding to the required stellar parameters ($\theta_{k}$) \citep{Kordopatis2011}. %The following expression is used to make this interpolation:
%the $B_{\theta}(\lambda)$ functions located locally about the $B_{\theta}(\lambda)$ function that is associated with the synthetic spectrum on the grid, $S_{0}(\lambda)$, which is closest to the required stellar parameters. The $B_{\theta}(\lambda)$ functions are used to find the variations in flux from $S_{0}(\lambda)$ which correspond to the difference between the parameters of $S_{0}(\lambda)$ ($\theta_{{_0}k}$) and the required stellar parameters ($\theta_{k}$). this results in the interpolated synthetic spectra, $S(\lambda)$ \citep{Kordopatis2011}. The following expression is used to make this calculation:
%\begin{equation}
% S_{0}(\lambda) - S(\lambda) = \sum\limits_{k=1,K} (\theta_{{_0}k} - \theta_k) \sum\limits_{k'=1,K}, B^{-1}_{kk'}B_{\theta_k'}(\lambda)
%\end{equation}

%\noindent where $B_{kk'}$ is the correlation matrix between the basis vectors, defined as:

%\begin{equation}
% B_{kk'} = \sum\limits_{\lambda} B_{\theta_k}(\lambda)B_{\theta_k'}(\lambda).
%\end{equation}

This interpolated spectrum is then used to calculate a $\chi^2$ between the interpolated and input normalised spectrum. This provides a measure of the goodness of the fit of the derived stellar parameters to the observed spectrum. Stellar parameters, with $\log \chi^2$, and corresponding interpolated synthetic spectra are generated for every spectrum that is analysed in MATISSE.

\subsection{Grid of synthetic spectra for the classification algorithm}\label{sec:ambre_synthgrid}
Within the AMBRE Project, the adopted procedure for the automatic classification of stellar spectra relies on a library of reference spectra in order to derive their atmospheric parameters and their chemical abundances. Due to the lack of a library of observed spectra that covers a large range of atmospheric parameters and chemical abundances over a very large spectral domain and resolution, the only solution was to compute large grids of synthetic spectra.

For the AMBRE application, the computed grid of theoretical stellar spectra has to cover the whole optical spectral range at very high resolution in order to be used for the analysis of the majority of the spectroscopic data. For that reason, we have computed a synthetic spectra grid covering the wavelength range between 300 and 1\,200~nm with a wavelength step of 0.001~nm (900,000 pixels in total) \citep{deLaverny2012}. Since this project is mostly devoted to the analysis of FGKM stars, this grid is based on the latest generation of MARCS model atmospheres presented in \citet{Gustafsson2008}. An extension to hotter effective temperatures with Kurucz stellar atmosphere models \citep{Kurucz1979ModelGrid} is planned for the near future. The considered parameters of the spectral grid are the effective temperature ($T_{\textrm{eff}}$ in K), the stellar surface gravity ($\log g$ in dex), the mean metallicity ([M/H] in dex) and the enrichment in $\alpha$-elements with respect to iron ([$\alpha$/Fe] in dex). 

The metallicity definition here is different to the classical use of [Fe/H] to denote the metallicity of a star. For [Fe/H] the metallicity is defined using the derived abundances from Fe lines only. MATISSE provides the opportunity to use all the available metal lines (those atoms heavier than He) to define a metallicity providing a global metallicity designated as [M/H]. The opportunity also exists with MATISSE to derive a global $\alpha$ element abundance using as many $\alpha$ element spectral lines as possible, where we assumed that in the generation of the synthetic spectra the abundances of the different $\alpha$ elements vary in lockstep. Throughout the AMBRE Project, the following chemical species are assumed to be $\alpha$-elements: O, Ne, Mg, Si, S, Ar, Ca and Ti, although for any selected wavelength region spectral features for all of these elements may not necessarily be present.

From the selected MARCS model atmospheres, the synthetic spectra were computed with the turbospectrum code (\citealt{Alvarez1998}, and further improvements by Plez) in plane-parallel and spherical geometry assuming hydrostatic and local thermodynamic equilibrium. Atomic lines have been recovered from the Vienna Atomic Line Database (in August 2009; \citealt{Kupka1999}). The molecular line list has been provided by B. Plez. It includes transitions from ZrO, TiO, VO, OH, CN, C2, CH, SiH, CaH, MgH and FeH with their corresponding isotopic variations (see \citet{Gustafsson2008}, for a list of references). %Collisional broadening by atomic hydrogen of several atomic lines is computed as in \citet{Barklem2000}. Other lines are broadened with the classical theory.  They have been computed by assuming the same solar abundances \citep{Grevesse2007} as the ones adopted by the MARCS group. 

This grid covers the following ranges of atmospheric parameters: $T_{\textrm{eff}}$ between 2\,500~K and 8\,000~K, $\log g$ from $-0.5$ to $+5.5$~dex, and [Fe/H] from $-5.0$ to $+1.0$~dex, although not all combinations of the parameters are available within the grid. The selected MARCS models have [$\alpha$/Fe]=$0.0$ for [M/H] $\ge$ $0.0$, [$\alpha$/Fe]=$+0.4$ for [M/H] $\le$ $-1.0$ and, in between, [$\alpha$/Fe]=$-0.25$x[M/H]. For the spectra computation from each of these MARCS models, we considered an [$\alpha$/Fe] enrichment from $-0.4$ to $+0.4$~dex with respect to the canonical values that correspond to the original abundances of the MARCS models. The final AMBRE synthetic spectra grid consists of 16783 flux normalized spectra \citep[See][for a complete description of this grid]{deLaverny2012}.

The microturbulence ($\xi$) is not a free parameter in this synthetic spectra grid. For the atmospheric models with high $\log g$ ($+3.5 \leq \log g \leq +5.5$) $\xi$ is set at 1.0~kms$^{-1}$. For low $\log g$ ($\log g < +3.0$) $\xi$ is set at 2.0~kms$^{-1}$, being typical values for dwarfs and giants respectively. These values reflect the MARCS model atmospheres configurations, and further details on the model selection is given in \citet{deLaverny2012}. Due to the microturbulence being hardwired into the synthetic grid it was not possible to carry out tests on the effects of variations in $\xi$ on the stellar parameter determination across the whole synthetic spectra grid. However the effects of changes in $\xi$ are most prominent for strong lines. By using the global metallicity [M/H] rather than [Fe/H] we expect the contribution from strong lines on the [M/H] to be negligible due to the significantly larger quantity of non-$\xi$ sensitive small metallic lines.

Section~\ref{sec:ext_error} describes the extensive testing of the pipeline that was carried out using observed spectra from which an external error on the resulting parameters is derived by comparison to literature values. To gauge the effect of microturbulence on the derived stellar parameters a key sample of 66 Main Sequence (MS) stars \citep{Bensby2003}, and two key samples of 16 Red Giant Branch (RGB) stars \citep{McWilliam1990,Hekker2007} were considered from within the test sample. 

The most noticeable effect within the MS sample was an underestimation of the $T_{\textrm{eff}}$ for the hotter MS stars, for which the constant $\xi$ of 1.0~kms$^{-1}$ underestimates the accepted value. Hence at $T_{\textrm{eff}} \sim 6500$~K, where $\Delta\xi \sim +0.8$kms$^{-1}$ the effect was $\Delta T_{\textrm{eff}} \sim -100$~K. This is well within the derived $T_{\textrm{eff}}$ external error of 120~K (see Section~\ref{sec:ext_error}).

The most noticeably effect in the RGB sample was on the derivation of $\log g$. At the base of the RGB, where the constant $\xi$ of 2.0~kms$^{-1}$ is an overestimation, the derived $\log g$ was observed to be overestimated by $\sim 0.3$~dex for $\Delta\xi \sim +0.7$kms$^{-1}$. At the RGB tip, where $\xi = 2.0$~kms$^{-1}$ is an underestimation, $\log g$ was observed to be underestimated by $\sim 0.3$~dex for $\Delta\xi \sim -0.7$~kms$^{-1}$. These variations are within the derived $\log g$ external error of 0.37~dex. Hence based on these samples, variations of $\xi$ from the assumed constant values do have an effect on the derived parameters that varies in magnitude with stellar evolutionary stage. However the external error derived from the entire reference sample takes account of them in a global sense (see Section~\ref{sec:ext_error}). 
%Regarding the assumed geometry in these calculations (model atmosphere and synthetic spectra), massless plane-parallel models have been considered for $+3.5$ $\leq$ log g $\leq$ $+5.5$ (in these cases, the atmospheric extension is negligible relative to the radius, and thus sphericity effects can be neglected). They have a microturbulent-velocity parameter of $1.0$~km/s. Spherical geometry has been considered when log g $\leq$ $3.0$ since sphericity effects may be important for low masses and/or low log g (see Heiter \& Eriksson, 2006). For these gravity, a mass of $1.0$~M$_\odot$ and a microturbulence parameter of $2.0$~km/s has been considered.
 
%Finally, it has to be noted that many models are missing (all the combination of the atmospheric parameters were not available) due to the approach to the Eddington flux limit or poor convergence (see \citealt{Gustafsson2008}).

\subsection{AMBRE:FEROS subgrid}
The FEROS spectra cover most of the optical wavelengths but only a subset of these wavelengths was required for the MATISSE analysis. In order to carry out the MATISSE training phase, where the $B_{\theta}(\lambda)$ vectors are generated, for the FEROS analysis the optimum resolution, wavelength regions, and sampling of the FEROS spectra were determined. The full AMBRE synthetic spectra grid was then adapted to the same specifications creating the AMBRE:FEROS subgrid that was used in the training phase to generate the AMBRE:FEROS $B_{\theta}(\lambda)$ vectors.

In this analysis, only $B_{\theta}(\lambda)$ functions computed from the direct numerical inversion of the correlation matrix \citep[see][]{Recio-Blanco2006} were considered. These functions are thus not optimized for very low SNR spectra. However, due to the large number of spectral features available in the selected spectral domains, it was not necessary to use the approximated $B_{\theta}(\lambda)$ functions that are computed with the Landweber algorithm as in \citet{Kordopatis2011}. As will be shown in Section~\ref{sec:int_error} the estimated internal errors based on the AMBRE:FEROS grid are indeed already very small (see Figure~\ref{fig:snr_interr}).

\subsubsection{Selection of wavelength regions}\label{sec:feros_wavelengths}
A detailed analysis of the full wavelength range of the FEROS spectra was undertaken in order to select wavelength regions that would provide the greatest amount of information for each of the derived parameters while also minimising the number of pixels in order to reduce computing time.

%The archived sample analysed in the AMBRE project contains a total of 21551 spectra ($\sim$7000 stars) obtained from October 2005 until December 2009. The archived sample has been reduced using the ESO FEROS pipeline (wavelength calibrated, merged and unmerged line spectra) prior to delivery to OCA. 
 
\begin{figure*}[!th]
%\centering
\begin{minipage}{90mm}
%\centering
\includegraphics[width=90mm]{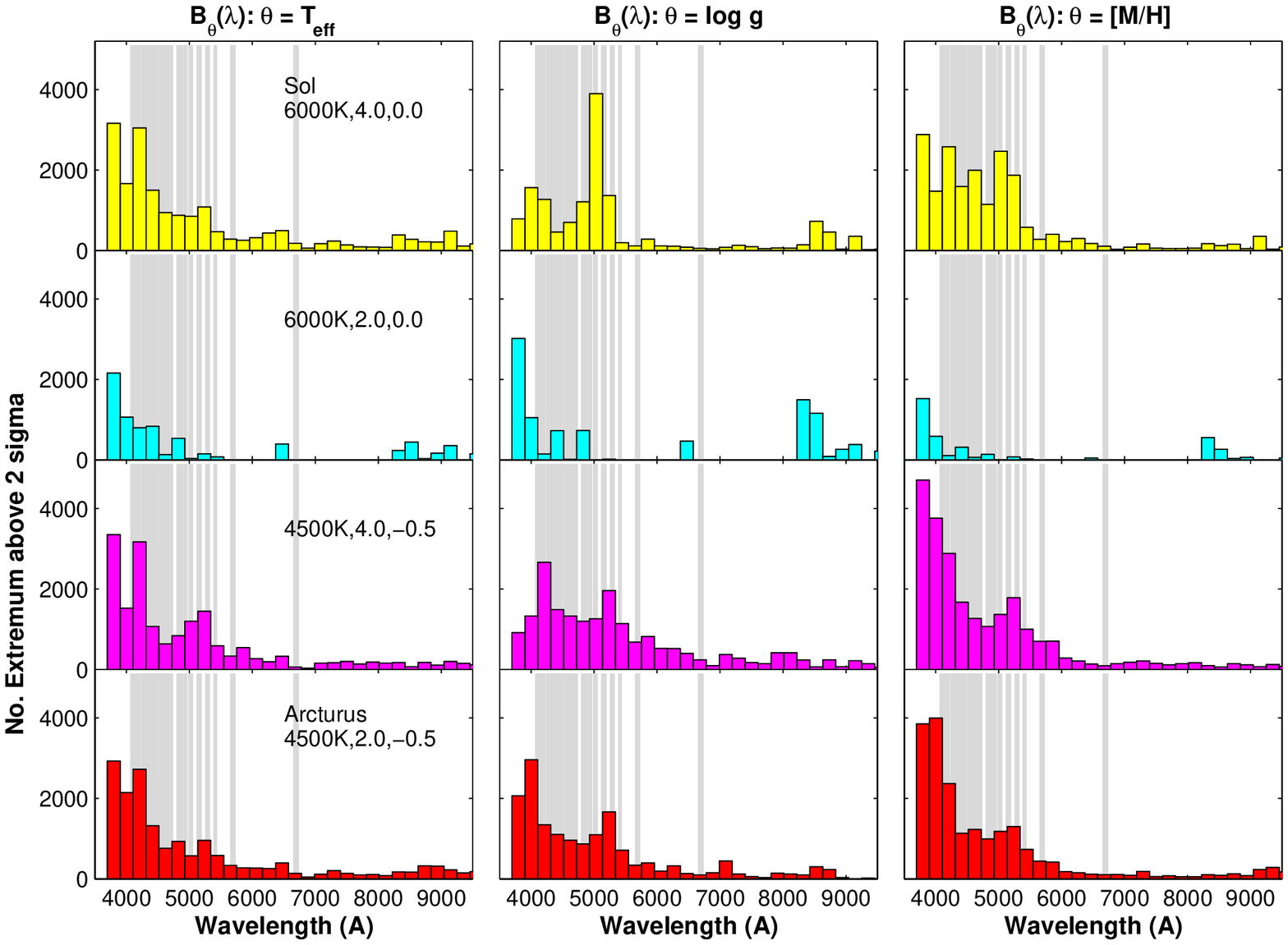}
%\vspace{-2cm}
\caption{$B_{\theta}(\lambda)$ Sensitivity: Number of extrema above 2~$\sigma$ in $\sim$200~\AA\ bins against wavelength for four sets of stellar parameters: the Sun; $T_{\textrm{eff}} = 6000$~K, $\log g = 2$, [M/H]~=~0.0; Arcturus; and $T_{\textrm{eff}} = 4500$~K, $\log g = 4$, [M/H]~=~--0.5. The $B_{\theta}(\lambda)$ vectors from left to right are: $\theta$ = $T_{\textrm{eff}}$, $\log g$ and [M/H]. The grey stripes represent the final wavelength regions selected for the AMBRE:FEROS analysis (see Table~\ref{tab:feros_wavelengths}).}\label{fig:bf_sensitivity}
\end{minipage}
%\end{figure}
\hfill
%\begin{figure}[!h]
%\centering
\begin{minipage}{90mm}
\vspace{-1.1cm}
%\centering
\includegraphics[width=90mm,height=67mm]{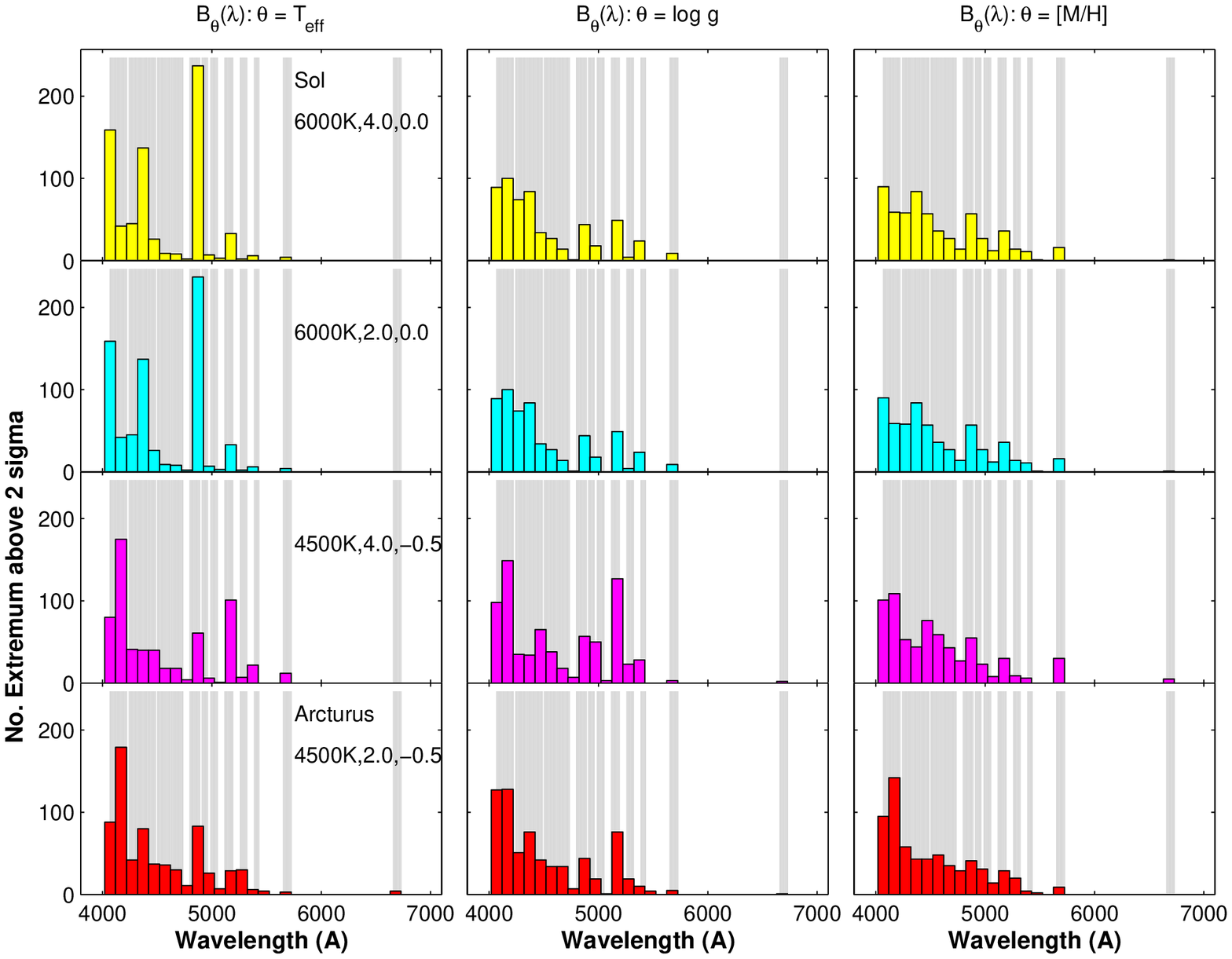}
\caption{As for Figure~\ref{fig:bf_sensitivity} but for the corresponding $B_{\theta}(\lambda)$ vectors of the AMBRE:FEROS synthetic spectra grid. The y-axis is scaled down compared to Figure~\ref{fig:bf_sensitivity}.}\label{fig:afbf_sensitivity}
\end{minipage}
\end{figure*}

FEROS disperses the spectra into 39 orders of varying wavelength interval from $\sim$3500~\AA\ to $\sim$9200~\AA\ at high resolution (R$\sim$48,000). The optimum spectral regions for the determination of the stellar parameters were selected taking into account the need to avoid excessive computing time and to avoid low spectral information region. To this end, the signal-to-noise (SNR) per pixel (0.03 or 0.06~\AA\ per pixel for FEROS) as a function of the wavelength was derived for each archived spectra by determining the SNR profile of each spectral order and thus providing a SNR profile over the entire spectral domain. Wavelengths regions with low SNR were rejected, typically the start and end of each order. The wavelengths regions affected by sky absorption and telluric features were also rejected, and also regions where continuum placement proved too difficult owing to wide spectral features in the region. Some key spectral features, such as H$_{\alpha}$ and the Ca~II H \& K lines, were also discarded as these features were found to be poorly synthesised based on our current understanding of stellar atmospheres as well as being difficult to normalise automatically.

The remaining spectral regions were then investigated for their intrinsic sensitivity to three of the four stellar parameters to be determined by MATISSE: $T_{\textrm{eff}}$; $\log g$; and [M/H]. This was carried out using preliminary $B_{\theta}(\lambda)$ vectors that covered the full optical domain at very high resolution generated using a reduced synthetic grid. Due to the underlying equation for calculating the $B_{\theta}(\lambda)$ vectors, the high resolution $B_{\theta}(\lambda)$ vectors could not be simply degraded to lower resolutions to investigate the sensitivity, as this would not accurately reflect the information at that resolution. However the high resolution $B_{\theta}(\lambda)$ vectors are useful to give a general sense of the sensitivity.
   
For stellar parameters at: 

\begin{enumerate}
 \item $T_{\textrm{eff}}$=6000~K, $\log g$=4.0~dex, [M/H]=0.0~dex (Sun);
 \item $T_{\textrm{eff}}$=6000~K, $\log g$=2.0~dex, [M/H]=0.0~dex;
 \item $T_{\textrm{eff}}$=4500~K, $\log g$=2.0~dex, [M/H]=--0.5~dex (Arcturus);
 \item $T_{\textrm{eff}}$=4500~K, $\log g$=4.0~dex, [M/H]=--0.5~dex,
\end{enumerate}

\noindent the $B_{\theta}(\lambda)$ vectors at the same $T_{\textrm{eff}}$, $\log g$, and [M/H] have been analysed for their extrema distribution. This selection explores the sensitivity of both dwarfs and giants  that are similar to the two standard stars, the Sun and Arcturus. The panels in Figure~\ref{fig:bf_sensitivity} show how the spectral domain changes in sensitivity with the different stellar parameters. The standard deviation ($\sigma$) for each $B_{\theta}(\lambda)$ vector was calculated as the spread of the pixel values in the $B_{\theta}(\lambda)$ vector. Then for each $B_{\theta}(\lambda)$ vector the pixels outside 2~$\sigma$ were binned in $\sim$200~\AA\ bins over the wavelength range from $\sim$4000~\AA\ to $\sim$9500~\AA. Hence the bars in Figure~\ref{fig:bf_sensitivity} show the number of pixels per bin which have a high sensitivity to the respective $\theta$. Clearly the bluer wavelengths show the greatest sensitivity to all three parameters, reflecting the greater quantity of spectral features in the blue.

\begin{table}[htbp]
\caption{FEROS \'{e}chelle order, starting wavelength and finishing wavelength for each region used in the analysis of the FEROS spectra.}\label{tab:feros_wavelengths}
\begin{center}
\begin{tabular}{ccccccc}
\hline\hline
%\'{E}chelle & $\lambda$ Min & $\lambda$ Max &  & \'{E}chelle & $\lambda$ Min & $\lambda$ Max \\ 
%Order & \AA & \AA &  & Order & \AA & \AA \\ \hline
%34 & 6660 & 6730 &  & 49 & 4585 & 4670 \\ 
%40 & 5655 & 5725 &  & 50 & 4505 & 4580 \\ 
%42 & 5390 & 5430 &  & 51 & 4405 & 4492 \\ 
%43 & 5260 & 5320 &  & 52 & 4300 & 4405 \\ 
%44 & 5120 & 5190 &  & 53 & 4245 & 4300 \\ 
%45 & 4990 & 5055 &  & 54 & 4140 & 4220 \\ 
%46 & 4910 & 4960 &  & 55 & 4075 & 4140 \\ 
%47 & 4800 & 4890 &  & 56 & 4019 & 4075 \\ 
%48 & 4674 & 4740 &  &  &  &  \\ 
Region & $\lambda$ min & $\lambda$ max &  & Region & $\lambda$ min & $\lambda$ max \\ 
 & \AA & \AA &  &  & \AA & \AA \\ \hline
1 & 6660 & 6730 &  & 10 & 4585 & 4670 \\ 
2 & 5655 & 5725 &  & 11 & 4505 & 4580 \\ 
3 & 5390 & 5430 &  & 12 & 4405 & 4492 \\ 
4 & 5260 & 5320 &  & 13 & 4300 & 4405 \\ 
5 & 5120 & 5190 &  & 14 & 4245 & 4300 \\ 
6 & 4990 & 5055 &  & 15 & 4140 & 4220 \\ 
7 & 4910 & 4960 &  & 16 & 4075 & 4140 \\ 
8 & 4800 & 4890 &  & 17 & 4019 & 4075 \\ 
9 & 4674 & 4740 &  &  &  &  \\ 
\end{tabular}
\end{center}
\end{table}

This sensitivity to $\theta$ for different ranges in $\theta$ was used to select the wavelengths regions in the FEROS spectra to be used in the AMBRE analysis. As the greatest sensitivity was located towards the blue, regions were selected, where possible, to capture this sensitivity. The grey regions in Figure~\ref{fig:bf_sensitivity} represent the final wavelength regions selected for AMBRE:FEROS which are listed in Table~\ref{tab:feros_wavelengths}. The wavelengths include key spectral features such as the magnesium triplet at $\sim$5160~\AA, H$_{\beta}$ (4861~\AA), H$_{\gamma}$ (4340~\AA), H$_{\delta}$ (4101~\AA), a CH bandhead (4305~\AA) and a CN bandhead (4142~\AA) as well as the spectral features for many atomic elements. This wealth of information provided sufficient sensitivity over the range of required stellar parameters in anticipation of the range of spectral types that were potentially included within the FEROS archived sample.

For confirmation of the $B_{\theta}(\lambda)$ sensitivity, Figure~\ref{fig:afbf_sensitivity} was generated by the same process as Figure~\ref{fig:bf_sensitivity} but using the corresponding $B_{\theta}(\lambda)$ vectors for the final AMBRE:FEROS grid. Therefore Figure~\ref{fig:afbf_sensitivity} reflects the sensitivity of the $B_{\theta}(\lambda)$ vectors at the actual resolution of the AMBRE:FEROS analysis. Due to the reduced number of points in sampling and resolution the 2~$\sigma$ limit is significantly reduced. While some regions show there are subtle differences in the sensitivity of some regions, overall the distribution of greater sensitivity in the blue is in good agreement with the sensitivity at high resolution.  

\subsubsection{Convolution \& sampling of synthetic spectra grid}\label{sec:feros_gridconvolution}
At the high resolution of R$\sim$48,000, the FEROS spectra have typical wavelength sampling ($\Delta\lambda$) of 0.03~\AA\ (1$\times$1 binning) or 0.06~\AA\ (1$\times$2 binning). The final selected wavelengths corresponded to a total of $\sim$1500~\AA\ and at this sampling this translated to a (maximum) total of $\sim$50,000 pixels. Creating the FEROS synthetic spectra subgrid of $\sim$16,000 spectra at this resolution and sampling would result in excessively large memory and computing requirements when creating the $B_{\theta}(\lambda)$ vectors in the training phase, and also when loading these vectors during the MATISSSE analysis. It was necessary to reduce the resolution and sampling to a more reasonable pixel total, but without sacrificing the key spectral information. The most reasonable number of pixels per spectra was determined to be $\sim$15,000 in order to ease the use of the available computing power. 

By optimising the memory and spectral information requirements, the final specifications for the FEROS synthetic spectra subgrid was determined to be 11890 pixels with R$\sim$15,000 at $\lambda \sim 4500$~\AA\ with a sampling of 0.1~\AA\ over $\sim$1500~\AA. Hence the full AMBRE synthetic spectra grid \citep{deLaverny2012} was then sliced to the wavelengths specified in Table~\ref{tab:feros_wavelengths}, convolved by a Gaussian kernel with constant $\sigma$ and resampled to the optimised FEROS pixels. The training phase was then carried out whereby the $B_{\theta}(\lambda)$ vectors were generated for use in the MATISSE analysis of the FEROS archived spectra.

\section{FEROS analysis pipeline}\label{sec:FEROSpipeline}
\begin{figure}[!ht]
\centering
\begin{minipage}{90mm}
\centering
\includegraphics[width=85mm]{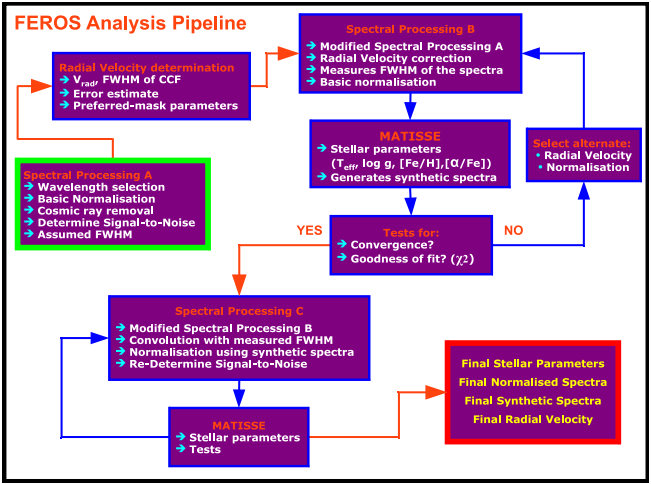}
\caption{The FEROS Analysis Pipeline: The key stages are displayed in order of analysis. Spectral Processing A and Spectral Processing B carry out testing of spectra quality and preliminary parameter determination, including calculation of the radial velocity and spectral FWHM. In Spectral Processing C robust iterative procedures are carried out resulting in the final stellar parameters and normalised spectra. See text for detailed explanation of each stage.}\label{fig:ferospipe}
\end{minipage}
\end{figure}

A complex analysis pipeline (written in shell, Java and IDL) has been built to wavelength slice, radial velocity correct, normalise and convolve the FEROS spectra and then feed them into MATISSE for the determination of their stellar parameters. Figure~\ref{fig:ferospipe} shows a flowchart of the key stages in the analysis pipeline. The pipeline has been developed so that it can be easily adapted to the remaining three instrument datasets of ESO archived spectra. The following sections outline the procedures carried out for each of the key stages in the pipeline.

\subsection{Spectral Processing A: Spectra processing for radial velocity determination}\label{sec:SPA}
Spectral Processing A (SPA) is the initial stage that prepares the FEROS spectra in order to determine the radial velocities. The procedures used here are also used in the later stages of the pipeline. Key flags were defined within the pipeline to be attached to any spectra that satisfied the conditions of the flag. The flags were for `Faulty Spectra' (e.g. missing spectral orders), `Extreme Emission Features' (spectral emission features in the blue with width greater than 50~\AA), `Excessive Noise' (extremely noisy spectra with the number of negative flux pixels $> 25\%$) and `Poor Normalisation' (spectra with a total flux above $1.2 > 10\%$). A spectrum may satisfy multiple flags, particularly in the case of excessive noise which naturally led to poor normalisation. Such types of spectra were unlikely to produce valid results in the MATISSE analysis and as such were rejected from the full analysis. Flags were also implemented to identify `Large Emission Features' (width greater than 1.2~\AA) and the presence of odd features that were potentially `Instrumental Relics'. However these flags alone were not enough to reject a spectrum from the full analysis. All of the quality flags were accounted for in the generation of the final dataset of results for ESO, which is explained in Section~\ref{sec:AFstellarparams}. 

The first reduction process extracts the required wavelengths (Table~\ref{tab:feros_wavelengths}) from the spectra and then normalises them. However no convolution is carried out so the original resolution is maintained. A hot star spectrum (almost featureless) in the FEROS sample was identified and used to determine a continuum profile for each wavelength region. This profile is divided out of the spectra removing the residual instrument profile and performing a first rough normalisation to place the continuum near unity. At this stage the spectra must necessarily be considered as unknown in stellar parameters and hence the normalisation process must be carried out assuming that key pre-selected continuum regions need to be normalised to unity in intensity. This assumption may not in fact correctly normalise each spectrum. However in these initial stages it is the best, albeit crude, estimate available.

Up to three continuum regions for each wavelength region were pre-selected by an investigation of synthetic spectra and a sample of FEROS spectra covering a range of stellar parameters. For the normalisation, a simple linear profile is constructed for each wavelength region by determination of the mean signal in the respective continuum regions. This profile is divided out of the wavelength region and then the region undergoes cleaning for cosmic rays. The region is scanned for emission features greater than a factor of 1.5 above the noise on the continuum (unity). These features are tested for width and those less than 1.2~\AA\ (40~pixels) are cleaned to unity. Features larger than 1.2~\AA\ (40~pixels) in width are flagged as large emission features and are not cleaned. This limit of 1.2~\AA\ (40~pixels) was determined by an examination of many examples of cosmic rays in the FEROS spectra to get a sense of the number of pixels over which they could be dispersed.

Each wavelength region is treated individually for each spectrum, and the regions are then combined into one vector. This merged spectrum is tested for the goodness of the normalisation and flagged if the normalisation is poor. The resulting spectrum is used in the determination of the radial velocity. At this stage the original resolution is retained. None of the FEROS spectra were rejected at this stage in case later testing and processing recovered a previously rejected spectra (i.e. mis-calculated radial velocity is corrected). 

\subsection{Radial velocity determination}\label{sec:vrad_determination}
A completely automatic radial velocity (V$_{rad}$) programme has been established that can analyse spectra across a very wide range of stellar parameters and is based on a cross-correlation algorithm that compares the normalised observed spectrum to binary masks \citep[C. Melo, private communication,][]{Melo2001}. To complement this programme a procedure was developed at OCA that computes a set of binary masks to match the observations in terms of wavelength range and resolution, and are used as input to the radial velocity programme. Masks covering the parameter space of the AMBRE:FEROS analysis were computed using synthetic spectra taken from the high resolution, full optical domain, AMBRE synthetic spectra grid (see Section~\ref{sec:ambre_synthgrid}). Six masks for stars hotter that 8000~K were also computed using synthetic spectra from the POLLUX database \citep{Palacios2010}. For the AMBRE:FEROS analysis 51 masks were computed at the stellar parameters given in Table~\ref{tab:vrad_masks}. Also listed are 5 standard masks for dwarf stars that were supplied with the radial velocity programme (C. Melo, private communication) and used in the AMBRE:FEROS analysis.

\begin{table*}[bht]
\caption{List of the stellar parameters for the binary masks used in the AMBRE:FEROS radial velocity determination that were built from the full AMBRE synthetic grid (45 masks) and the POLLUX database (6 masks). Five standard masks for dwarf stars provided by C.~Melo are also listed.}\label{tab:vrad_masks}
\begin{center}
{\tiny
\begin{tabular}{cccc|cccc|cccc|cccc}
\hline\hline
% \multicolumn{16}{c}{Stellar Parameters for AMBRE:FEROS Masks built from the AMBRE Synthetic Grid} \\ 
$T_{\textrm{eff}}$ & $\log g$ & [M/H] & [$\alpha$/Fe] & $T_{\textrm{eff}}$ & $\log g$ & [M/H] & [$\alpha$/Fe] & $T_{\textrm{eff}}$ & $\log g$ & [M/H] & [$\alpha$/Fe] & $T_{\textrm{eff}}$ & $\log g$ & [M/H] & [$\alpha$/Fe] \\ 
\hline
2500 & 0.5 & -2.50 & 0.40 & 4000 & 4.5 & -2.00 & 0.40 & 6500 & 1.0 & 0.00 & 0.00 & 9000 & 4.5 & 0.00 & 0.00 \\ 
2500 & 0.5 & 0.00 & 0.00 & 4000 & 4.5 & 0.00 & 0.00 & 6500 & 2.0 & -2.00 & 0.40 & 10000 & 4.5 & 0.00 & 0.00 \\ 
2500 & 2.5 & -2.00 & 0.40 & 5000 & 0.5 & -2.00 & 0.40 & 6500 & 2.0 & 0.25 & 0.00 & 12000 & 4.5 & 0.00 & 0.00 \\ 
2500 & 2.5 & 0.00 & 0.00 & 5000 & 0.5 & 0.00 & 0.00 & 6500 & 4.5 & -2.00 & 0.40 & 15000 & 4.5 & 0.00 & 0.00 \\ 
2500 & 4.5 & 0.00 & 0.00 & 5000 & 2.5 & -2.00 & 0.40 & 6500 & 4.5 & 0.00 & 0.00 & 30000 & 3.2 & 0.00 & 0.00 \\ 
3100 & 0.5 & -2.00 & 0.40 & 5000 & 2.5 & 0.00 & 0.00 & 7500 & 1.5 & -0.25 & 0.10 & 37500 & 3.5 & 0.00 & 0.00 \\ 
3100 & 0.5 & 0.00 & 0.00 & 5000 & 4.5 & -2.00 & 0.40 & 7500 & 1.5 & 0.25 & 0.00 &  &  &  &  \\ 
3100 & 2.5 & -2.50 & 0.40 & 5000 & 4.5 & 0.00 & 0.00 & 7500 & 2.5 & -2.00 & 0.40 &  &  & &  \\ 
3100 & 2.5 & 0.00 & 0.00 & 5750 & 0.5 & -0.25 & 0.10 & 7500 & 2.5 & 0.00 & 0.00 &  &  &  &  \\ 
3100 & 4.5 & -2.00 & 0.40 & 5750 & 0.5 & 0.50 & 0.00 & 7500 & 4.5 & -2.00 & 0.40 &   \multicolumn{2}{r}{Std Masks} & $T_{\textrm{eff}}$ &  \\ 
3100 & 4.5 & 0.00 & 0.00 & 5750 & 2.5 & -2.00 & 0.40 & 7500 & 4.5 & 0.00 & 0.00 &  & R37M4 & 4048 &  \\ 
4000 & 0.5 & -2.00 & 0.40 & 5750 & 2.5 & 0.00 & 0.00 & 8000 & 2.5 & -2.00 & 0.40 &  & R00K0 & 4900 &  \\ 
4000 & 0.5 & 0.00 & 0.00 & 5750 & 4.5 & -2.00 & 0.40 & 8000 & 2.5 & 0.00 & 0.00 &  & R37K0 & 4900 &  \\ 
4000 & 2.5 & -2.00 & 0.40 & 5750 & 4.5 & 0.00 & 0.00 & 8000 & 4.5 & -2.50 & 0.40 &  & R50G2 & 6300 &  \\ 
4000 & 2.5 & 0.00 & 0.00 & 6500 & 1.0 & -1.50 & 0.40 & 8000 & 4.5 & 0.00 & 0.00 &  & R37F0 & 7400 &  \\ 
\end{tabular}
}
\end{center}
\end{table*}

For each spectrum a cross-correlation function (CCF) ranging from $-500$ to $+500$~kms$^{-1}$ was calculated for each of the 56 masks. The CCF was calculated in radial velocity steps ($\Delta$V$_{rad}$) based on the specified wavelength step ($\Delta\lambda$) of the spectrum. For spectra with $\Delta\lambda = 0.03 \AA$ the CCF step was calculated to be $\Delta$V$_{rad} \sim 1.8$~kms$^{-1}$, and for $\Delta\lambda = 0.06 \AA$ the CCF step was $\Delta$V$_{rad} \sim 3.6$~kms$^{-1}$. For each CCF two separate gaussian fits were made in order to determine the minimum of the profile, and hence the radial velocity. The first fit used the majority of the profile, while the second fit used a section of the profile centred near the mininum with a width equal to the full-width-at-half-maximum (FWHM) of the profile.

\begin{figure}[h]
\centering
\begin{minipage}{90mm}  %for one column use 90, 90
%\centering
%\hspace{-0.3cm}
\includegraphics[width=95mm]{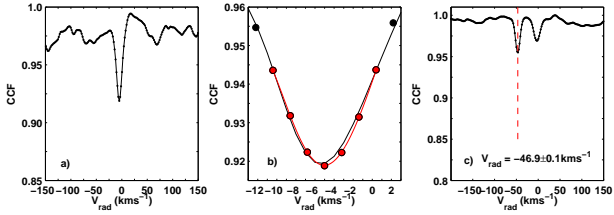}
\caption{CCFs produced for calculating radial velocities. a) Noisy CCF with an asymmetric profile. b) Core of profile showing skewed full profile gaussian fit (black). The secondary gaussian fit (red) provides a better estimate. c) CCF of the spectroscopic binary, HD135728, showing two profiles.}\label{fig:skew_ccf}
\end{minipage}
\end{figure}

Figure~\ref{fig:skew_ccf}a. shows an example of a noisy CCF although the key profile is distinct. However, as seen in Figure~\ref{fig:skew_ccf}b., the fit to the full profile (black) provides a minimum skewed away from the actual minimum due to asymmetries of the full profile. The use of the central core of the profile provides a more accurate fit (red). Figure~\ref{fig:skew_ccf}c. shows an example of a CCF for which two profiles are observed. This is the spectroscopic binary HD135728. The pipeline selects the most prominent of the two profiles in the determination of the radial velocity. The CCF seems the most likely tool with which to be able to identify the spectra of spectroscopic binaries. However at this stage we have not developed such a routine and so binaries are not specifically detected in the pipeline. The stellar parameters for such spectra are likely to be poorly determined and so will be at least identified as having large associated errors from the radial velocity and stellar parameter analysis. Quality control flags, such as the $\chi^2$, may also be used to help identify spectroscopic binaries.

\begin{figure}[!bh]
\centering
\begin{minipage}{90mm}  %for one column use 90, 90
\centering
\includegraphics[width=95mm]{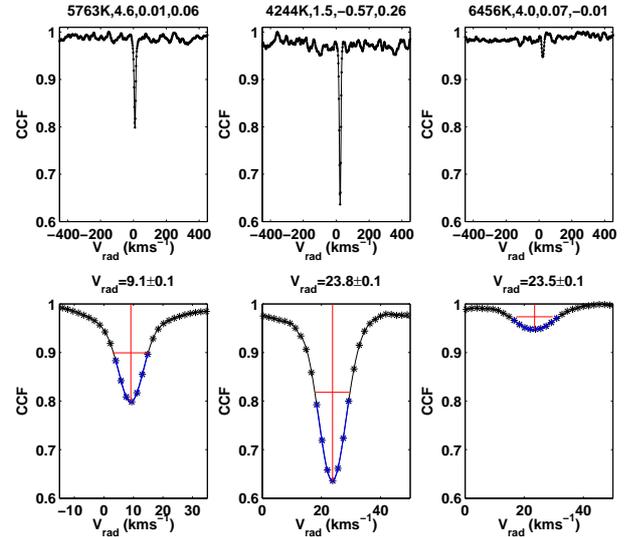}
\caption{Examples of cross-correlation functions of the best fit masks for three of the FEROS spectra. The first row shows the full CCFs while the second row shows the profiles from which the radial velocities were determined. The final AMBRE:FEROS stellar parameters ($T_{\textrm{eff}}, \log \ g$, [M/H], [$\alpha$/Fe]) and radial velocities are stated for each spectrum. The red lines are used to calculate the contrast and FWHM for each profile.}\label{fig:sap_vradccf}
\end{minipage}
\end{figure}

Figure~\ref{fig:sap_vradccf} shows the CCF corresponding to the best fit mask for three FEROS spectra at different stellar parameters. The full CCF for each spectrum is shown in the first row, while the second row shows more clearly the profile used to calculate the radial velocity. The full CCF can show varying degrees of noise and gradient depending on the quality of the spectra as is shown in each of these three examples. The respective profiles used to determine the radial velocity reflect the characteristics of the different spectra. The two cooler stars have very prominent inverted peaks, reflecting the numerous spectral lines available with which to make the cross-correlation. The hotter star, with fewer spectral lines, has a much less pronounced profile. These differences can be quantified by the measurement of the contrast of the profile.

%\vspace{-0.5cm}

The contrast was defined here as the amplitude of the profile multiplied by 100 (hence a percentage) where the amplitude of the profile is the length of the vertical red line as shown in each example profile in Figure~\ref{fig:sap_vradccf}. The amplitude of the CCF, the continuum placement of the CCF and their associated errors are calculated automatically when using the GAUSSFIT routine in IDL. The sign of the amplitude does change depending on whether the CCF has an absorption or an emission profile. The expected profile in this analysis is an absorption profile.

The two conditions that were required for a profile to have been well fitted by a gausssian were first: the contrast must be less than zero ensuring an absorption profile; and second, the error on the amplitude must be less than 20$\%$ of the amplitude ensuring that the CCF profile dominates above the noise. CCFs that satisfied these conditions were considered to be well-defined. Finally the error on the radial velocity ($\sigma_{Vrad}$) was calculated using the prescription outlined in \citet{Tonry1979}. This prescription makes use of the relative heights of the primary and secondary peaks in the CCF profile hence $\sigma_{Vrad}$ reflects the goodness of the definition of the primary peak.

The radial velocity programme carried out the calculations for each of the 56 masks, resulting in 56 determinations of the radial velocity for each spectrum. The errors derived from the quality of the gaussian fit and $\sigma_{Vrad}$ were used to select the best fit radial velocity. Typically the majority of these determinations were in good agreement at the correct radial velocity. However for masks with parameters far from the true parameters of the star, a radial velocity would be determined that was incorrect but sometimes had sufficiently small errors such that it would be incorrectly selected as the best fit, if the smallest error was the only selection criteria. To avoid this a binning procedure was implemented as follows that discarded such outliers.

For a single spectrum, all of the V$_{rad}$ determinations with well-defined CCFs were binned by V$_{rad}$ into a maximum of 3 bins in correlation with the overall spread of the V$_{rad}$ values. The bin size was set in each instance by the difference in the maximum and minimum V$_{rad}$ values divided by the number of bins. The bin with the largest number of members was retained and the outlying bins were discarded. The V$_{rad}$ within accepted bin with the smallest $\sigma_{Vrad}$ was selected as the final V$_{rad}$. In the case of an equal number of members within 2 or more bins, the bin with the smallest spread in V$_{rad}$ values (so best agreement) was selected. Typically, most of the radial velocities for a spectrum were in good agreement, so the maximum bin was easily identified. Otherwise the radial velocities were highly dispersed, and this would be reflected in the $\sigma_{Vrad}$ of the final selected value. 

The errors on the gaussian fit of the CCF and the $\sigma_{Vrad}$ values were used later in the analysis pipeline to identify problematic spectra and select alternate radial velocities where necessary, and also as quality criteria for the final set of results to be delivered to ESO.

The S$^4$N \citep{Allende-Prieto2004} and \citet{Crifo2010} libraries provided good samples of stars with which to validate the radial velocities determined in the AMBRE:FEROS pipeline. The S$^4$N library comprises of 118 F and G dwarf stars for which a detailed analysis was carried out using high resolution high SNR spectra. All four parameters ($T_{\textrm{eff}}, \log \ g$, [Fe/H] and [$\alpha$/Fe]) as well as radial velocities are available for each of the S$^4$N stars. \citet{Crifo2010} is the preliminary list of 1420 candidate standard stars to be used to calibrate the radial velocities obtained with Gaia RVS. Section~\ref{sec:ext_error} gives further discussion on the use of these libraries to validate the AMBRE:FEROS results.  

\begin{figure}[h]
\centering
\begin{minipage}{90mm}  %for one column use 90, 90
\centering
\includegraphics[width=98mm]{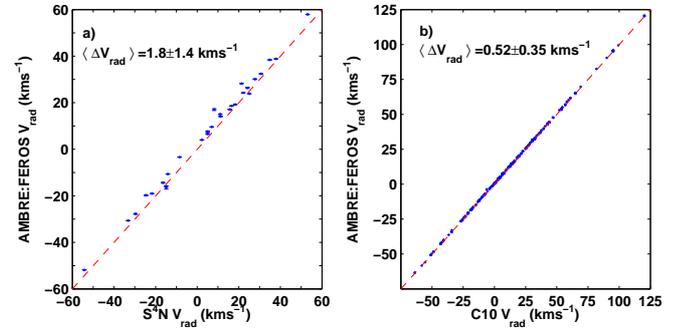}
\caption{Comparison of AMBRE:FEROS radial velocities: a) S$^4$N library for 29 stars (338 spectra); and b) \citet{Crifo2010} for 158 stars (318 spectra).}\label{fig:s4ncrifo_vrad}
\end{minipage}
\end{figure}

Within the FEROS archived dataset 30 stars (338 spectra) were found that are also within the S$^4$N library, and 183 stars (411 spectra) were found that are also within \citet{Crifo2010}. Figure~\ref{fig:s4ncrifo_vrad}a and b compare the S$^4$N and \citet{Crifo2010} radial velocities values with those determined for AMBRE:FEROS. The mean and standard deviation of the differences between the two sets for each sample are also shown. There is a small offset in both sets with the S$^4$N having a greater offset of 1.8$\pm$1.4~kms$^{-1}$ compared to \citet{Crifo2010} of 0.52$\pm$0.35~kms$^{-1}$. The FEROS spectra were sampled at either 0.03~\AA\ or 0.06~\AA\ from the ESO:FEROS reduction pipeline corresponding to expected V$_{rad}$ accuracies of 1.6~kms$^{-1}$ and 3.3~kms$^{-1}$ respectively. Hence these offsets indicate very good agreement for both samples, particularly for the \citet{Crifo2010} values.

\begin{figure}[h]
\centering
\begin{minipage}{90mm}  %for one column use 90, 90
\centering
\includegraphics[width=80mm]{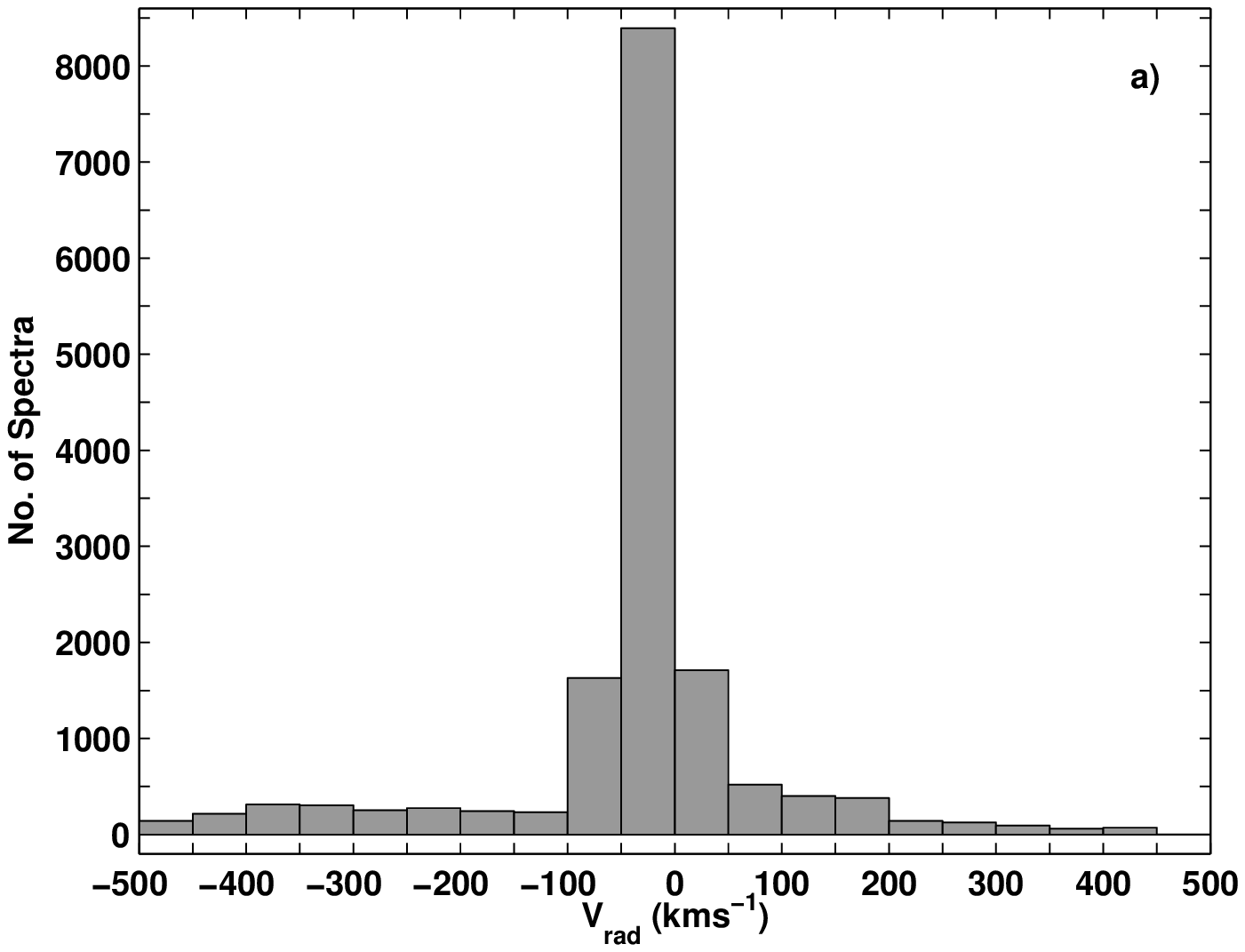}
\includegraphics[width=80mm]{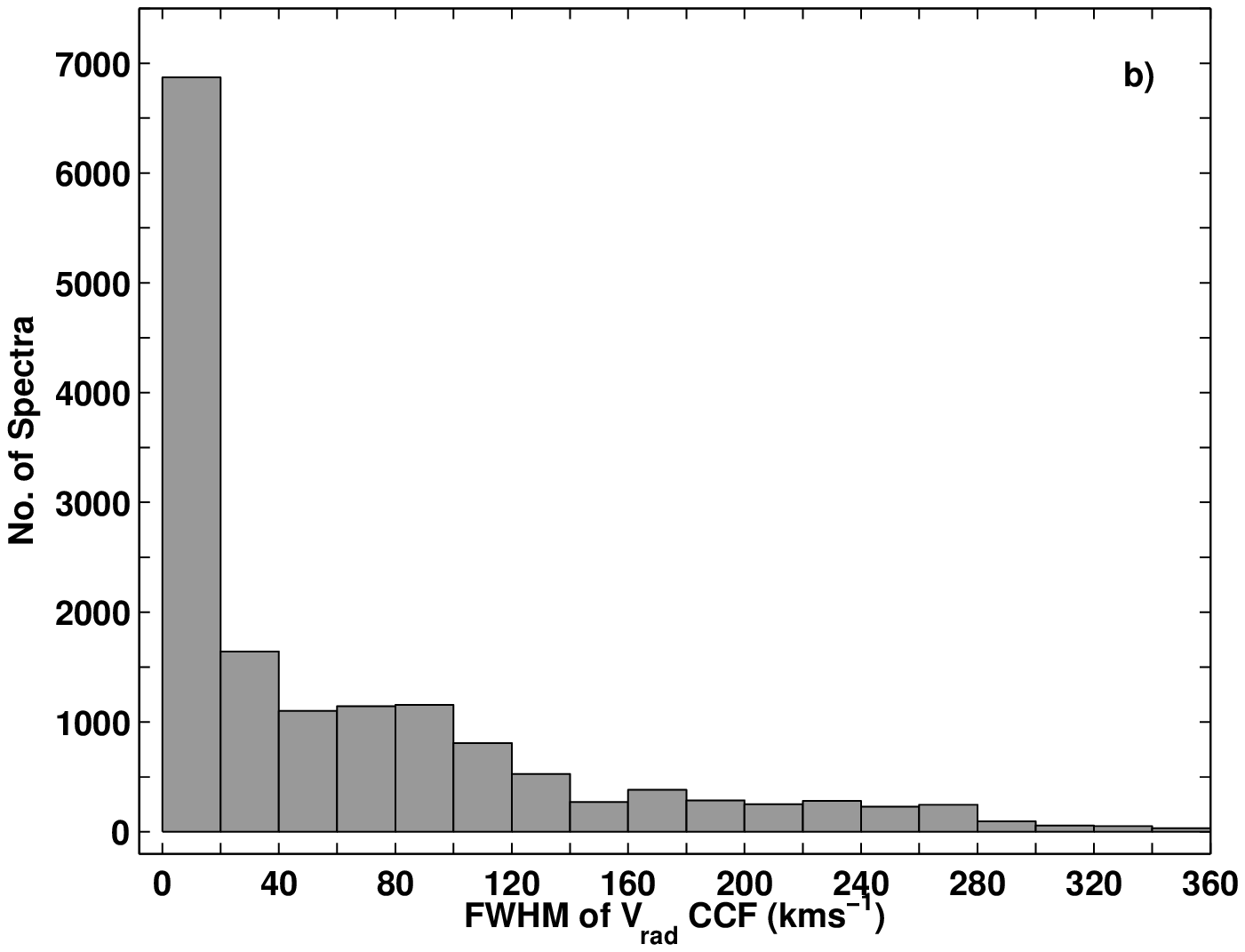}
\caption{Histogram of the radial velocity values calculated for all of the FEROS archived spectra defined as good spectra with well-defined CCFs: a) V$_{rad}$, and b) FWHM of the V$_{rad}$ CCF.}\label{fig:hist_vradvals}
\end{minipage}
\end{figure}

All of the FEROS spectra were analysed for their radial velocity. However for 4217 spectra ($19.6\%$) poorly defined or non-standard cross-correlation functions lead to unreliable estimates of the radial velocity. This is most likely due to some peculiar or non-stellar properties of the spectra (i.e. nova). Of these 4217 spectra, 549 also failed the conditions of the rejection flags as defined in Section~\ref{sec:SPA}. 

Figure~\ref{fig:hist_vradvals}a is a histogram of the radial velocities calculated for the FEROS spectra defined as good quality with well-defined CCF (15513 spectra). The majority of the spectra have radial velocities between $-100$ and $+50$~kms$^{-1}$. Figure~\ref{fig:hist_vradvals}b is a histogram of the FWHM of the CCF, calculated from the full CCF profile, for the same sample of spectra. It shows that the majority of spectra returned a FWHM of less than 50~kms$^{-1}$. The effects of the CCF on the selection of the final dataset will be discussed in Section~\ref{sec:AFstellarparams}. %Figures~\ref{fig:hist_vradvals}c and d show histograms for the calculated error \citep{Tonry1979} on the V$_{rad}$ values, split into V$_{rad}$ error~$<10$~kms$^{-1}$ and V$_{rad}$ error~$>10$~kms$^{-1}$ respectively. The majority of the spectra have a low error ($<0.5$~kms$^{-1}$). However there are a significant number with errors greater than 10. The effect of the V$_{rad}$ error on the selection of the final set of parameter will also be discussed in Section~\ref{sec:AFstellarparams}.

\subsection{Spectral Processing B: Spectra reduction, convolution and first parameter estimation}\label{sec:SPB}
Spectral Processing B (SPB) proceeds under two iterations through the normalisation procedure and the MATISSE analysis. These are necessary as key tests on the normalisation and radial velocity correction are carried out between the two iterations allowing alterations to be made that provide better first estimates of the stellar parameters for the remaining pipeline procedures. For both iterations the original archived spectra are re-analysed using the same procedures as outlined in SPA in terms of wavelength selection, cosmic ray cleaning and normalisation. However in both iterations of SPB the measured V$_{rad}$ correction is automatically applied to each spectrum prior to normalisation. This allows each spectrum to be shifted to the laboratory rest frame at which the synthetic grid, and so the $B_{\theta}(\lambda)$ vector functions, have been calculated. 

The AMBRE:FEROS synthetic grid was set to a resolution lower than the resolution of the observed FEROS spectra. In order to convolve the observed spectra to the appropriate resolution three factors needed to be taken into account. First, the observed spectra are observed with constant resolving power, which corresponds to increasing full-width-at-half-maximum (FWHM) of the spectral features with wavelength. Second, the synthetic spectra grid was convolved using a gaussian profile with constant FWHM hence the synthetic spectra have constant spectral FWHM with wavelength, not a constant resolution. Third, the construction of the full AMBRE synthetic spectra grid did not include astrophysical broadening such as those due to $Vsini$ and $\xi$ \citep{deLaverny2012}. Hence a straight-forward convolution using the nominal instrument FWHM and the FWHM applied to obtain the AMBRE:FEROS synthetic spectra grid would result in convolved observed spectra that had increasing FWHM with wavelength in disagreement with the synthetic spectra. Also for stars with astrophysical broadening greater than the broadening due to the resolving power of the instrument the spectra would be too broad for the appropriate parameter range of synthetic spectra.

Therefore a procedure was implemented that measured the spectral FWHM for each of the 17 selected wavelength regions (see Table~\ref{tab:feros_wavelengths}) for each spectrum and then applied an appropriate gaussian profile to smooth each wavelength region in order that the convolution resulted in observed spectra that had spectral FWHM as close as possible to the constant FWHM of the synthetic spectra grid.

The FWHM values were measured as part of SPB and this was found to significantly increase the processing time of this stage of the pipeline. Hence, due to it being primarily a testing phase, for the first iteration of SPB the observed spectra are convolved using predetermined `mean' spectral FWHM values that were determined for each of the 17 wavelength regions. These default spectral FWHM values were determined by a statistical analysis of a subsample of 384 of the FEROS archived spectra, whereby the FWHM for as many weak and medium strength spectral lines as possible were measured in each of the sample spectra. 

The FWHM was measured by fitting a gaussian function to each spectral feature and calculating the width of the gaussian profile at the midpoint of the depth of the profile. For each spectrum the FWHM values increased with wavelength hence assuming one FWHM for the entire spectrum was not reasonable across the wavelength domain of the AMBRE:FEROS analysis. Hence a mean FWHM value for each of the 17 wavelength sections was determined and these were used in the convolution of the corresponding wavelength section. The convolution was carried out using a transformation gaussian profile for which the standard deviation ($\sigma_X$) was defined as:
\begin{align}\label{equ:fwhm_convolve}
 \sigma_X &= \sqrt{\frac{FWHM_{grid}}{\sigma_{FWHM}}^2 - \frac{FWHM_{obs}}{\sigma_{FWHM}}^2}  \\
 \sigma_{FWHM} &= \sqrt{2.0 \ln(2.0)}
\end{align} 

\noindent where FWHM$_{obs}$ is the default FWHM for the region, FWHM$_{grid}$ = 0.33~m\AA\ and $\sigma_{FWHM}$ transforms the FWHMs to $\sigma$s. Therefore each region of the observed spectrum was convolved such that the FWHM of the convolved spectrum matched the FWHM of the synthetic spectrum. 

Part of the convolution process is to resample the observed spectra to the same wavelength bins as the FEROS synthetic spectra grid. Also for SPB, the same quality tests outlined in SPA are used but again no spectra were rejected based on these tests in this first iteration.

The convolved FEROS spectra were then analysed using MATISSE to obtain the first estimate of the stellar atmospheric parameters ($T_{\textrm{eff}}$, $\log g$, [M/H] and [$\alpha$/Fe]). MATISSE also outputs synthetic spectra interpolated to the derived stellar parameters. Within the MATISSE algorithm the determination of the stellar parameters undergoes 10 iterations during which the algorithm converges on the final parameters for each spectrum. This number is set empirically and the majority of spectra converge to their parameters in much less than 10 iterations. After the MATISSE analysis is complete each spectrum is tested for whether convergence on the final $B_{\theta}(\lambda)$ function occurred within the 10 iterations and also the spectrum is tested as to the goodness of the fit between the normalised and the interpolated synthetic spectra in a $\log(\chi^2)$ calculation. In the case of non-convergence within MATISSE, the parameters determined at the ninth and tenth iterations are compared and the solution with the lowest $\log(\chi^2)$ is selected as the final set of parameters. If there is no convergence  and/or a high $\log(\chi^2)$, the spectra are flagged for further investigation. The majority of these instances were attributable to incorrect selection of the radial velocity. A separate routine was developed that carries out an automated visual inspection of these spectra in order to select, where possible, an improved radial velocity correction from the full list of radial velocity masks. In some instances the normalisation of the spectra was poor due to noisy spectra, strange spectral features or an invalid radial velocity correction. In the latter case the adjustment to the radial velocity would correct the poor normalisation. In the former cases these spectra were captured by quality/rejection flags in the normalisation process or in the construction of the final ESO dataset.

After these adjustments were made the spectra were reprocessed in the second iteration of SPB. This process is exactly the same as the first iteration except the mean spectral FWHM of the absorption lines for each wavelength section is now measured directly for each individual spectrum. This second iteration of SPB provides the first scientific estimate of the parameters, and hence convolving the spectrum with spectral FWHM values measured specifically for each spectrum provides a more robust analysis. The measured spectral FWHM values were used to convolve each region of each spectrum to match the FWHM of the AMBRE:FEROS synthetic spectra grid. Therefore in Equation~\ref{equ:fwhm_convolve} the FWHM$_{obs}$ is the measured spectral FWHM for each wavelength section for each spectrum. 

Extensive testing was carried out which compared spectral lines that were in common between the observed and synthetic spectra both before and after convolution in order to ensure that the transformation calculation resulted in appropriately convolved observed spectra. This included testing the procedure on the original high resolution synthetic spectra to ensure the AMBRE:FEROS resolution was obtained after convolution by this procedure. The FWHM$_{grid}$ (0.33~m\AA) was calculated based on a sample of the AMBRE:FEROS synthetic spectra for which the FWHM values were determined by the same method as for the observed spectrum and were found to be constant with wavelength as expected due to how the synthetic spectra were generated. This ensured consistency in the calculation and application of the FWHM in the convolution process.

Part of the calculation of the spectral FWHM was to classify each spectrum using the number of spectral lines and the widths that were measured. Three categories were used: `Weak' for less than 25 lines measured which were typically found to have FWHM~$<0.11$ (FWHM$_{weak}$) and were generally noisy spectra; `Medium' for more than 25 lines measured, where the line depths of the measured lines were between 0.5 and 0.95 in normalised intensity, and no more than 1 strong line was identified (FWHM$_{medium}$); and `Strong' for more than 1 strong line identified where the line depth of the `Strong' line was between 0.35 and 0.5 of the normalised intensity (FWHM$_{strong}$). The spectral lines were identified by locating the minima and maxima in the spectra, fitting a gaussian at each minima and then discarding those lines which failed the quality tests of the gaussian fit. For a spectrum typically classified as FWHM$_{medium}$ 1000 to 2000 lines were identified which were reduced to $\sim$400 lines with good gaussian fits. 

The `Strong' classification was an attempt to identify the spectra with large, broad spectral features as this provided another estimate of the spectral classification of the star. This was accomplished by applying a high FWHM smoothing function to each spectrum and then testing for which spectral features still remained. A spectrum without `Strong' lines would be reduced to a line near the continuum, while `Strong' features would otherwise still be prominent. This method had to be tailored to the FEROS wavelength region in order to place relevant limits on the possible number of `Strong' lines and their location within the wavelength regions. If, by this method, more than 1 `Strong' line were found then the spectrum was classifed as FWHM$_{strong}$ and the mean spectral FWHM of the `Strong' lines was recorded.

In the cases where the spectrum was designated as FWHM$_{weak}$ or FWHM$_{strong}$, then default spectral FWHM values were used instead. The spectral FWHM values for each spectrum that were used in the convolution, measured or mean values, were saved to an external file to be used in the next stage of the pipeline. Extensive testing was carried out using corresponding spectral lines in the original, convolved and synthetic spectra to confirm that the spectral FWHM procedure correctly convolved the archived spectra to the resolution of the synthetic grid across the wavelength range.

At this stage the spectra which failed the tests identifying un-analysable spectra (extreme emission features, extreme noise-dominated spectra etc) were rejected from the analysis process. This reduced list was analysed once more in MATISSE in order to determine the first estimate of the stellar parameters to be used in the next stage of the analysis pipeline. The convergence and $\log \chi^2$ were tested again to catch any further mis-identified radial velocity corrections. At the end of the second iteration of SPB 2370 FEROS spectra were rejected, 11\% of the total number of FEROS archived spectra.

\begin{figure*}[!th]
\centering
\begin{minipage}{180mm}  %for one column use 90, 90
\centering
\includegraphics[width=170mm]{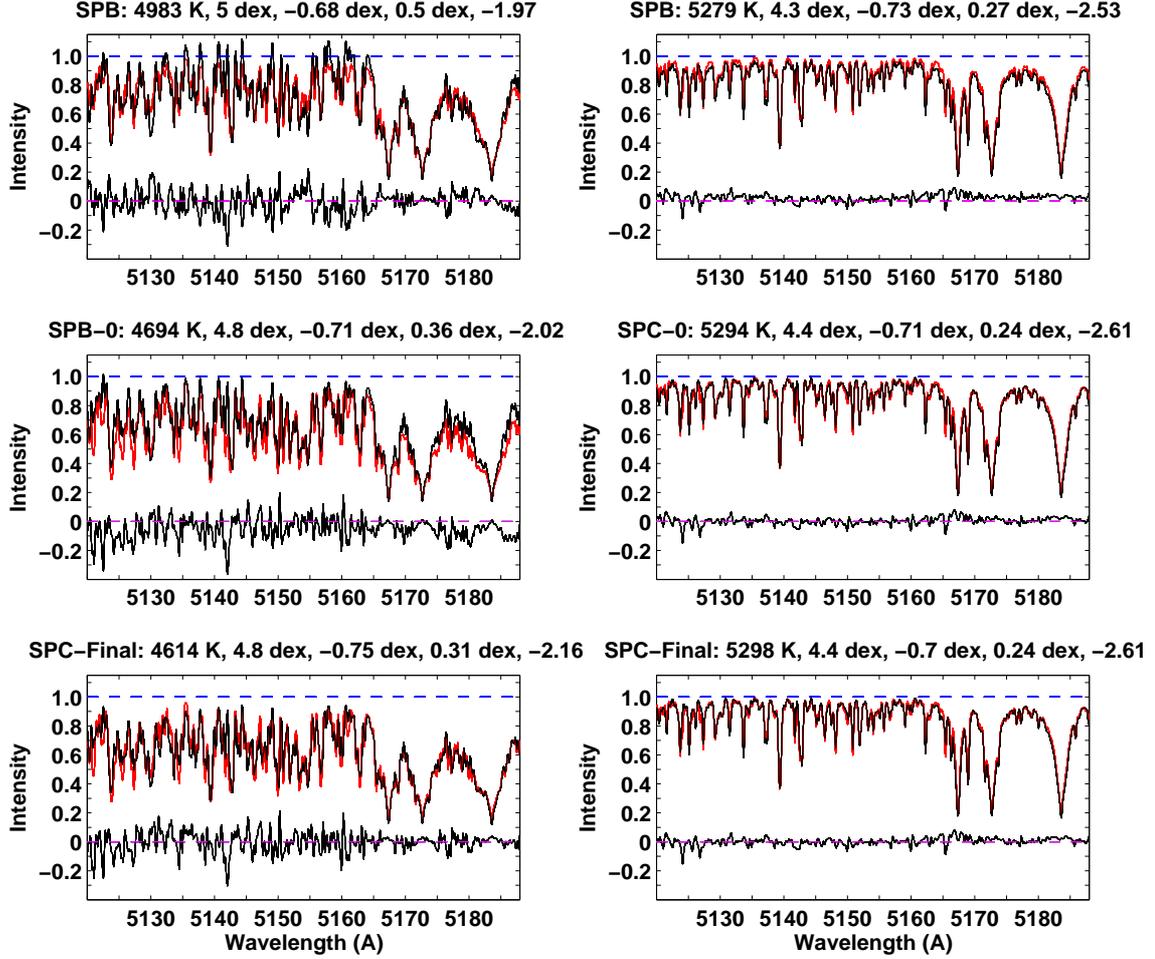}
\caption{Normalisation procedure for a two stars with $T_{\textrm{eff}} \sim 4600$~K (LEFT) and $T_{\textrm{eff}} \sim 5300$~K (RIGHT) shown by the wavelength region about two lines of the magnesium triplet. TOP: the second iteration of SPB, MIDDLE: first iteration of SPC, and BOTTOM: final solution from SPC. The stellar parameters ($T_{\textrm{eff}}$, $\log g$, [M/H] and [$\alpha$/Fe]) and goodness of fit over all the AMBRE:FEROS wavelengths ($\log \chi^2$) are shown. The synthetic spectrum interpolated to these parameters is shown in red, while the normalised observed spectrum is shown in black. The difference between observed and synthetic spectra is also shown at 0.0.}\label{fig:norm_pro}
\end{minipage}
\end{figure*}

\subsection{Spectral Processing C: Iterative spectra normalisation \& parameterisation}
The final number of FEROS spectra that were passed to the final stage of the analysis pipeline, Spectral Processing C (SPC), was 19181. SPC again uses the same procedures as SPA and SPB with two differences. The first difference is that the spectra are convolved using the spectral FWHM values determined in the second iteration of SPB rather than re-measuring the FWHM. The second difference is the use of a more robust method for the normalisation of the spectra. 

Previously, in SPA and SPB, a `rough' normalisation process was carried out that used pre-selected continuum regions to normalise the spectra to unity. But this is an invalid assumption for many stars, in particular for cool stars which have depressed continuum regions due to molecular bandheads that should not be normalised to unity. However at the beginning of SPC we have an estimate of the stellar parameters from the second iteration of SPB that gives us some valid information about each spectra that we did not previously possess.

We take advantage of this information by using the interpolated synthetic spectra generated for each set of stellar parameters as a better estimate of the continuum placement for the normalisation of the observed spectra. In an iterative process that discards absorption and emission features by polynomial fitting and sigma clipping to leave only `continuum' regions, the synthetic spectrum is divided out of the corresponding observed spectrum leaving a residual of continuum points. A continuum profile is fitted to this residual which is then divided out of the observed spectrum, normalising it to a pseudo-continuum that better represents the parameters of the star. This process overrides any residual curvature that remains after, or was introduced by, the normalisation to the hot star continuum.

These pseudo-continuum normalised spectra are reanalysed in MATISSE to obtain a second estimate of the stellar parameters and another set of synthetic spectra. The `normalisation treatment \& MATISSE analysis' cycle is repeated 9 times, which had been determined to be a sufficient number of iterations within which the normalised spectra and stellar parameters could converge to their optimum state. This typically occurred within five iterations. This approach provides a robust incremental adjustment of the parameters and normalisation process that hones in on a realistic estimate of the stellar parameters of each spectra, a process which is not possible using the normalisation method of SPA and SPB.

Figure~\ref{fig:norm_pro} shows the normalisation procedure for two FEROS spectra in the wavelength region about two of the magnesium triplet at 5160\AA\ as an example of depressed continuum regions. The two stars correspond to temperatures of $T_{\textrm{eff}} \sim 4600$~K (SNR = 115) and $T_{\textrm{eff}} \sim 5300$~K (SNR = 204) respectively. For each spectrum the normalisation by SPB, the first iteration of SPC and the final solution that was found in SPC are shown. The stellar parameters and $\log \chi^2$ (which is calculated over all the AMBRE:FEROS wavelengths) determined at each stage are shown as well as the synthetic spectrum interpolated to the stellar parameters. The difference between the observed and synthetic spectra is also included to show how it converges.

For the cooler star this rough normalisation of SPB is not a good fit as the observed spectrum is set too high compared to the synthetic spectrum generated at the corresponding stellar parameters. This shows how inadequate normalisation to unity is for cool star spectra. However the rough normalisation does provides a reasonable fit to the synthetic spectrum for the hot star although the placement of the continuum is still not ideal. 

At the initial iteration of SPC the spectrum undergoes its first normalisation to the previous solution synthetic spectrum. For the cooler star the observed spectrum now sits below unity in better agreement with the new synthetic spectrum, but there are still mismatches in the relative placement of the two spectra and the line depths. For the hotter star there is very good agreement between the observed and synthetic spectra. In both cases the improvement in the match are shown by the lower $\log \chi^2$ values.

The final solution for the cooler star ($i=6$) shows excellent agreement between the observed and the synthetic spectra, which is reflected in the much lower $\log \chi^2$ value. Indeed the visual inspection of the spectra show that they are in good agreement in this section of wavelength. For the hotter star the solution ($i=7$) is very close to the solution determined in the first iteration of SPC, indeed the $\log \chi^2$ does not change. Visually this good fit is obvious between the observed and synthetic spectra. The comparison of these two stars illustrates the greater difficulty there is in the normalisation and parameterisation of cool stars. However the procedure developed for the AMBRE pipeline can successfully manage both cases and the normalisation procedure in SPC is particularly effective in the case of cool stars where the continuum regions are more heavily obscured by spectral features. 

Typically convergence occured by the fourth or fifth iteration of the normalisation and stellar parameter determination cycle. When convergence has occurred the final set of parameters are selected as those from the converged solutions with the lowest $\log(\chi^2)$. In the case of non-convergence within the 9 iterations of SPC, it is assumed that for the final six iterations any bias introduced by the initial rough normalisation has been erased. Of these final six iterations the solution with the lowest $\log(\chi^2)$ is selected as the final stellar parameters for the spectra along with the corresponding normalised and synthetic spectra.

Hence the radial velocity, stellar parameters ($T_{\textrm{eff}}$, $\log g$, [M/H] and [$\alpha$/Fe]), normalised spectra and corresponding inteprolated synthetic spectra were determined in SPC for 19181 of the 21551 FEROS spectra.

Extensive testing of the FEROS analysis pipeline was carried out in order to optimise the procedures and to validate the MATISSE results by making comparison to literature values. The following sections describe the different tests and validations that were carried out. 

\section{AMBRE:FEROS internal error analysis}\label{sec:int_error}
The internal errors can be used to determine how well MATISSE will derive the stellar parameters of similar stellar types which have different levels of noise and different uncertainties in radial velocity. In order to test for noise and radial velocity effects a sample of 500 synthetic spectra were generated at random stellar parameters covering the entire stellar parameter range, spectral resolution and wavelength domains of the AMBRE:FEROS synthetic spectra grid.

\subsection{Internal errors: SNR}
To test the effects of noise the sample of synthetic spectra was reproduced five times by adding differing levels of noise per pixel (0.1~\AA\ per pixel for AMBRE:FEROS synthetic grid) at SNR = 10, 20, 50, 100. Each sample was then re-analysed in MATISSE and the derived stellar parameters were compared to the original values. 
 
\begin{figure}[!h]
\centering
\begin{minipage}{90mm}  %for one column use 90, 90
\centering
\includegraphics[width=93mm]{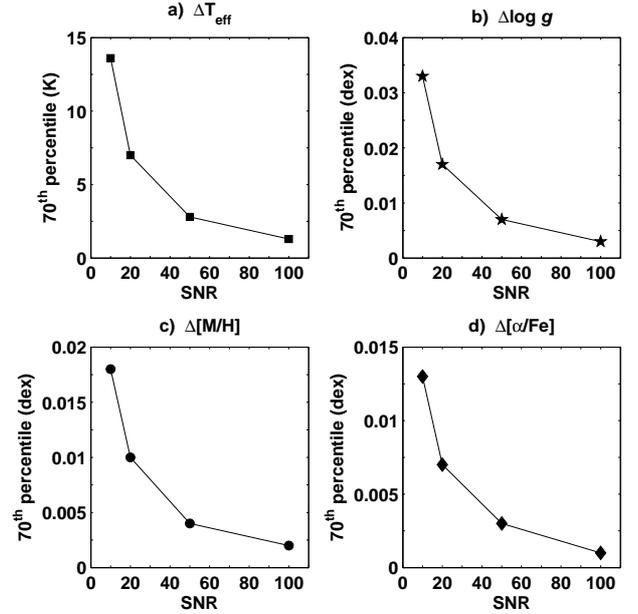}
\caption{Internal error for each parameter with changes in SNR. a) The $\Delta T_{\textrm{eff}}$ that 70\% of the synthetic sample were less than or equal to when calculating the difference between the nominal $T_{\textrm{eff}}$ and the $T_{\textrm{eff}}$ determined for the respective SNR. b) As for a) but for $\log \ g$. c) As for a) but for [M/H]. d) As for a) but for [$\alpha$/Fe].}\label{fig:snr_interr}
\end{minipage}
\end{figure}

\begin{figure}[h]
\centering
\begin{minipage}{90mm}
\centering
\includegraphics[width=93mm]{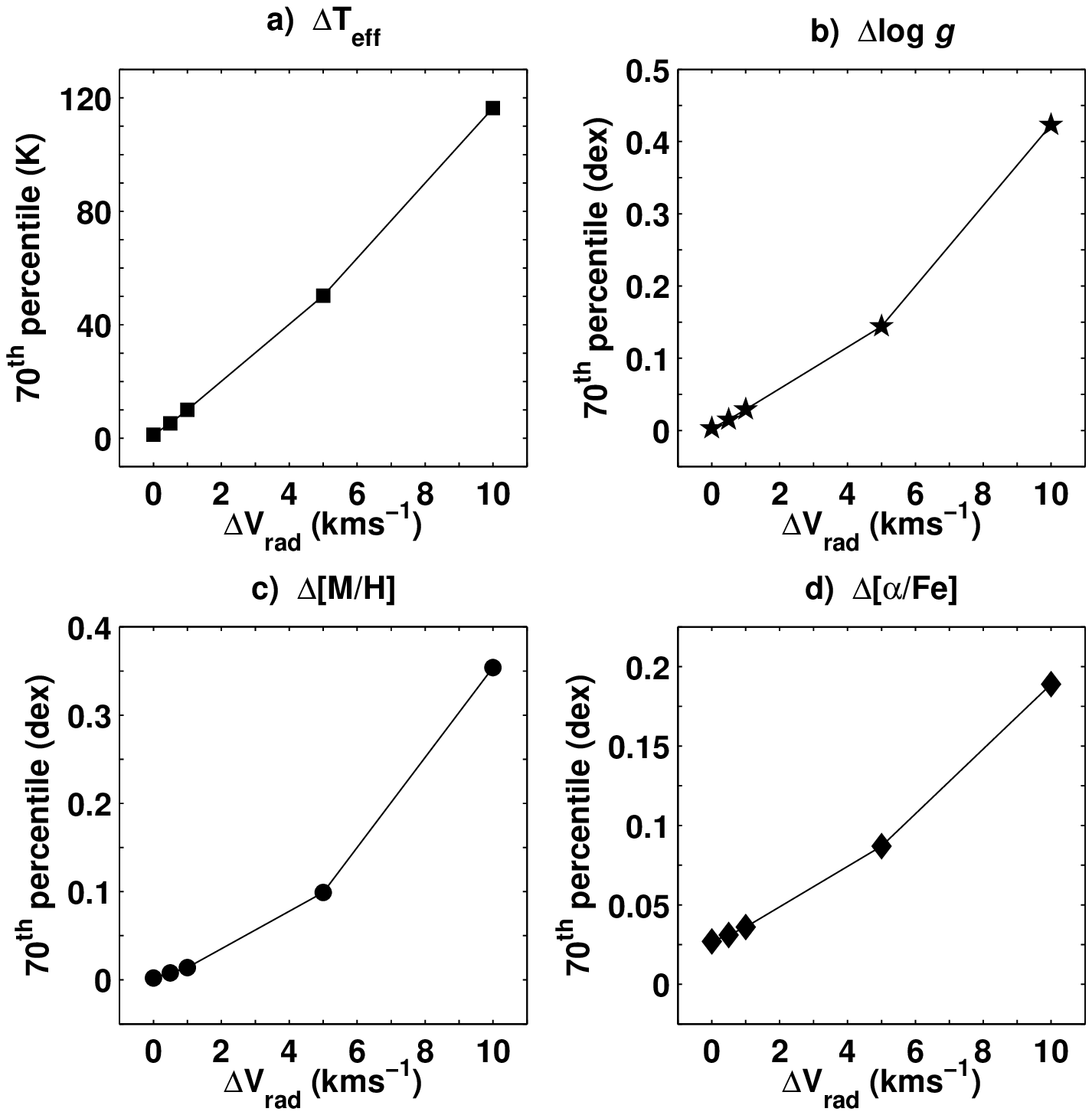}
\caption{Internal error for each parameter with changes in the V$_{rad}$ uncertainty (kms$^{-1}$). a) The $\Delta T_{\textrm{eff}}$ that 70\% of the synthetic sample were less than or equal to when calculating the difference between the nominal $T_{\textrm{eff}}$ and the $T_{\textrm{eff}}$ determined for the respective V$_{rad}$ uncertainty. b) As for a) but for $\log \ g$. c) As for a) but for [M/H]. d) As for a) but for [$\alpha$/Fe].}\label{fig:vrad_interr}
\end{minipage}
\end{figure}

Figures~\ref{fig:snr_interr}a. to d. show how the difference in stellar parameters changes with increasing SNR for $T_{\textrm{eff}}$, $\log g$, [M/H] and [$\alpha$/Fe] respectively. For each noise-added sample the difference between the original and derived parameters was calculated and then the difference value at the 70th percentile ($\sim 1 \sigma$) was determined. For example, in Figure~\ref{fig:snr_interr}a. at a SNR of 10 (the noisiest), 70\% of the synthetic spectra returned $T_{\textrm{eff}}$ values within 14~K of the original $T_{\textrm{eff}}$ value. At a SNR of 100, 70\% of the sample returned values within 2~K of the original $T_{\textrm{eff}}$ value. Similarly small errors were also found for $\log g$, [M/H] and [$\alpha$/Fe] that diminished to negligible levels at SNR of 100 in all cases. This analysis shows that for two synthetic spectra with very similar intrinsic stellar parameters, where the differences in the spectra are only attributable to noise, MATISSE will determine close to the same stellar parameters indicating that the internal errors of MATISSE are very small. This can be attributed to the large wavelength coverage and high resolution of the FEROS spectra which contain a great deal of information on the stellar parameters in terms of spectral lines, even at low SNR, of which MATISSE takes advantage.

Overall this analysis shows that the internal error due to SNR is negligible in the FEROS results. The internal ($int$) error on each parameter due to SNR ($\sigma_{int,snr}$) is calculated as part of the MATISSSE program. It was used to calculate the total internal error as explained at the end of this section.

\subsection{Internal errors: V$_{rad}$ \& $Vsini$}\label{sec:int_errV}
%\vspace{-0.5cm}
A similar investigation was carried out in order to determine the effect of uncertainties in radial velocity on the determination of the stellar parameters by MATISSE. The sample of interpolated spectra were reproduced this time undergoing radial velocity shifts of $\Delta$V$_{rad}$ = 0, 0.5, 1.0, 5 and 10~kms$^{-1}$ with an SNR of 100. Figures~\ref{fig:vrad_interr}a to d show how the difference in stellar parameters changes with increasing radial velocity uncertainty for $T_{\textrm{eff}}$, $\log g$, [M/H] and [$\alpha$/Fe] respectively. Again the difference between the original and derived stellar parameters was obtained and the difference value at the 70$^{th}$ percentile for each sample was calculated. At a radial velocity uncertainty of 5~kms$^{-1}$ ($\sim$~0.09 pixels) the error in stellar parameter determination becomes significant ($>$1$\%$) for each of the stellar parameters. However if the uncertainty in the radial velocity is less than 5~kms$^{-1}$ the MATISSE parameters compare well to the true values. 

The internal uncertainty in each stellar parameter due to radial velocity was calculated using the equations that connect the points in Figure~\ref{fig:vrad_interr}. The error on the radial velocity, $\sigma(\textrm{V}_{rad})$, determined in the radial velocity programme was used as the input to determine the internal error due to the V$_{rad}$ uncertainty ($\sigma_{int,\textrm{V}_{rad}}$) for each parameter for each spectrum according to the following conditions:

{\small
\begin{equation*}
{\footnotesize
\sigma(T_{eff})_{int,\textrm{V}_{rad}}= 
\begin{cases} 5.2 & \text{if $\sigma(\textrm{V}_{rad}) < 0.5$,}
\\
9.60 \sigma(\textrm{V}_{rad}) + 0.40 &\text{if $0.5 \leq \sigma(\textrm{V}_{rad}) < 1.0 $,}
\\
10.05 \sigma(\textrm{V}_{rad}) - 0.05 &\text{if $1.0 \leq \sigma(\textrm{V}_{rad}) < 5.0$,}
\\
13.24 \sigma(\textrm{V}_{rad}) - 16.0 &\text{if $5.0 \leq \sigma(\textrm{V}_{rad}) \leq 10.0$.}
\end{cases} }
\end{equation*}
\begin{equation*}
{\footnotesize
\sigma(\log \ g)_{int,\textrm{V}_{rad}}= 
\begin{cases} 0.015 & \text{if $\sigma(\textrm{V}_{rad}) < 0.5$,}
\\
0.028 \sigma(\textrm{V}_{rad}) + 0.001 &\text{if $0.5 \leq \sigma(\textrm{V}_{rad}) < 1.0 $,}
\\
0.0287 \sigma(\textrm{V}_{rad}) + 0.00025 &\text{if $1.0 \leq \sigma(\textrm{V}_{rad}) < 5.0$,}
\\
0.0558 \sigma(\textrm{V}_{rad}) + 0.135 &\text{if $5.0 \leq \sigma(\textrm{V}_{rad}) \leq 10.0$.}
\end{cases} }
\end{equation*}
\begin{equation*}
{\footnotesize
\sigma([\text{M/H}])_{int,\textrm{V}_{rad}}= 
\begin{cases} 0.008 & \text{if $\sigma(\textrm{V}_{rad}) < 0.5$,}
\\
0.012 \sigma(\textrm{V}_{rad}) + 0.002 &\text{if $0.5 \leq \sigma(\textrm{V}_{rad}) < 1.0 $,}
\\
0.0213 \sigma(\textrm{V}_{rad}) - 0.0073 &\text{if $1.0 \leq \sigma(\textrm{V}_{rad}) < 5.0$,}
\\
0.0510 \sigma(\textrm{V}_{rad}) - 0.156 &\text{if $5.0 \leq \sigma(\textrm{V}_{rad}) \leq 10.0$.}
\end{cases} }
\end{equation*}
\begin{equation*}
{\footnotesize
\sigma([\alpha\text{/Fe}])_{int,\textrm{V}_{rad}}= 
\begin{cases} 0.006 & \text{if $\sigma(\textrm{V}_{rad}) < 0.5$,}
\\
0.010 \sigma(\textrm{V}_{rad}) + 0.001 &\text{if $0.5 \leq \sigma(\textrm{V}_{rad}) < 1.0 $,}
\\
0.0128 \sigma(\textrm{V}_{rad}) - 0.0018 &\text{if $1.0 \leq \sigma(\textrm{V}_{rad}) < 5.0$,}
\\
0.0204 \sigma(\textrm{V}_{rad}) - 0.0400 &\text{if $5.0 \leq \sigma(\textrm{V}_{rad}) \leq 10.0$.}
\end{cases} }
\vspace{0.5cm}
\end{equation*}
}

The rotational velocity ($Vsini$) of a star can have an impact the broadening of the observed spectral features depending on the magnitude of the rotational velocity and the resolution of the instrument. The AMBRE synthetic spectra grid was generated with no variations in $Vsini$, assuming all stars to be slow rotators. However it is important to consider whether variations in $Vsini$ will have an effect on the determined stellar parameters. \citet{Gazzano2010} carried out a robust investigation into the $Vsini$ limits of reliable stellar parameter determination using MATISSE and found that spectra with $Vsini < 11$~kms$^{-1}$ produced good results in the MATISSE analysis of a sample of FGK dwarfs. Based on this, \citet{Gazzano2010} accepted all CCF FWHM $\leq 20$~kms$^{-1}$ for spectra at a resolution of R$\sim$26,000. The FEROS sample and the AMBRE synthetic spectra grid are of much broader range in stellar parameters and wavelength, and of lower resolution (R$\sim$15,000) than the \citet{Gazzano2010} study. Hence the effects of $Vsini$ on the range of stars for which the synthetic grid has been optimised, would be well-masked for the AMBRE:FEROS configuration. In Section~\ref{sec:highccffwhm} we examine the FEROS sample in the context of the CCF FWHM to identify spectra which are not well-represented by the AMBRE synthetic spectra grid.

\subsection{Internal errors: Normalisation}
The effects of normalisation on the MATISSE determination were measured using the iterative normalisation process (SPC) of the FEROS analysis pipeline. Using a test sample of 384 FEROS spectra ($\langle$SNR$\rangle$=100$\pm$45) the changes in normalisation between iterations of SPC was explored. 

\begin{figure}[h]
\centering
\begin{minipage}{90mm}  %for one column use 90, 90
\centering
\includegraphics[width=85mm]{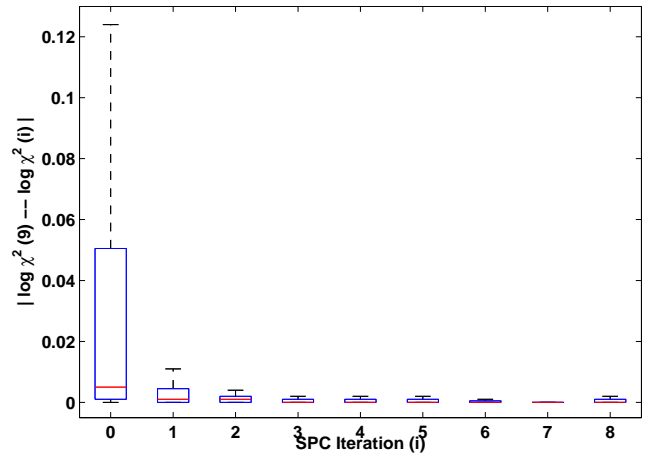}
\caption{Box \& Whisker graph of the absolute difference in the $\log \chi^2$ of the reconstructed to the normalised spectra for the 384 test spectra between the $9^{th}$ SPC iteration and each of the previous iterations ($| \log \chi^2(9) - \log \chi^2(i) |$). The box is constrained by the 25$^{th}$ and 75$^{th}$ percentiles, and the red lines are the medians. The whiskers extend to the furthest data points not considered to be outliers. }\label{fig:abslcs_interr}
\end{minipage}
\end{figure}

Figure~\ref{fig:abslcs_interr} shows the progression of the goodness of fit between the reconstructed spectra and the normalised spectra at each iteration of SPC as a Box \& Whisker graph. Assuming that by the $9^{th}$ iteration convergence has occured, the absolute difference in the $\log \chi^2$ between the $9^{th}$ and i$^{th}$ iteration was calculated for each of the 384 test spectra. %The edges of the boxes are the 25$^{th}$ and 75$^{th}$ percentiles, the red lines are the medians, and the whiskers extend to the furthest points not considered to be outliers. 
The spread in values is largest at $i=0$ and there is a large decrease in the spread for $i=1$. The following iterations show a more sedate decrease in spread and appear to have converged from $i=6$. 

\begin{figure*}[!t]
\centering
\begin{minipage}{180mm}  %for one column use 90, 90
\centering
\includegraphics[width=150mm]{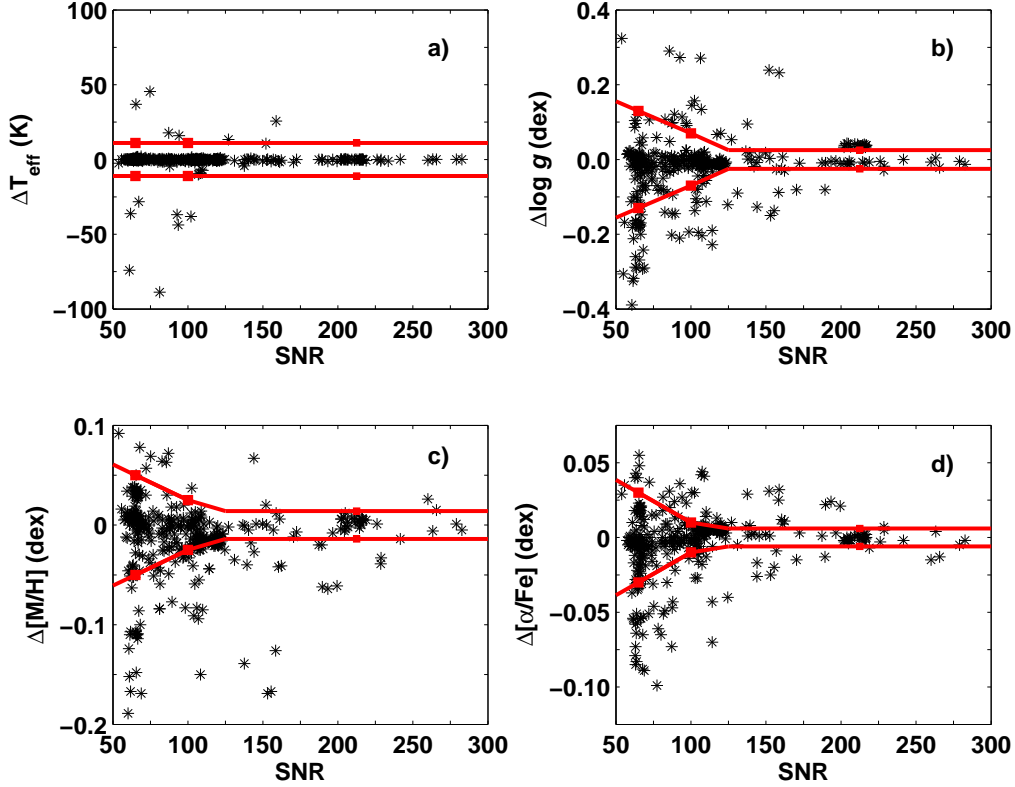}
\caption{Internal error with changes in Normalisation for a) $T_{\textrm{eff}}$; b) $\log g$; c) [M/H]; and d) [$\alpha$/Fe]. The red lines are determined as the $2 \sigma$ uncertainty on values in two SNR bins ($\Delta$SNR~$\approx30$) centred at SNR = 65 \& 100 and constant uncertainty for SNR $>$ 125.}\label{fig:norm_interr}
\end{minipage}
\end{figure*}

To measure the internal error associated with the normalisation process we investigated the difference in stellar parameters between the $9^{th}$ and $7^{th}$ iteration for the test sample. Hence we assumed that the solutions have converged from at least the $7^{th}$ iteration and the subsequent variations in the stellar parameters are due to minor adjustment of the normalisation of the observed spectra. The variation from convergence ($i=7$) to the final iteration ($i=9$) was used to give the internal error due to the normalisation process.

Figure~\ref{fig:norm_interr} shows the difference in each stellar parameter between the $9^{th}$ and $7^{th}$ iterations against SNR for the 384 test spectra as black points. These values were binned into two SNR bins ($\Delta$SNR~$\approx30$) centred at SNR = 65 \& 100, and a single bin for all points with SNR $>$ 125. The 2~$\sigma$ uncertainty for each bin is shown as red squares at the bin centre. Red lines connect the bin uncertainty values. For $T_{\textrm{eff}}$ the spread is constant for each bin and so a constant value for the internal uncertainty due to normalisation was adopted. For $\log g$, [M/H] and [$\alpha$/Fe] the spread diminishes with increased SNR until SNR = 125 after which a constant spread is found. The adopted internal errors for these three parameters were therefore a piecewise function of SNR. The following equations define the internal errors due to normalisation for each parameter ($\sigma(\theta)_{int,norm}$):

%\begin{align*}
%\sigma(T_{eff})_{int,norm} &= 11 \text{~K} \\
%\sigma(\log \ g)_{int,norm} &= 0.025 \text{~dex} \\
%\sigma(\text{[M/H]})_{int,norm} &= 0.014 \text{~dex} \\
%\sigma([\alpha/\text{Fe]})_{int,norm} &= 0.006 \text{~dex} \\
%\end{align*}

%\begin{equation*}
%\begin{cases} 0.006 & \text{if $\sigma(\textrm{V}_{rad}) < 0.5$,}
%\\
%0.010 \sigma(\textrm{V}_{rad}) + 0.001 &\text{if $0.5 \leq \sigma(\textrm{V}_{rad}) < 1.0 %$,}
%\end{equation*}

{\tiny
\begin{equation*}
%\begin{align*}
{\footnotesize
\sigma(T_{\textrm{eff}})_{int,norm}= 11.0 \text{~K}} \\
\end{equation*}
\begin{equation*}
%\sigma(\log \ g)_{ext} &= 0.2 \text{~dex} \\
{\footnotesize
\sigma(\log \ g)_{int,norm}= % &= 
\begin{cases}
 -0.0017 \text{~SNR} + 0.2414 \text{~dex} & \text{if SNR $< 100$,}\\
 -0.0018 \text{~SNR} + 0.2500 \text{~dex} & \text{if $100 \leq$ SNR $< 125$,}\\
 -0.025 \text{~dex} & \text{if SNR $\geq 125$,}\\
\end{cases}\\}
\end{equation*}
\begin{equation*}
{\footnotesize
\sigma(\text{[M/H]})_{int,norm}= % &= \text{~dex} \\
\begin{cases}
 -0.0007 \text{~SNR} + 0.0964 \text{~dex} & \text{if SNR $< 100$,}\\
 -0.0004 \text{~SNR} + 0.0690 \text{~dex} & \text{if $100 \leq$ SNR $< 125$,}\\
 -0.014 \text{~dex} & \text{if SNR $\geq 125$,}\\
\end{cases}\\}
\end{equation*}
\begin{equation*}
{\footnotesize
\sigma([\alpha/\text{Fe]})_{int,norm}= % &= \text{~dex} \\
\begin{cases}
 -0.0006 \text{~SNR} + 0.0671 \text{~dex} & \text{if SNR $< 100$,}\\
 -0.0002 \text{~SNR} + 0.0260 \text{~dex} & \text{if $100 \leq$ SNR $< 125$,}\\
 -0.006 \text{~dex} & \text{if SNR $\geq 125$,}\\
\end{cases}\\}
%\end{align*}
\end{equation*}
}

\subsection{Total internal error}
The internal errors due to the SNR, V$_{rad}$ and Normalisation calculated above were combined in quadrature to give the total internal error for each parameter ($\sigma(\theta)_{int}$) for each spectrum. This was calculated as follows:

\begin{equation}
\sigma(\theta)_{int} = \sqrt{\sigma^2(\theta)_{int,snr} + \sigma^2(\theta)_{int,\textrm{V}_{rad}} + \sigma^2(\theta)_{int,norm}}
\end{equation}

These values were reported for each spectrum in the ESO dataset as described in Section~\ref{sec:AFstellarparams}.

\section{AMBRE:FEROS external error analysis}\label{sec:ext_error}
The external error was quantified by comparing the AMBRE:FEROS stellar parameter values to literature values for key reference stars that exist within the AMBRE:FEROS dataset.

%From the complete ESO:FEROS archive dataset that was received, a total of 21551 scientific spectra have been identified, covering the period 2005 to 2010. These spectra correspond to 6285 different stars. Key information (target name and coordinates, exposure time, etc) from the fits header of each spectra were extracted and combined with any corresponding stellar properties found in the SIMBAD and 2MASS. These literature values of radial velocity, stellar parameters and photometric properties were used in the quality control of the AMBRE:FEROS results.

A list of stars that had been observed with FEROS but were also part of other high quality spectroscopic studies were identified and this reference sample was used to provide quality criteria for each derived stellar parameter. Unfortunately, reference stars covering the whole range of possible parameters could not be found in the literature. In several cases, estimates of the mean stellar metallicity were lacking and the situation was even worse for the [$\alpha$/Fe] chemical contents since too few reference stars with known [$\alpha$/Fe] content could be found in the literature.

\subsection{Building the FEROS reference sample}
In order to compare the MATISSE stellar atmospheric parameters to the literature and to define some quality criteria, a list of reference stars was built using key spectral atlases and libraries.

\subsubsection{Stellar Atlases: Sun, Arcturus, Procyon}
Table~\ref{tab:refstars_atlas} lists the standard stars and the corresponding stellar atlases that were used to test the FEROS analysis pipeline. Of these atlases only the S$^4$N solar atlas was observed using FEROS. However the adaptability of the pipeline meant that the use of spectra from other instruments presented no difficulty as long as the required wavelength ranges were available. The accepted stellar parameters for each standard star are listed in bold in Table~\ref{tab:refstars_atlas}. The final stellar parameters %and typical external errors, $\sigma(\theta)_{ext}$ (see end of this section) 
that have been derived in the AMBRE:FEROS analysis with bias corrections (see end of this section) for each atlas are also given.

\begin{table}[!hb]
\caption{List of standard stars and the corresponding spectral atlases used to test the FEROS analysis pipeline. The accepted parameters for the standard stars are listed (bold) as well as the final stellar parameters with bias corrections derived for each atlas in the AMBRE:FEROS analysis.}
\begin{tabular}{ccccc}
\hline\hline
Standard star/ATLAS & $T_{\textrm{eff}}$ & $\log g$ & [Fe/H] & [$\alpha$/Fe] \\ 
 & (K) & (dex) & (dex) & (dex) \\ 
\hline
{\bf SUN} & {\bf 5770} & {\bf 4.44} & {\bf 0.00} & {\bf 0.00} \\ 
\citet{SolarAtlasHinkle}\tablefootmark{a} & 5732 & 4.42 & -0.03 & 0.05 \\ 
\citet{BNSolAtlas}\tablefootmark{b} & 5674 & 4.32 & 0.00 & 0.08 \\ 
BASS2000\tablefootmark{c} & 5841 & 4.49 & -0.01 & 0.06 \\ 
S$^4$N\tablefootmark{d} & 5735 & 4.43 & -0.03 & 0.05 \\ 
 &  &  &  &  \\ 
{\bf ARCTURUS\tablefootmark{*}} & {\bf 4300} & {\bf 1.70} & {\bf -0.60} & {\bf 0.20} \\ 
\citet{ArcturusAtlasHinkle}\tablefootmark{a} & 4537 & 2.12 & -0.49 & 0.28 \\ 
UVES-POP\tablefootmark{e} & 4468 & 2.03 & -0.47 & 0.29 \\ 
 &  &  &  &  \\ 
{\bf PROCYON\tablefootmark{*}} & {\bf 6600} & {\bf 4.00} & {\bf 0.00} & {\bf 0.00} \\ 
UVES-POP\tablefootmark{e} & 6538 & 4.00 & -0.25 & 0.07 \\ 
 &  &  &  &  \\
%Dwarfs $\sigma(\theta)_{ext}$ & 120 & 0.20 & 0.10 & 0.10 \\ 
%Giants $\sigma(\theta)_{ext}$ & 120 & 0.40 & 0.10 & 0.10 \\ 
\hline
% &  &  &  &  \\
\multicolumn{5}{l}{\tablefoottext{a}{ftp://ftp.noao.edu/catalogs/arcturusatlas/visual/}} \\
\multicolumn{5}{l}{\tablefoottext{b}{ftp://ftp.hs.uni-hamburg.de/pub/outgoing/FTS-Atlas/}}\\
\multicolumn{5}{l}{\tablefoottext{c}{\citet{BASS2000}; http://bass2000.obspm.fr/}}\\
\multicolumn{5}{l}{\tablefoottext{d}{\citet{Allende-Prieto2004}; http://hebe.as.utexas.edu/s4n/}}\\
\multicolumn{5}{l}{\tablefoottext{e}{\citet{UVESPOP}; http://www.sc.eso.org/santiago/uvespop/}}\\
\multicolumn{5}{l}{\tablefoottext{*}{Mean of parameters listed on SIMBAD}} \\
\end{tabular}
\label{tab:refstars_atlas}
\end{table}

These atlases were combined in a test sample that included high resolution synthetic spectra for the Sun and Arcturus, as well as synthetic spectra for each grid point in the AMBRE:FEROS synthetic grid corresponding to the Sun and Arcturus. These were used to check the internal consistency of the pipeline in terms of normalisation and convolution of the input spectra. The results for the Sun are in very good agreement with the accepted values. For Arcturus we over-estimate the $T_{\textrm{eff}}$ and $\log g$ values however we are within 2~$\sigma$ of the accepted values for $T_{\textrm{eff}}$. For Procyon there is good agreement in $T_{\textrm{eff}}$ and $\log g$, although the $T_{\textrm{eff}}$ is slightly under--estimated within the stated external errors. However the metallicity is under-estimated by --0.25~dex which is outside the 2~$\sigma$ limit. The under--estimation of the $T_{\textrm{eff}}$ may account for some of this discrepancy in metallicity. However further investigation into understanding the variation in the stellar parameters for these standard stars is ongoing.

\begin{sidewaystable*}
\caption{List of spectral libraries used to select comparison stars from the FEROS archived spectra. The number of stars from each library and the corresponding number of FEROS spectra are also listed.}
\begin{center}
\begin{tabular}{lllccl}
\hline\hline
Source & Papers & Abbrv. & \multicolumn{1}{l}{No. Stars} & \multicolumn{1}{l}{No. Spectra} & Methods applied \\ 
\hline
S$^4$N & \citet{Allende-Prieto2004} ($T_{\textrm{eff}}$,$\log g$,[Fe/H],[$\alpha$/Fe],V$_{rad}$) & AP04 & 30 & 338 & \citet{Alonso1996,Alonso1999} calibration, isochrones,  \\ 
 &  &  &  &  & \citet{Allende-Prieto2003} GA, spectroscopic inversion \\ 
Gaia RVS Standards & \citet{Crifo2010} (V$_{rad}$) & C10 & 158 & 318 & Convolution, CORAVEL cross-correlation \\ 
 & \citet{Bensby2003} ($T_{\textrm{eff}}$,$\log g$,[Fe/H],[$\alpha$/Fe])\tablefootmark{a} & B03 & 68 & 68 & Fe I EP balance, parallax, equivalent widths  \\ 
 &  &  &  &  &  \\ 
PASTEL & \textbf{$T_{\textrm{eff}}$,$\log g$,[Fe/H]} &  &  &  &  \\ 
\citet{Soubiran2010} & \citet{Fuhrmann1997,Fuhrmann1998,Fuhrmann1998b,Fuhrmann1998a,Fuhrmann2004,Fuhrmann2008} & F97-08 & 49 & 206 & Balmer line wings, Fe IP balance, Fe I/Fe II line profiles  \\ 
 & \citet{Gratton1982,Gratton1989,Gratton1996,Gratton2003} & G82-03 & 23 & 118 & Fe I EP balance,Fe I/Fe II IP balance, equivalent widths\\ 
 & \citet{Hekker2007} & H\&M07 & 6 & 27 & Fe I EP balance, Fe I/Fe II IP balance, equivalent widths  \\ 
 & \citet{Luck1983,Luck1991,Luck2005,Luck2006} & L\&H83-06 & 41 & 224 & Fe I EP balance, Fe I/Fe II IP balance, equivalent widths\\ 
 & \citet{McWilliam1990} & McW90 & 10 & 41 & IRFM+colour calibration, parallax, equivalent widths \\ 
 & \citet{Mishenina2001} & M\&K01;M04-08 & 42 & 227 & Wings of H$_{\alpha}$, Fe I EP balance,   \\ 
 & \citet{Mishenina2003,Mishenina2004,Mishenina2006,Mishenina2008} &  &  &  & Fe I/Fe II IP balance, equivalent widths \\ 
 & \citet{Ramirez2007} & R07 & 26 & 151 & Colour calibration, parallax, equivalent widths \\ 
 & \citet{Valenti2005} & V\&F05 & 81 & 310 & `Spectroscopy Made Easy' \\ 
 &  &  &  &  &  \\ 
 & \textbf{$T_{\textrm{eff}}$ only} &  &  &  &  \\ 
 & \citet{Alonso1996,Alonso1999} & A96,99 & 22 & 147 & Infra-Red Flux Method (IRFM) \\ 
 & \citet{Blackwell1998} & B\&LG98 & 20 & 152 & IRFM \\ 
 & \citet{diBenedetto1998} & dB98 & 18 & 163 & Surface brightness and $V-K$ calibration \\ 
 & \citet{Gonzalez2009} & GH\&B09 & 26 & 162 & Line depth ratios \\ 
 & \citet{Kovtyukh2003,Kovtyukh2004,Kovtyukh2006} & K03-06 & 48 & 354 & SED fit + $V-K$ calibration \\ 
 & \citet{Masana2006} & M06 & 57 & 306 & IRFM \\ 
 & \citet{Ramirez2005} & R\&M05 & 37 & 197 & IRFM + colour calibration \\ 
 &  &  & \multicolumn{1}{l}{} &  &  \\ 
 & \textbf{PASTEL Total} &  & 148 & 618 &  \\ 
 &  &  &  &  &  \\ 
\hline
 & \tablefoottext{a}{FEROS Observations 2000-2001} &  &  &  &  \\ 
 & \multicolumn{3}{l}{GA = Genetic Algorithm, EP = Excitation Potential, IP = Ionisation Potential} &  &  \\ 
\end{tabular}
\end{center}
\label{tab:refstars_lib}
\vspace{-20cm}
\end{sidewaystable*}

\subsubsection{Spectral libraries reference sample}
Preliminary investigations showed that merely making comparison between the MATISSE results and all of the stellar parameter values that were available in SIMBAD for the FEROS stars resulted in a great deal of disparity. There was no quality determination on the values with which we made the comparison. Homogeneously analysed large sample datasets were necessary in order to ensure that a valid comparison was being made. To this end key spectral libraries were investigated in order to locate within the FEROS archive dataset stars that had been analysed in high quality studies. This list of spectral libraries is given in Table~\ref{tab:refstars_lib}.

The primary spectral library that was used was the PASTEL database \citep{Soubiran2010}\footnote{http://pastel.obs.u-bordeaux1.fr}, which provides a selection of key papers that carried out detailed spectroscopic studies at high resolution and high SNR to derive high quality stellar parameters. From these papers we were able to identify a total of 148 stars corresponding to 618 FEROS archived spectra (Table~\ref{tab:refstars_lib}). There was significant crossover of stars between the studies which provided another level of comparison for this work. From the PASTEL sample $\sim$120 stars ($\sim$300 spectra) had values for three stellar parameters, $T_{\textrm{eff}}, \log g$ and [Fe/H], while the remaining stars had $T_{\textrm{eff}}$ values only. In the following comparison we assume the literature values to be correct. However we note that the literature values, while they are high quality studies, were determined using a range of techniques \citep{Soubiran2010}, in particular, several studies determined photometric $T_{\textrm{eff}}$ values using the Infra-Red Flux Method \citep{Alonso1996,Alonso1999,Ramirez2005,Gonzalez2009}.

\begin{figure*}[!th]
%\centering
\begin{minipage}{90mm}  %for one column use 90, 90
\centering
\includegraphics[width=85mm]{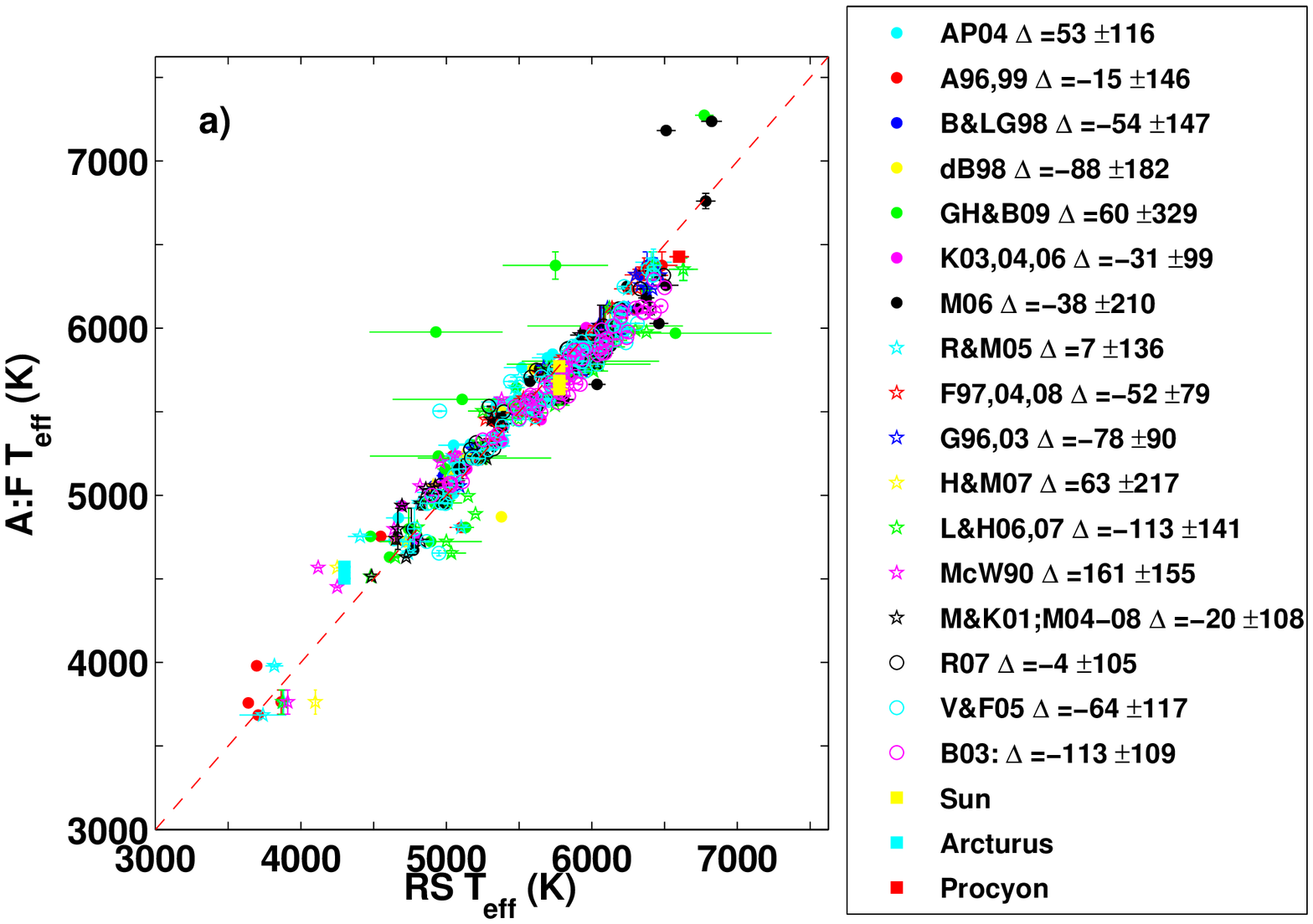}
%\caption{$T_{\textrm{eff}}$ results comparison between AMBRE:FEROS (A:F) and reference sample (RS) from PASTEL and specific test spectra. The legend identifies the individual papers or test spectra and the associated bias. The dashed line indicates a 1:1 agreement.}
%\label{fig:feros_teff}
\end{minipage}
%\end{figure}
%\hspace{0.2cm}
\hfill
%\begin{figure}[h]
%\centering
\begin{minipage}{90mm}
\centering
\includegraphics[width=85mm]{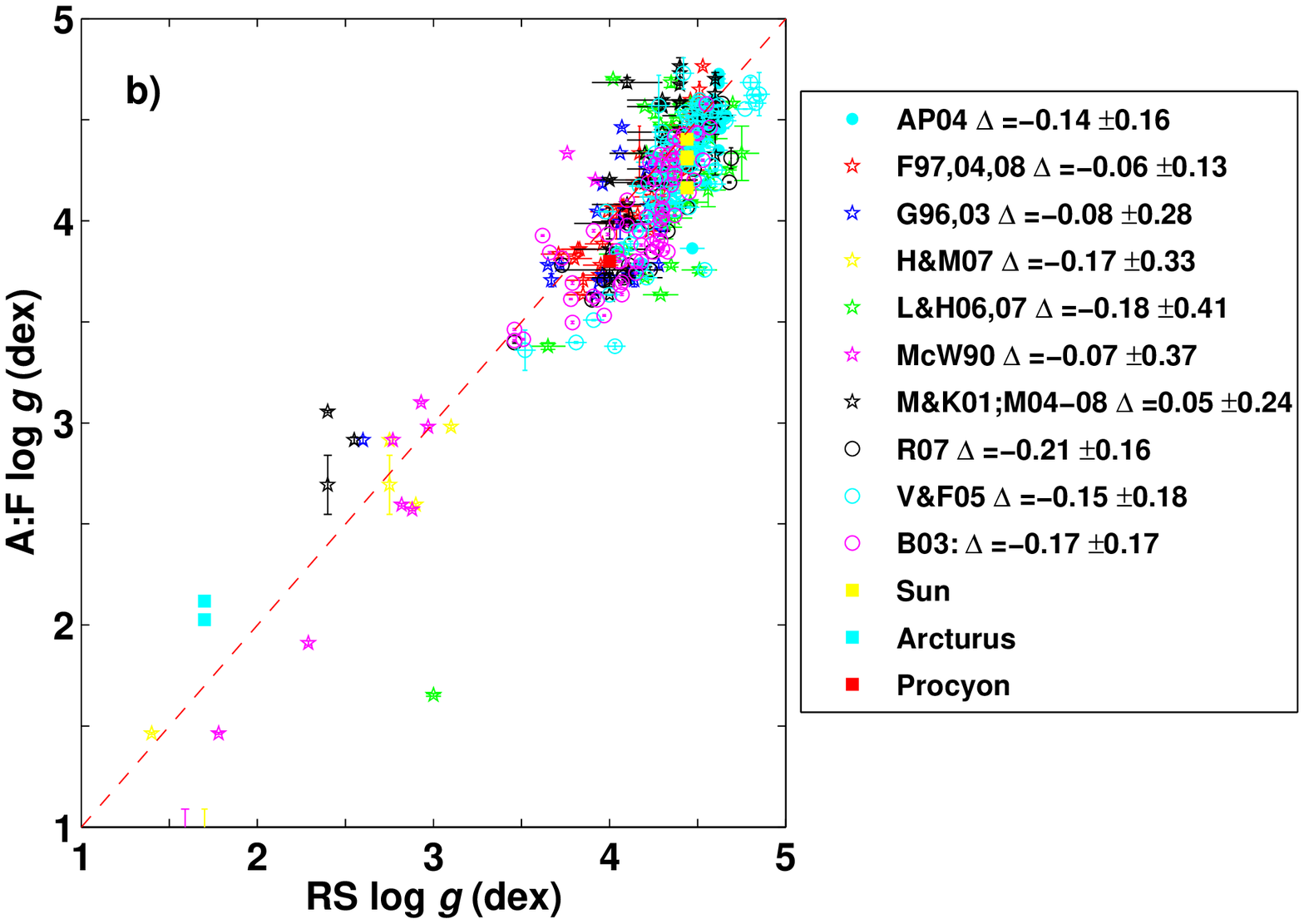}
%\caption{As for Figure~\ref{fig:feros_teff} but comparing the $\log \ g$ values. The legend describes the relevant samples and their biases.}
%\label{fig:feros_logg}
\end{minipage}
%\end{figure}
%\begin{figure}[h]
%\centering
\begin{minipage}{90mm}
%\vspace{-1cm}
\centering
\includegraphics[width=85mm]{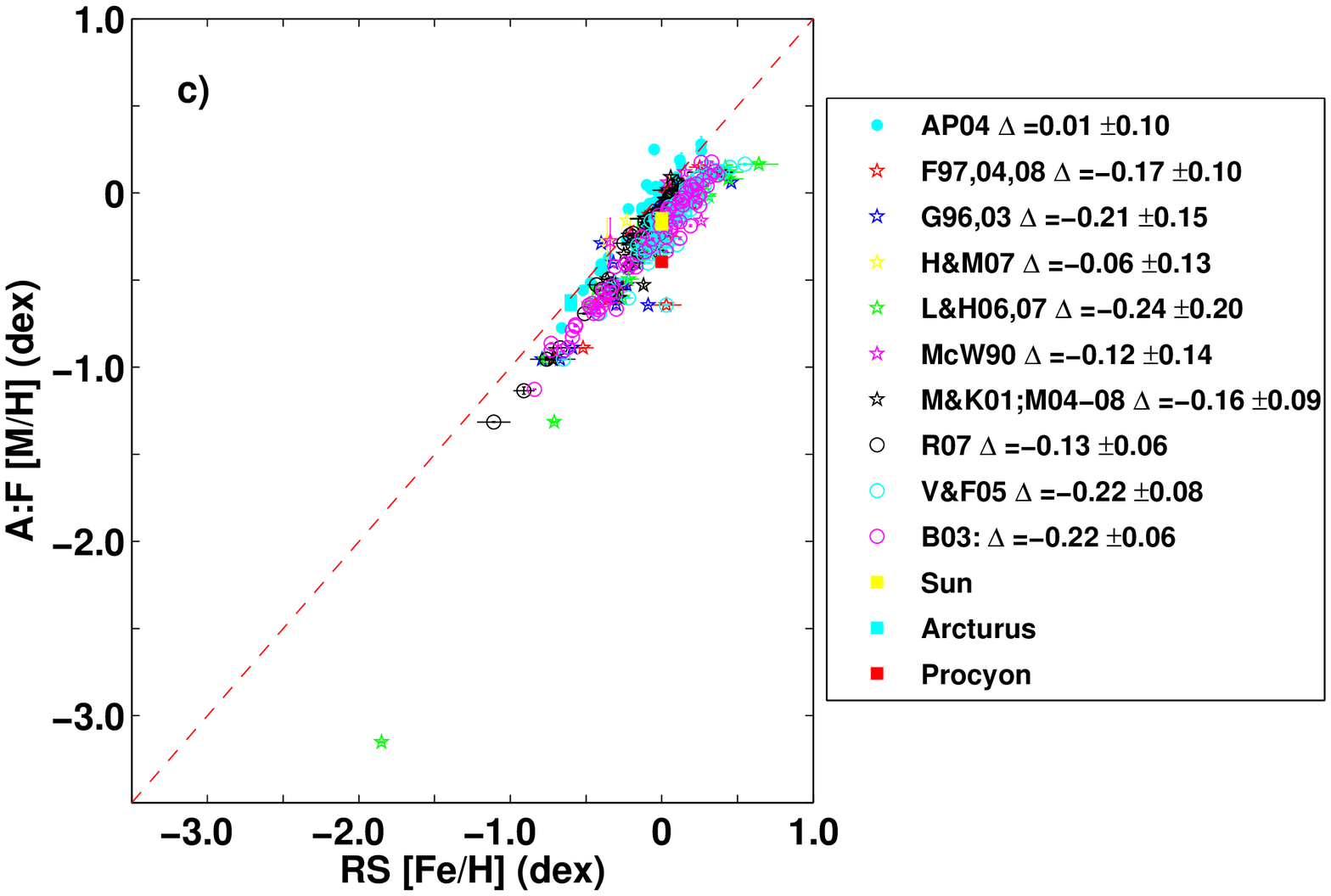}
%\caption{As for Figure~\ref{fig:feros_teff} but comparing the AMBRE:FEROS [M/H] values with the literature [Fe/H] values. The legend describes the relevant samples and their biases.}
%\label{fig:feros_mh}
\end{minipage}
%\end{figure}
%\hspace{0.2cm}
\hfill
%\begin{figure}[h]
%\centering
\begin{minipage}{90mm}
%\vspace{-1cm}
\centering
\includegraphics[width=85mm]{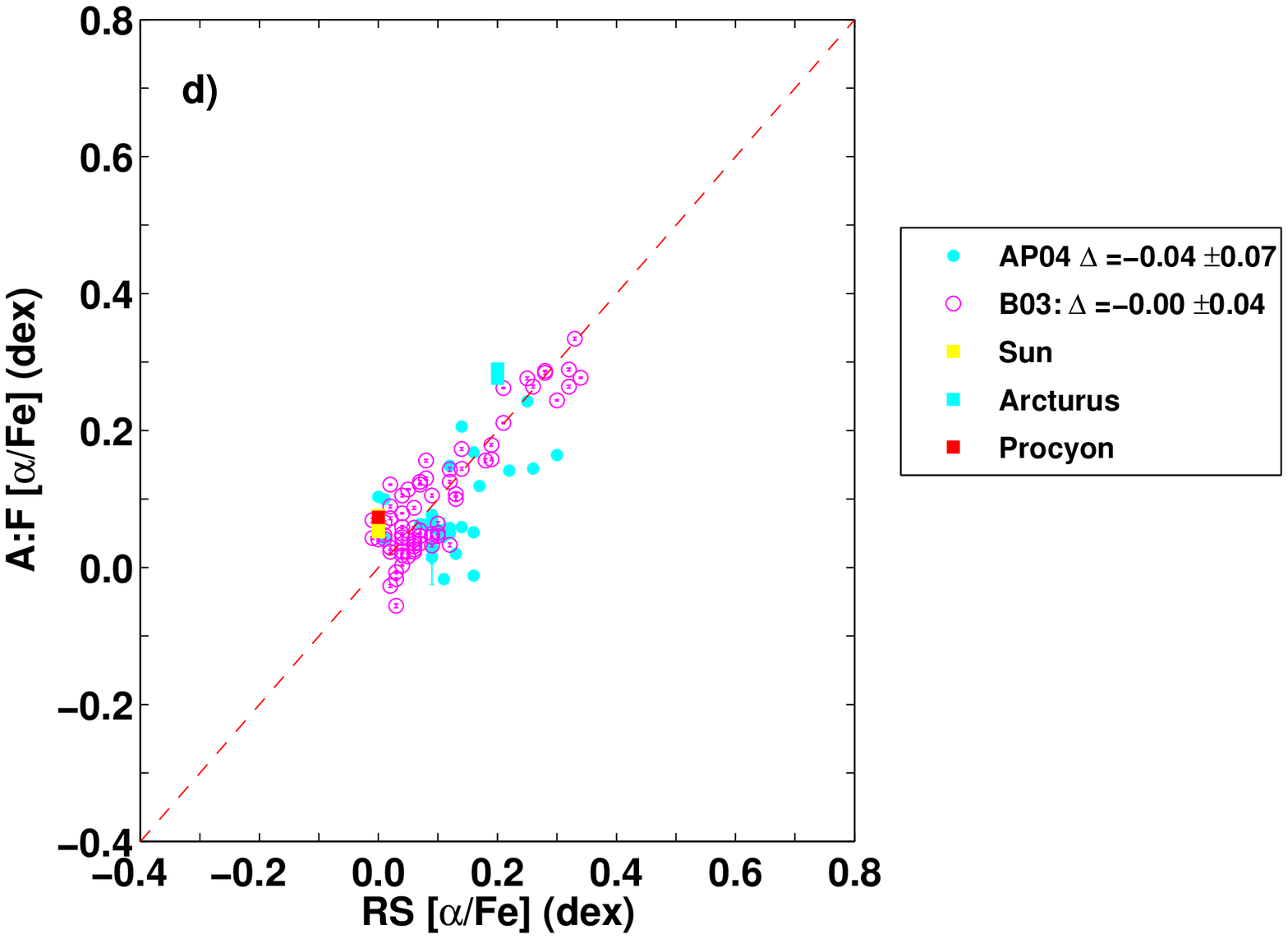}
%\caption{As for Figure~\ref{fig:feros_teff} but comparing the [$\alpha$/Fe] for the S$^4$N sample, \cite{Bensby2003} and the specific test spectra.}
%\label{fig:feros_alphafe}
\end{minipage}
\caption{Comparison of stellar parameters determined in AMBRE:FEROS (A:F) and the reference sample (RS) from PASTEL: a) $T_{\textrm{eff}}$,  b) $\log g$, c) [M/H] compared with [Fe/H], d) [$\alpha$/Fe]. The legend identifies the relevant individual papers and the associated sample bias$\pm\sigma$. The stellar atlases are also shown. The dashed line indicates a 1:1 agreement. The axis limits represent the limits on the final accepted parameters.}\label{fig:feros_pastel}
\end{figure*}

Figure~\ref{fig:feros_pastel}a. to b compare the $T_{\textrm{eff}}$, $\log g$, [M/H] and [$\alpha$/Fe] values determined by AMBRE:FEROS with the literature values of the reference sample and stellar atlases. The legend in each provides the bias and $\sigma$ for each subsample. The axes represent the limits of the final accepted parameters and clearly the reference sample does not cover the entire grid as would be ideal. Despite this limitation the reference sample was sufficient to define bias corrections which were necessary to apply to the AMBRE:FEROS stellar parameters.

\subsubsection{Reference sample: $T_{\textrm{eff}}$}
Figure~\ref{fig:feros_pastel}a. compares the $T_{\textrm{eff}}$ for each star. It is important to note that many of the literature $T_{\textrm{eff}}$ values were not determined spectroscopically, as stated above. Despite the differences in measurement technique, overall there is excellent agreement between the AMBRE:FEROS results and the PASTEL values, and also good agreement for the stellar atlases. There is a slight turning over of the distribution from $T_{\textrm{eff}}\gtrsim\sim5750$~K for which a small bias correction was applied that will be quantified at the end of this section.

\subsubsection{Reference sample: $\log g$}
Figure~\ref{fig:feros_pastel}b. shows the comparision of the $\log g$ values for which there is reasonably good agreement. However the majority of the sample are dwarfs (high $\log g$), and the giants (low $\log g$) are under-represented. There is a slight gradient within the sample which resulted in a small bias correction being applied at high $\log g$ for the final results (see end of the section). The lack of a significant sample of giants made it difficult to accurately define a bias at low $\log g$, hence the correction was only applied at high $\log g$. However the uncertainty on the giants is definitely higher.

\subsubsection{Reference sample: [M/H]}
Figure~\ref{fig:feros_pastel}c. shows the comparison of the AMBRE:FEROS [M/H] values with the literature [Fe/H] values. The difference in definition between [M/H] and [Fe/H], as discussed previously, means that this is not an accurate comparison, as so many more elements (albeit of lesser contribution) than just Fe are included in the MATISSE metallicity. The comparison shows an overall systematic offset between the AMBRE:FEROS and literature values reflected in the subsample biases. The sources of this bias are difficult to quantify so we assumed a direct comparison between [Fe/H] and [M/H] then made a systematic bias correction to the final [M/H] results as outlined at the end of this section.

\subsubsection{Reference sample: [$\alpha$/Fe]}
Unfortunately very few comparison stars with published values of their [$\alpha$/Fe] exist in the literature. The stellar parameters in the S$^4$N library \citep{Allende-Prieto2004} also include $\alpha$ element abundances where possible, which provided some comparisons with the AMBRE:FEROS [$\alpha$/Fe] results. The S$^4$N $T_{\textrm{eff}}$ values were determined using photometric calibrations while the $\log g$ were determined from {\it Hipparcos} parallaxes. The chemical abundances were determined using $\chi^2$ minimisation of the spectral line profile between the observed spectrum and a grid of synthetic spectra \citep{Allende-Prieto2004}.

A key study that was investigated specifically in order to validate the AMBRE:FEROS [$\alpha$/Fe] values was \cite{Bensby2003}. This study was a detailed analysis of 66 F and G dwarf stars in the galactic disk, and accurate abundances were determined for the $\alpha$ elements, Mg, Si, Ca, Ti. The mean of these abundances was taken as the value for the global [$\alpha$/Fe] with which to compare the AMBRE:FEROS [$\alpha$/Fe] values. This study was particularly useful because the spectra analysed in \cite{Bensby2003} were observed with FEROS. However the observations took place in 2000 and 2001 and so the spectra were not part of the archived sample delivered to AMBRE. The original spectra were obtained (Bensby, private communication) and analysed in the AMBRE:FEROS pipeline. Hence a direct comparison between AMBRE:FEROS and \cite{Bensby2003} could be made for all four parameters (See Figures~\ref{fig:feros_pastel}a to c). 

Figure~\ref{fig:feros_pastel}d. shows the comparison of the AMBRE:FEROS [$\alpha$/Fe] results with the values from the S$^4$N library, \cite{Bensby2003} and the stellar atlases. The biases are also listed for each sample. There is reasonably good agreement with these samples. In particular there is excellent agreement in the results for the \cite{Bensby2003} stars, which as stated above, is a study where the $\alpha$ element abundances were carefully determined, thereby providing an excellent validation of the AMBRE:FEROS [$\alpha$/Fe] results.

\subsection{Bias corrections}
The sample of reference stars and the spectral atlases provided a crucial comparison at all stages of the development of the AMBRE:FEROS analysis pipeline. Although it was not possible to cover the entire parameter space, and the reference sample was ultimately biased towards metal-rich dwarfs over a small temperature range, the results provided sufficient information with which to identify biases within the analysis, assuming the literature values to be correct. The following corrections were made in order to remove these biases:
{\small
\begin{minipage}{80mm}
\begin{equation*}
{\footnotesize
T_{\textrm{eff}}(cor)= 
\begin{cases} T_{\textrm{eff}} - 35 & \text{if $T_{\textrm{eff}} < 5300$~K,}
\\
T_{\textrm{eff}} + 0.21 \times T_{\textrm{eff}} - 1141.4 &\text{if $5300 \leq T_{\textrm{eff}} \leq 6000$~K,}
\\
T_{\textrm{eff}} + 110 &\text{if $T_{\textrm{eff}} > 6000$~K.}
\end{cases} }
\end{equation*}
\begin{equation*}
{\footnotesize
\log g(cor)= 
\begin{cases} \log \ g - 0.296 \times \log g + 1.388 & \text{if $\log g \geq 4.0$,}
\\
\log g  + 0.204 &\text{if $3.75 \leq \log g < 4.0$,}
\\
\log g  + 0.817 \times \log g - 2.860 &\text{if $3.5 \leq \log g < 3.75$,}
\\
\log g &\text{if $\log g < 3.5$.}
\end{cases} }
\end{equation*}
\begin{align*}
\text{[M/H]}(cor) &= \text{[M/H]} + 0.15 \\
[\alpha\text{/Fe}](cor) &= [\alpha\text{/Fe}]
\end{align*}
\end{minipage}
}
\vspace{0.2cm}

There are several potential sources of the bias corrections for the AMBRE:FEROS analysis. First, the differing techniques, as well as differing spectral domains and resolutions, used in the determination of the stellar parameters by the reference studies compared with AMBRE:FEROS analysis may make a significant contribution to the difference in derived parameters. Second, while the normalisation process within the AMBRE analysis pipeline was designed to be as robust as possible, the normalisation of stellar spectra is an inherently complex problem, particularly over large wavelength ranges with many spectral features. For spectra of particular spectral types (e.g. cool, low gravity, metal-rich) the normalisation procedure may not be as robust as for less detailed spectra. Third, due to the length of the atomic and molecular linelist that was required to synthesise the AMBRE:FEROS wavelength regions, it was inefficient to carry out a calibration of the linelist to standard stars (i.e. the Sun). It was assumed that the number of lines was statistically sufficient to dampen the noise from ill-fitted spectral features but non-calibration of the line list may play some role in the degree of the bias corrections. Individually, and in combination, these are the most likely sources of the bias corrections for the AMBRE:FEROS stellar parameters.   

\subsection{External error}
The external error for each stellar parameter was determined from the above reference sample analysis. For each individual study the mean difference and spread in differences ($\sigma$) between the literature and AMBRE:FEROS values was calculated. The mean of these $\sigma$ values for each stellar parameter was taken as the global external error ($\sigma(\theta)_{ext}$) for all of the spectra as follows:

\begin{minipage}{85mm}
\begin{align*}
\sigma(T_{\textrm{eff}})_{ext} &= 120 \text{~K} \\
%\sigma(\log \ g)_{ext} &= 0.2 \text{~dex} \\
\sigma(\log \ g)_{ext} &= 
\begin{cases}
 0.20 \text{~dex} &\text{if $\log g \geq 3.2$,}\\
 0.37 \text{~dex} &\text{if $\log g < 3.2$,}\\
\end{cases}\\
\sigma(\text{[M/H]})_{ext} &= 0.10 \text{~dex} \\
\sigma([\alpha/\text{Fe]})_{ext} &= 0.10 \text{~dex} \\
\end{align*}
\end{minipage}

\section{ESO Table: Rejection criteria}\label{sec:esotab_rej}                                                                                                                                                                                                                     
The construction of the ESO Table of stellar parameters that would be delivered to the ESO Archive was the final stage of the pipeline. At this stage the quality flags and tests included throughout the pipeline were drawn together to provide the final set of stellar parameters. The columns headings, their definitions, range of values, null values and rejection conditions used to construct the ESO Table are listed in Table~\ref{tab:esotab_descrip}.

There were two phases of rejecting spectra in the AMBRE:FEROS analysis. First, prior to analysis in SPC spectra were rejected due to spectral quality issues. This will be described in the Section~\ref{sec:nonstanspe}. Second, the remaining spectra were all analysed in SPC for their stellar parameters and then another set of rejection criteria were applied to construct the final table. These criteria are described and discussed in Sections~\ref{sec:rejcritafterspc} to \ref{sec:outsidegrid}.

\subsection{Pre-SPC: Non-standard spectra}\label{sec:nonstanspe}
Prior to SPC, 11\% of the FEROS archived spectra were rejected as being non-analysable primarily due to being non-standard spectra. The rejection flags, as defined in Section~\ref{sec:SPA}, were for `Faulty Spectra', `Extreme Emission Features',`Poor Normalisation' and `Excessive Noise'. Often the nature of a spectrum rejected at this stage meant the conditions for two or more of these flags were met. For instance, an extremely noisy spectrum was also consequently poorly normalised. Quality flags were also attached for `Large Emission Features' and `Instrumental Relics' but these spectra were not rejected based solely on these flags. The number of spectra identified for each rejection flag and for each quality flag are listed in Table~\ref{tab:esotab_rejflags}.

\begin{table}[htbp]
\caption{The number of spectra rejected (Rej) prior to SPC based on the defined rejection flags and the number of spectra accepted (Acc) for SPC but identified by the quality flags.}\label{tab:esotab_rejflags}
\begin{center}
\begin{tabular}{cccc}
\hline\hline
 & No. & \% of & \% of \\ 
 & Spectra & FEROS & Rej/Acc Sample \\ 
\hline
FEROS & 21551 &  &  \\ 
{\bf Rejected before SPC} & {\bf 2370} & {\bf 11} &  \\ 
Faulty Files & 71 & 0.33 & 3.0 \\ 
Extreme Emission Features & 1455 & 6.75 & 61.4 \\ 
Excessive Noise & 109 & 0.51 & 4.6 \\ 
Poor Normalisation & 735 & 3.41 & 31.0 \\
\\ 
{\bf Accepted for SPC} & {\bf 19181} & {\bf 89} &  \\ 
Large Emission Features & 3177 & 14.7 & 16.6 \\ 
Instrumental Relics & 58 & 0.27 & 0.30 \\ 
\hline
\end{tabular}
\end{center}
\end{table}

\subsection{Rejection criteria post-SPC}\label{sec:rejcritafterspc}
\begin{figure*}[!th]
\centering
\begin{minipage}{180mm}  %for one column use 90, 90
\centering
\includegraphics[width=180mm]{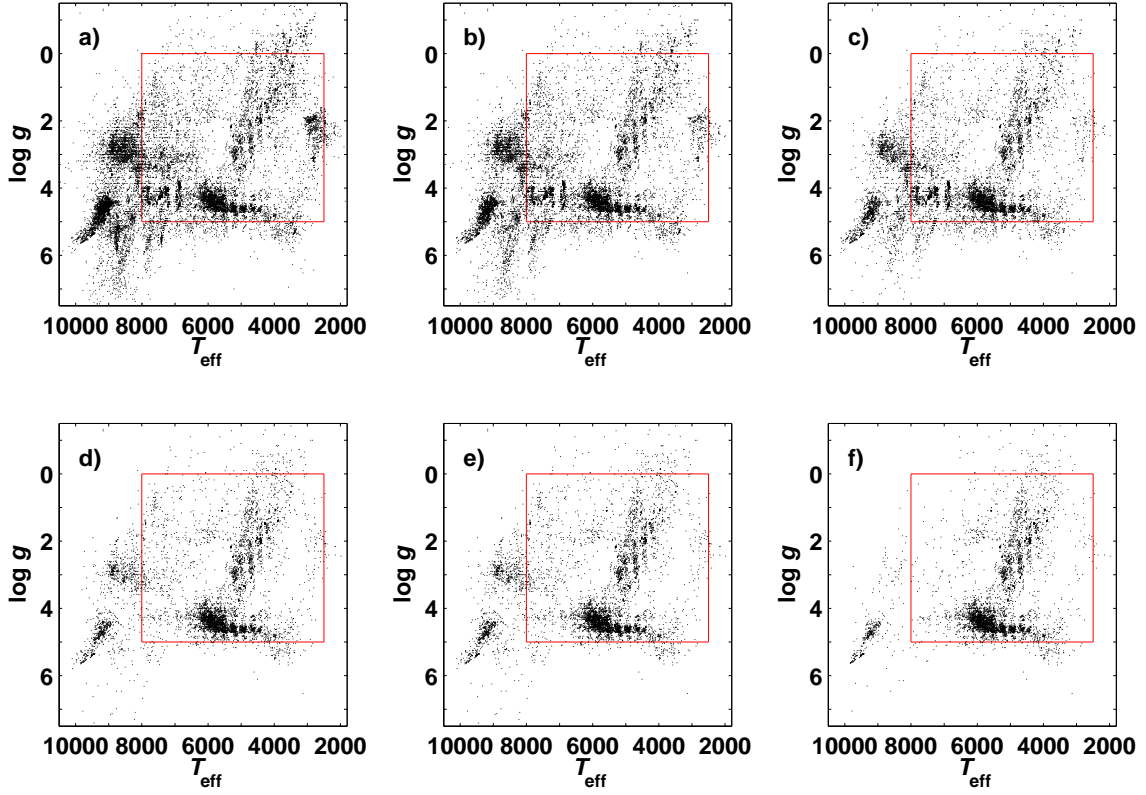}
\caption{HR diagrams of the FEROS stellar parameters as each rejection criterion is applied as follows: a) All spectra analysed in SPC; b) Spectra with V$_{rad}$ CCF with positive contrast, $\frac{\sigma_{Amp}}{Amp} \leq 0.2$ and $\frac{\sigma_{Cont}}{Cont} \leq 0.1$; c) $\sigma_{\textrm{V}_{rad}} \leq 10$~kms$^{-1}$; d) CCF FWHM $\leq$ 40~kms$^{-1}$; e) FWHM of medium strength spectral lines $\leq$ 0.8~m\AA; f) FWHM of strong spectral lines $\leq$ 8~m\AA. The $T_{\textrm{eff}}$ and $\log g$ boundaries of the synthetic grid are indicated in red.}\label{fig:hr_criteria}
\end{minipage}
\end{figure*}

The remaining 89\% of the FEROS spectra were analysed in SPC and so stellar parameters were obtained for each of these spectra. Figure~\ref{fig:hr_criteria}a shows the HR diagram for all the stellar parameters obtained in SPC. The giant branch and main sequence can be observed but there is a great deal of mis-classification at hot temperatures for $\log g \gtrsim 2$~dex, and an overdensity at $T_{\textrm{eff}} \approx 3000$~K and $\log g \approx 2$~dex. Most noticeable is that there are many spectra (8947) for which the derived parameters lie outside the $T_{\textrm{eff}}$ and $\log g$ grid limits which are indicated in red. In the construction of the ESO Table it would have been very straightforward to just reject all spectra whose parameters lay outside the limits of the synthetic grid and so only report to ESO those spectra with parameters inside the grid. However to have the synthetic grid limits as the sole reason to reject so many spectra ($\sim 47\%$) without further explanation seemed inadequate. Thus we further investigated the observational and astrophysical characteristics (indicators) of the spectra in order to better understand the dataset and to be able to impose more robust quality control criteria on the final dataset to be reported to ESO.

The SNR and $\log \chi^2$ were investigated as potential indicators but both proved to be functions of spectral type (the more spectral features the lower the SNR, the poorer the observed to synthetic spectra fit) and the criteria rejected RGB stars rather than the high temperature mis-classifications. 

Indicators relating to the measurement of the spectral FWHM and the radial velocity proved to be the least affected by spectral type and better identified the mis-classifications. The key indicators that were explored were: the V$_{rad}$ CCF contrast, the error on the amplitude and the error on continuum of the V$_{rad}$ CCF; the V$_{rad}$ error; the FWHM of the V$_{rad}$ CCF; the spectral FWHM; and the AMBRE:FEROS synthetic grid boundaries in $T_{\textrm{eff}}$, $\log g$, [M/H], [$\alpha$/Fe].

Table~\ref{tab:postspc_rejcrit} summarises the number of spectra that satisfied the conditions of each rejection criteria for the spectra rejected before and after SPC. The post-SPC rejection criteria were applied to the spectra rejected before SPC to further characterise that sample. To illustrate the effects of the rejection criteria on the sample of spectra analysed in SPC, Figure~\ref{fig:hr_criteria} shows progressively the HR diagram of the AMBRE:FEROS stellar parameters as each criteria is applied. The limits of the synthetic grid in $T_{\textrm{eff}}$ and $\log g$ are also shown. This process is discussed in the following sections.

\begin{table}[!h]
\centering
\caption{Summary of the number of spectra flagged as satisfying the conditions of the post-SPC rejection criteria. The spectra that were rejected prior to the SPC analysis were rejected due to other quality issues (see Section~\ref{sec:nonstanspe}). The spectra analysed in SPC were accepted or rejected based on the listed criteria.}
\begin{tabular}{ccc}
\hline\hline
 & \multicolumn{2}{l}{Spectra rejected} \\
 & pre-SPC & post-SPC \\
\hline
Total & 2370 & 12673 \\
PD CCF$^a$ & 549 & 3668 \\
$\sigma_{\textrm{V}_{rad}}$ $^b$ & 309 & 6348 \\   
CCF$^c$ & 469 & 10062 \\
FWHM$^d$ & 7 & 1199 \\
Grid$^e$ &  $-$  & 8494 \\  
 &  & \\ 
\multicolumn{3}{l}{a. Poorly-Defined CCF (Section~\ref{sec:PDCCF})}  \\  
\multicolumn{3}{l}{b. $\sigma_{\textrm{V}_{rad}}>10$~km/s (Section~\ref{sec:highve})} \\
\multicolumn{3}{l}{c. FWHM CCF~$>$~40~km/s (Section~\ref{sec:highccffwhm})} \\ 
\multicolumn{3}{l}{d. FWHM$_{medium}$~$ > 0.8$~m\AA\ or} \\
\multicolumn{3}{l}{\hspace{0.2cm} FWHM$_{strong}$~$ > 8$~m\AA\ (Section~\ref{sec:highfwhm})} \\ 
\multicolumn{3}{l}{e. Parameters lie outside grid parameter space (Section~\ref{sec:outsidegrid})} \\ 
\hline
\end{tabular}
\label{tab:postspc_rejcrit}
\end{table}

\subsubsection{CCF contrast, amplitude \& continuum}\label{sec:PDCCF}
As described in Section~\ref{sec:vrad_determination}, the radial velocity was determined using binary masks to calculate a CCF. The contrast, the depth (amplitude) of the CCF, the continuum placement of the CCF and their associated errors (as calculated in the IDL:GAUSSFIT routine) are measures of the quality of the CCF. The rejection criteria were derived from these quantities as follows:

\begin{enumerate}
 \item A negative contrast means that the CCF profile is inverted, which is contrary to the expected result, hence all spectra with a negative contrast were rejected. \\
 \item The relative error on the amplitude of the CCF ($\frac{\sigma_{amp}}{Amp}$) is mainly a measure of how much noise there is in the CCF. If the $\frac{\sigma_{Amp}}{Amp} \leq 0.2$ then the noise makes up less than 20\% of the profile depth. We rejected spectra with $\frac{\sigma_{Amp}}{Amp} > 0.2$.\\
 \item The placement of the continuum is also an indication of how well the CCF was defined. The relative error on the continuum ($\frac{\sigma_{Cont}}{Cont}$) gives a measure of the noise in the continuum placement. We rejected all spectra with $\frac{\sigma_{Cont}}{Cont} > 0.1$.
\end{enumerate}

These three criteria identify spectra with poorly-defined CCFs. Combined they are the first criteria to be applied to the SPC dataset resulting in the rejection of a further 3668 spectra from the analysis. Figure~\ref{fig:hr_criteria}b shows the resulting HR diagram after these spectra were removed. The hot star mis-classifications and overdensity at $T_{\textrm{eff}} \approx 3000$~K are diminished.
%In total 4217 spectra were designated as having poorly-defined CCFs, hence 549 spectra had been previously rejected.

\begin{figure}[!h]
\centering
\begin{minipage}{90mm}  %for one column use 90, 90
\centering
\includegraphics[width=95mm]{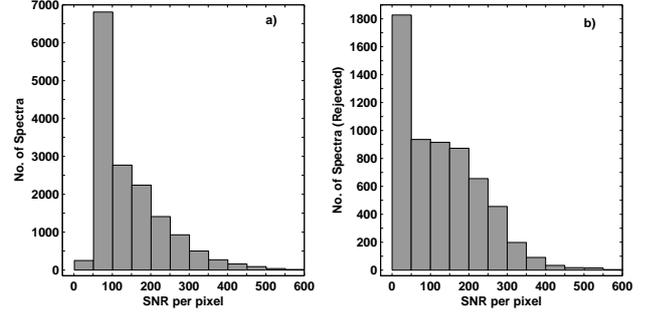}
\caption{Histogram of the measured SNR values of the FEROS archived spectra in bins of $\Delta$SNR~$=50$: a) Spectra defined as good quality and with well-defined CCFs (N=15513); b) Spectra rejected as poor quality and/or poorly defined CCFS (N=6038). The y-axes are to different scale to better display each sample.}\label{fig:snr_hist}
\end{minipage}
\end{figure}

Figure~\ref{fig:snr_hist}a and b show histograms of the SNR per pixel (0.03 or 0.06~\AA\ per pixel) for the full sample of FEROS archived spectra. The SNR has been calculated within the analysis pipeline using sigma clipping on either the predefined continuum regions in SPB or the pseudo-continuum residual in SPC. The highest SNR values do not necessarily mean optimum spectra and are identified by the quality flags within the pipeline as required. Figure~\ref{fig:snr_hist}a is the histogram of the spectra that were accepted as being of good quality and with a well-defined CCF, whereas Figure~\ref{fig:snr_hist}b is the histogram of the spectra rejected as being of poor quality or with poorly-defined CCF. Note that the y-axis scale is different between the figures. The histogram in Figure~\ref{fig:snr_hist}a peaks between 50 and 100 SNR. There are significantly fewer spectra with SNR$<$50 in Figure~\ref{fig:snr_hist}a compared with Figure~\ref{fig:snr_hist}b, indicating the cleaning thus far has indeed identified low SNR spectra.  However, as Figure~\ref{fig:snr_interr}a to d shows, even at SNR of 50 the internal error is negligible.

%\vspace{-0.5cm}

\subsubsection{V$_{rad}$ error}\label{sec:highve}
\begin{figure}[h]
\centering
\begin{minipage}{90mm}  %for one column use 90, 90
\centering
\includegraphics[width=95mm]{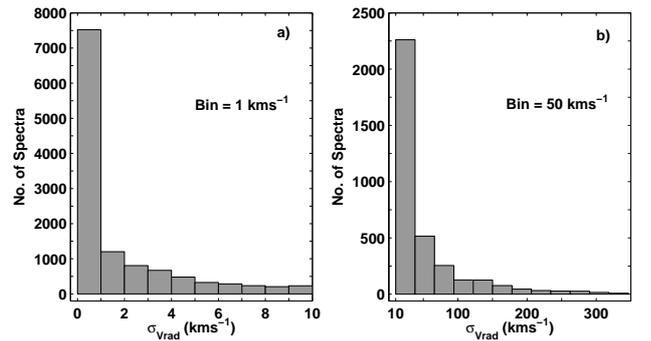}
\caption{Histogram of measured V$_{rad}$ uncertainty for each of the FEROS archived spectra of good quality with a well-defined CCF. a) The distribution of the sample with $\sigma_{\textrm{V}_{rad}} < 10$~kms$^{-1}$. b) The distribution of the sample with $\sigma_{\textrm{V}_{rad}} > 10$~kms$^{-1}$.}\label{fig:vrad_hist}
\end{minipage}
\end{figure}

The next criterion to be applied was the V$_{rad}$ error ($\sigma_{\textrm{V}_{rad}}$) which had been calculated using the prescription in \citet{Tonry1979}. Figure~\ref{fig:vrad_hist} shows histograms of the $\sigma_{\textrm{V}_{rad}}$ determined for each of the FEROS archived spectra of good quality with a well-defined CCF. Figure~\ref{fig:vrad_hist}a is a histogram of V$_{rad}$ error~$<10$~kms$^{-1}$ in bins of 1~kms$^{-1}$, and Figure~\ref{fig:vrad_hist}b is a histogram of $\sigma_{\textrm{V}_{rad}}$~$>10$~kms$^{-1}$ in bins of 10~kms$^{-1}$. The majority of the spectra have a low $\sigma_{\textrm{V}_{rad}}$ ($<0.5$~kms$^{-1}$), and based on the synthetic spectra analysis in Section~\ref{sec:vrad_determination}, Figure~\ref{fig:vrad_interr} shows that $\Delta$V$_{rad} \leq 5$~kms$^{-1}$  correspond to reasonable variations in the stellar parameters ($\theta$). However there are a significant number of spectra with $\sigma_{\textrm{V}_{rad}}$ greater than 10~kms$^{-1}$ (see Figure~\ref{fig:vrad_hist}b) which correspond to much larger uncertainties in the $\theta$ determination. 

%\vspace{-1cm}

Hence we decided to reject all spectra with $\sigma_{\textrm{V}_{rad}} > 10$~kms$^{-1}$. This resulted in a further 3550 spectra being rejected from the SPC dataset. The resulting HR diagram is shown in Figure~\ref{fig:hr_criteria}c. Again more of the mis-classifications at hot $T_{\textrm{eff}}$, particular for high $\log g$, have been removed and the overdensity at $T_{\textrm{eff}} \approx 3000$~K is considerably diminished.

%In total 6657 FEROS spectra were identified to have $\sigma_{V_{rad}} > 10$~kms$^{-1}$. This is 31\% of the FEROS sample and it was necessary to understand further why so many spectra had large $\sigma_{V_{rad}}$. This is discussed in detail in Section~\ref{sec:understand_rejspe}.

\begin{figure*}[!th]
\centering
\begin{minipage}{180mm}  %for one column use 90, 90
\centering
\includegraphics[width=170mm]{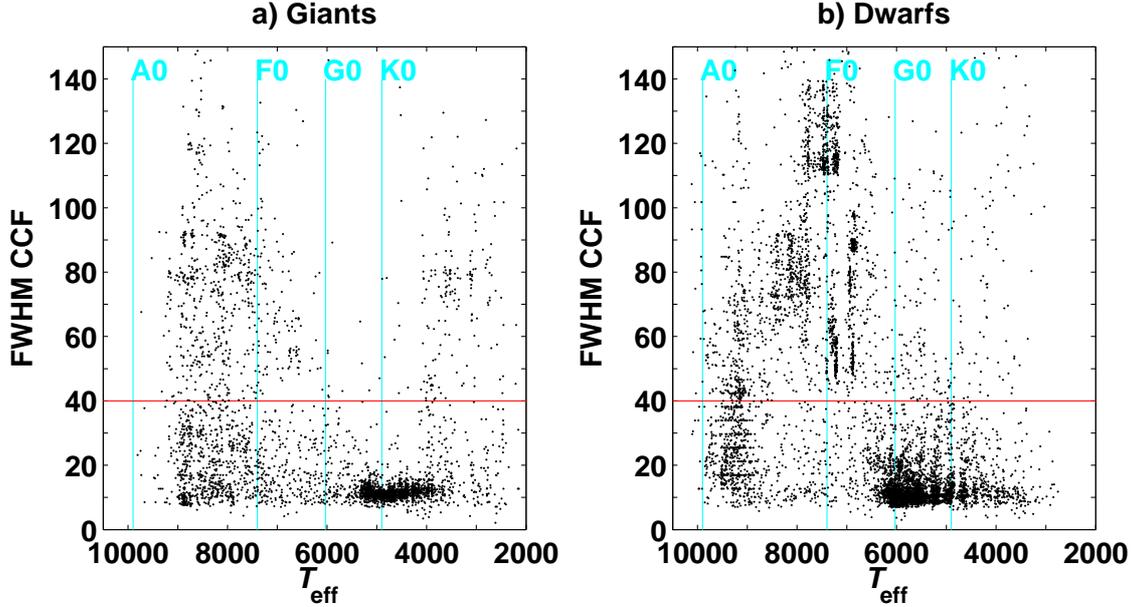}
\caption{$T_{\textrm{eff}}$ vs FWHM of the CCF for a) Giants ($\log g < 3.5$~dex), b) Dwarfs ($\log g \geq 3.5$~dex). The nominal temperature values for spectral types A0, F0, G0, K0 are shown as cyan lines. A CCF FWHM of 40~kms$^{-1}$ is indicated in red.}\label{fig:teff_ccffwhm}
\end{minipage}
\end{figure*}

\subsubsection{FWHM of V$_{rad}$ CCF}\label{sec:highccffwhm}
Astrophysically, the FWHM of the V$_{rad}$ CCF can be used to calculate the rotational velocity ($Vsini$) of a star. That particular calculation was not carried out here, for the purposes of expediency, but the FWHM of the CCF was used to identify spectra with broadened spectral features that would not be well represented by the synthetic grid. As discussed in Section~\ref{sec:int_errV}, the synthetic spectra grid was generated with no variations in $Vsini$, assuming all stars to be slow rotators. The analysis in \citet{Gazzano2010} found that for a sample of FGK dwarfs $Vsini < 11$~kms$^{-1}$ produced good results in the stellar parameter determination by MATISSE. At a resolution of R$\sim$26,000 an upper limit of CCF FWHM $= 20$~kms$^{-1}$ was set. Due to the lower resolution (R$\sim$15,000), and hence greater masking of the effects of $Vsini$, of the AMBRE:FEROS analysis we sought to relax that criterion.

The comparison of spectral type with $Vsini$ is used to show the increase in the number of fast rotators with hotter spectral type for both dwarfs and giants \citep[See Chapter 18,][]{Gray2005}. For dwarfs, fast rotators begin to appear at approximately F2 in spectral type, for giants at approximately G2. Using the stellar parameters derived in SPC and the spectral dataset that remained after the application of the $\sigma_{\textrm{V}_{rad}}$ rejection criterion, we replicated Spectral Type vs $Vsini$ using $T_{\textrm{eff}}$ and the CCF FWHM as proxies. The spectra were separated into subsamples of dwarfs and giants using the derived values for $\log g$. The comparisons are shown in Figure~\ref{fig:teff_ccffwhm}a and b.

The location of the Spectral Types in temperature are indicated (cyan lines). There is a clear dense grouping of spectra between 4000 and 6000~K with CCF FWHM less than 20 for both giants and dwarfs. Both samples have a great deal of scatter, particularly for $T_{\textrm{eff}}$ greater than 7000~K. Above the expected spectral type of F2 for the giants there is more dense scatter at higher CCF FWHMs. For the dwarfs the scatter becomes denser around 10000~K. A CCF FWHM of 40~kms$^{-1}$ is indicated in both cases (red line). This threshold was selected as being a factor of two greater than the threshold in \citet{Gazzano2010}, as the resolution of the AMBRE:FEROS synthetic grid is approximately a factor of two lower in resolution than that study. Hence the astrophysical broadening is masked to at least a factor of 2 in the convolved FEROS spectra. Above the 40~kms$^{-1}$ threshold the distribution of the points is sporadic while below this value the distribution is more coherent. However the majority of the points lie below 20~kms$^{-1}$ in line with \citet{Gazzano2010}. Ultimately we decided to reject all points with a CCF FWHM greater than 40~kms$^{-1}$.

The effects of this cleaning on the HR diagram are shown in Figure~\ref{fig:hr_criteria}d. A significant number of mis-classifications with $T_{\textrm{eff}} > 7000$~K have been removed as expected from the threshold shown in Figure~\ref{fig:teff_ccffwhm}. These rejected stars should be analysed with a hotter synthetic spectra grid with variations in $Vsini$.

\subsubsection{FWHM of the stellar spectrum}\label{sec:highfwhm}
\begin{figure*}[!th]
\centering
\begin{minipage}{180mm}  %for one column use 90, 90
\centering
\includegraphics[width=190mm]{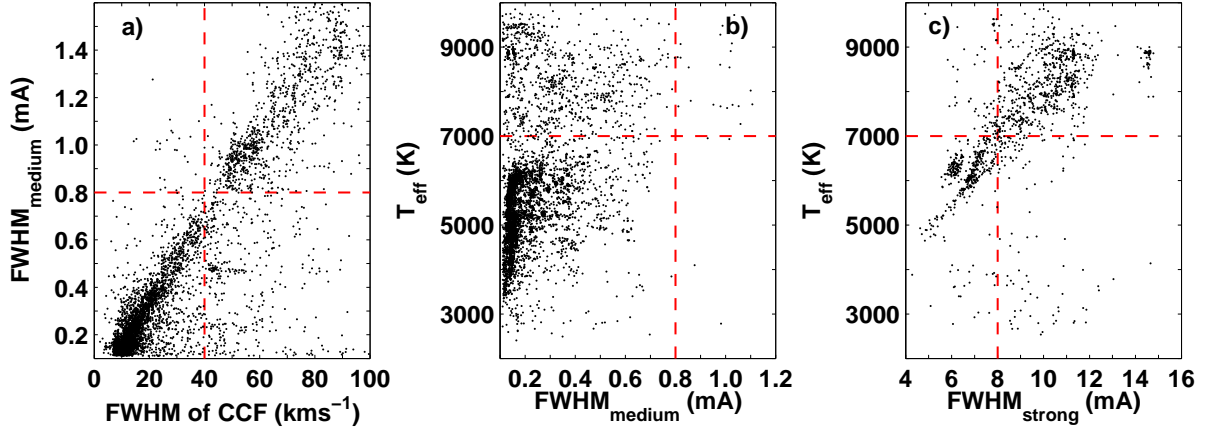}
\caption{Comparison of a) FWHM$_{medium}$ with the CCF FWHM, b) $T_{\textrm{eff}}$ for the dataset with CCF FWHM~$<$~40~kms$^{-1}$ compared to the FWHM$_{medium}$, and c) as for b) but for FWHM$_{strong}$. Thresholds at CCF FWHM~=~40~kms$^{-1}$, FWHM$_{medium}$~=~0.8~m\AA, FWHM$_{strong}$~=~8~m\AA, and  $T_{\textrm{eff}}=7000$~K are also shown.}\label{fig:fwhmspe_ccfteff}
\end{minipage}
\end{figure*}

The calculation of the spectral FWHM was necessary in order to convolve the observed spectra to the resolution of the synthetic grid. However the FWHM values also provide an extra degree of information with which to understand the spectral dataset. The FWHM measurements were separated into FWHM$_{weak}$, FWHM$_{strong}$ and FWHM$_{medium}$ classifications (Section~\ref{sec:SPB}). In particular the FWHM$_{strong}$ classification allowed us to identify spectra with broad features, independantly of the FWHM of the V$_{rad}$ CCF. For each FWHM$_{strong}$ spectra, the FWHM of the medium strength lines was measured (where possible) as well as the FWHM of the identified strong lines. Figure~\ref{fig:fwhmspe_ccfteff}a, b and c show the relationships between FWHM$_{medium}$ and CCF FWHM, FWHM$_{medium}$ and $T_{\textrm{eff}}$, and FWHM$_{strong}$ and $T_{\textrm{eff}}$ respectively for the dataset as of the application of the $\sigma_{\textrm{V}_{rad}}$ criterion. 

In Figure~\ref{fig:fwhmspe_ccfteff}a there is a clear trend of increasing CCF FWHM with FWHM$_{medium}$. The line at FWHM$_{medium}$ = 0.8~m\AA\ intersects the trend at the threshold of CCF FWHM = 40~kms$^{-1}$ where the trend becomes less dense. This provided the first criterion that we applied such that any spectra with a measured FWHM$_{medium}$ greater than 0.8~m\AA\ were rejected.

Figure~\ref{fig:hr_criteria}e shows the HR diagram after the removal of spectra based on this criterion. There is some improvement such that the density of the hot mis-classifications is decreased. However there are still a significant number of mis-classifications at temperatures greater than 7000~K. Further investigation showed that these remaining mis-classifications were connected to the `strong' FWHM classification (FWHM$_{strong}$). 

There was no clear trend between the CCF FWHM and the FWHM$_{strong}$ with which to derive a threshold of rejection. Instead a comparison was made to $T_{\textrm{eff}}$. Figure~\ref{fig:fwhmspe_ccfteff}b compares the $T_{\textrm{eff}}$ with the corresponding FWHM$_{medium}$. The lower boundary of possible values of $T_{\textrm{eff}}$ does increase with FWHM$_{medium}$ and at the derived threshold of FWHM$_{medium}$ = 0.8~m\AA\ this trend intersects with a $T_{\textrm{eff}} = 7000$~K. For greater FWHM$_{medium}$ values the trend of the lowest $T_{\textrm{eff}}$ disappears into greater scatter, essentially attributing a temperature value to the FWHM$_{medium}$ threshold. This temperature value was used to infer a FWHM$_{medium}$ threshold for the strong spectral lines. 

Figure~\ref{fig:fwhmspe_ccfteff}c shows the measured FWHM$_{strong}$ (where possible) with the $T_{\textrm{eff}}$. There is distinct trend of increasing temperature with increasing FWHM$_{strong}$. The nominal temperature threshold at 7000~K intersects with this trend at a FWHM$_{strong}$ of 8~m\AA. At greater FWHM$_{strong}$ there is an increase in scatter and the trend becomes more dispersed. The greater spread in values for spectra with FWHM$_{strong}$ $>$ 8~m\AA, and for FWHM$_{medium}$ $>$ 0.8~m\AA, may indicate a greater uncertainty in the parameterisation as the boundary of the grid is approached.

Based on this we can assume that spectra with a FWHM$_{strong}$ not more than 8~m\AA\ have stellar parameters that fall within the synthetic grid and thus are well-defined. Hence this provided the second criterion to apply such that spectra with a FWHM$_{strong}$ greater than 8~m\AA\ were rejected. Figure~\ref{fig:hr_criteria}f shows that indeed the mis-classifications between 7000~K and 8000~K are significantly reduced after the application of this criterion.

\subsection{Understanding the rejected spectra}\label{sec:understand_rejspe}
\begin{figure}[!b]
\centering
\begin{minipage}{90mm}  %for one column use 90, 90
\centering
\includegraphics[width=95mm]{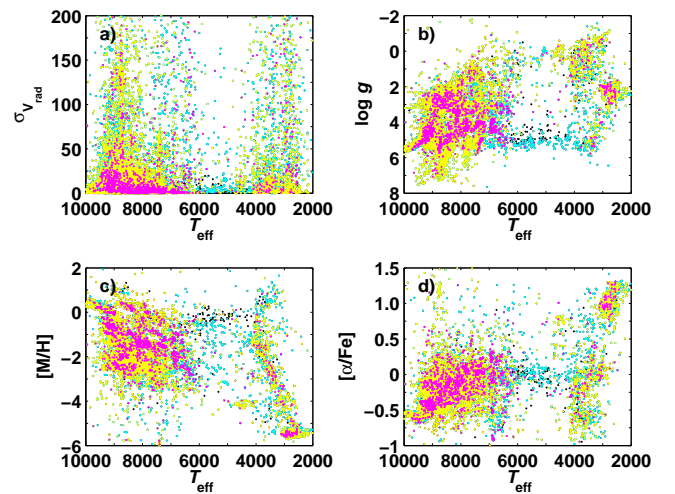}
\caption{Comparison of the $T_{\textrm{eff}}$ of the 11971 rejected spectra (black) with a) $\sigma_{\textrm{V}_{rad}}$, b) $\log g$, c) [M/H] and d) [$\alpha$/Fe]. The datasets for the FWHM$_{medium}$ rejections (pink), FWHM$_{strong}$ rejections (yellow) and CCF FWHM rejections (cyan) are shown.}\label{fig:variables_rej}
\end{minipage}
\end{figure}

Figure~\ref{fig:variables_rej} explores the types of stars to which the 11971 rejected spectra correspond. Figures~\ref{fig:variables_rej}a compares $T_{\textrm{eff}}$ with $\sigma_{\textrm{V}_{rad}}$ for the rejected sample. It is clearly bimodal with the primary peak at 8190$\pm$924~K (10068 spectra) and the secondary peak at 3351$\pm$679~K (1903 spectra). The datasets corresponding to the key rejection criteria are also indicated: FWHM$_{medium}$ rejections (pink), FWHM$_{strong}$ rejections (yellow) and CCF FWHM rejections (cyan). The spectra rejected on the basis of FWHM$_{medium}$ and FWHM$_{strong}$ are concentrated in the two peaks. This argues that at least 84\% of these spectra with broad features have returned parameters greater than the 8000~K temperature limit of the synthetic spectra grid.

%Figure~\ref{fig:variables_rej}a to d compares the $\sigma_{\textrm{V}_{rad}}$ of the 11971 rejected spectra with the corresponding values for the FWHM$_{medium}$, the FWHM$_{strong}$, the CCF FWHM and $T_{\textrm{eff}}$. Figure~\ref{fig:variables_rej}e is the HR diagram of these rejected spectra, while Figure~\ref{fig:variables_rej}f compares the $T_{\textrm{eff}}$ with [M/H].

Figure~\ref{fig:variables_rej}b is the HR diagram of the rejected spectra. The majority have $T_{\textrm{eff}} > 6000$~K, but also with $\log g > 2$~dex implying these are sub-giants and dwarf stars. It is possible that these spectra have parameters that lie outside the available synthetic grid parameter space and MATISSE has compensated by varying one or all of the stellar parameters away from the true values in order to converge on a solution. Combined with the high number of these spectra that have been measured to have broadened spectral features, it is possible that these are in fact hot and/or fast rotating stars for which the synthetic grid has not been designed.

In Figure~\ref{fig:variables_rej}a 16\% of the rejected sample are located near the cool temperature limit of the synthetic grid (3351$\pm$679~K). Are these truly such cool stars? Figure~\ref{fig:variables_rej}c compares the $T_{\textrm{eff}}$ with [M/H]. The cool temperature limit spectra show an interesting trend of decreasing [M/H] with decreasing temperature. For spectra with $T_{\textrm{eff}} < 3000$~K the [M/H] extends significantly outside the grid range. A possible answer is that these are hot or fast rotating stars (based on the FWHM measurements) for which MATISSE has converged to a significantly cooler temperature with significantly decreased metallicity. A similar but less dramatic argument can be made for the spectra rejected at the hot temperature limit. These lie within the [M/H] range for the synthetic grid although many are metal-poor to $-$3.5~dex. Combined with the high gravity values of these spectra the low metallicity emphasises the possiblity that these stars have been mis-classified due to their parameters lying outside the grid parameter space.

The comparison of [$\alpha$/Fe] to $T_{\textrm{eff}}$ in Figure~\ref{fig:variables_rej}d adds weight to this, as the cool temperature limit group have very high [$\alpha$/Fe] values. For the hot temperature limit group the 
[$\alpha$/Fe] are typically within the range of the synthetic grid.

From this examination of the key variables, the main argument that can be drawn as to why $\sim$66.1\% of the spectra analysed in SPC have been rejected by these very strict criteria is that they are most likely hot and/or fast rotating stars that are not well-represented by the current synthetic grid. Hence MATISSE has converged on solutions which mis-classify the spectra in the absence of a grid covering the required parameter range. This is not an unreasonable conclusion given that FEROS was designed and built by researchers within the hot star community \citep{Kaufer1999}. A visual inspection was carried out on large random samples of the spectra that were rejected based on each of the rejection criteria. This inspection supported the appropriate application of the criteria as the rejected spectra corresponded to noisy, peculiar or strong-lined spectra, and there was an over-representation of spectra comprising solely of strong balmer lines. However to confirm this conclusion these spectra can be re-analysed at a later date when a synthetic grid with $T_{\textrm{eff}} > 8000$~K and a synthetic grid with a range in $Vsini$ values will be available.

\subsection{Synthetic grid boundaries}\label{sec:outsidegrid}
The above rejection criteria succeeded in cleaning the stellar parameter sample without losing the well-defined branches of stellar evolution that are observed in the HR diagram. Finally, the stellar parameters determined near the boundaries of the synthetic grid are inherently less reliable due to there being less synthetic spectra present to aid in the definition of the parameters. Hence the final rejection criteria that were applied are based on the stellar parameters themselves. Taking into account boundary effects in the learning phase of MATISSE, the limits on the accepted parameters are defined as follows:

\begin{minipage}{85mm}
\begin{align*}
%{\footnotesize
3000 \ \leq \ &T_{\textrm{eff}} \ \leq 7625 \\
1 \ \leq \ &\log g \ \leq \ 5 \\
-3.5 \ \leq \ &\text{[M/H]} \ \leq \ 1 \\
-0.4 \ \leq \ &[\alpha/\text{Fe}] \ \leq \ 0.4 \hspace{0.5cm} \text{if [M/H] $\geq 0.0$} \\
%\\
-0.4 \ \leq \ &[\alpha/\text{Fe}] \ \leq \ 0.8 \hspace{0.5cm} \text{if $-1.0 < [M/H] < 0$} \\
%\\
0.0 \ \leq \ &[\alpha/\text{Fe}] \ \leq \ 0.8 \hspace{0.5cm} \text{if $[M/H] \leq -1$} \\
\end{align*}
\end{minipage}

Three categories were established using these definitions:
\begin{enumerate}
 \item TON: $T_{\textrm{eff}}$ values only within the accepted limits;
 \item TGM: $T_{\textrm{eff}}$, $\log g$ \& [M/H] values within accepted limits;
 \item TGMA: $T_{\textrm{eff}}$, $\log g$, [M/H] \& [$\alpha$/Fe] within accepted limits.
\end{enumerate}

The final count of spectra for these three categories were: TON = 394 spectra ($\sim$292 stars), TGM = 97 spectra ($\sim$87 stars), TGMA = 6017 spectra ($\sim$2780 stars). 

In total 12673 spectra were rejected from the SPC analysis. The final set of the stellar parameters accepted for delivery to ESO was comprised of 6508 spectra ($\sim$3087 stars). For the radial velocity analysis 11963 spectra ($\sim$4505 stars) were determined to have radial velocities with $\sigma_{\textrm{V}_{rad}} \leq 10$~kms$^{-1}$ and these values were also accepted for delivery to ESO. The full breakdown of the spectral number count is given in Table~\ref{tab:esotab_summary}.

\begin{table}[!h]
\centering
\caption{Number counts and percentages of spectra accepted (Acc.) or rejected (Rej.) during the AMBRE:FEROS analysis.}
\begin{tabular}{ccccc}
\hline\hline
 & No. Spectra & \% FEROS & \% SPC & No. Stars \\ 
\hline
FEROS & 21551 & $-$ & $-$ & 6285 \\ 
Rej. pre-SPC & 2370 & 11.0 & $-$ & 1118 \\ 
 &  &  &  &  \\ 
Acc. for SPC & 19181 & 89.0 & $-$ & 5500 \\ 
Rej. post-SPC & 12673 & 58.8 & 66.1 & 2549 \\ 
 &  &  &  & \\ 
\multicolumn{5}{c}{{\bf Accepted for ESO Archive}} \\
%{\bf Accepted for ESO Archive} &  &  &  &  \\ 
% &  &  &  & \\ 
V$_{rad}$ & 11963 & 55.5 & 62.4 & 4505 \\ 
 &  &  &  & \\ 
TON & 394 & 1.8 & 2.0 & 292 \\ 
TGM & 97 & 0.5 & 0.5 & 87 \\ 
TGMA & 6017 & 27.9 & 31.4 & 2780 \\ 
Total & 6508 & 30.2 & 33.9 & 3087 \\ 
% &  &  &  & \\ 
%\multicolumn{5}{l}{(a. $\sigma_{\textrm{V}_{rad}} \leq 10$~kms$^{-1}$)} \\
\hline
\end{tabular}
\label{tab:esotab_summary}
\end{table}

\subsection{Repeated observations}
\begin{figure}[h]
\centering
\begin{minipage}{90mm}  %for one column use 90, 90
\centering
\includegraphics[width=90mm]{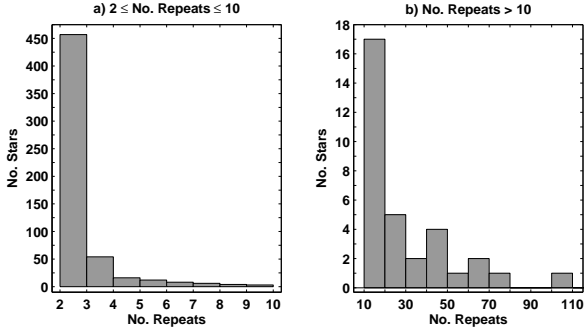}
\caption{Histogram of the humber of repeat observations in the TGMA sample for stars with: a) the number of observations less than or equal to 10; b) the number of observations greater than 10.}\label{fig:hist_repeat_obs}
\end{minipage}
\end{figure}

The FEROS dataset contains many stars with repeated observations. Figure~\ref{fig:hist_repeat_obs} shows histograms of the number of stars per number of observations for the TGMA sample. Figure~\ref{fig:hist_repeat_obs}a shows the histogram for the number of stars with observations less than or equal to 10, while Figure~\ref{fig:hist_repeat_obs}b shows the number of stars with observations greater than 10. There is a significant number of stars with 2 or more observations, and some stars have been observed a great many times. The repeated observations were used to assess whether the AMBRE analysis derived consistent stellar parameters for the same star from different spectra.

All objects with 2 or more observations were extracted from the TGMA sample in order to explore the agreement in parameters for a single object over a number of observations of similar SNR. We also extracted from SIMBAD a list of objects within the FEROS dataset (based on a 10'' coordinate search) that were designated as a binary and/or variable star. We found 17 binary stars and 111 variable stars in the final sample which we eliminated from the sample of repeat observations. Hence we removed those stars with possible astrophysical variations in the parameters to be left with variations that are solely due to the analysis process. (It must be noted that, as the list in SIMBAD is not exhaustive, it is likely that there are still stars for which the parameters vary astrophysically within the repeated observations sample.) The repeated observations sample was further constrained such that each set of repeats had a mean SNR less than 200, as the observations above SNR=200 were sparcely sampled. Due to the prior application of the rejection criteria there were no spectra present with SNR~$<$~40. Finally, the sample of repeats comprised of 584 stars equating to 3438 spectra. The maximum number of repeats was 73 for a single star in the repeats sample.

This sample represents the internal error of the AMBRE analysis hence we compared it to the internal error ($\sigma_{int}$) that we calculated for each spectra based on the SNR, V$_{rad}$ and Normalisation (Section~\ref{sec:int_error}).

\begin{figure}[h]
\centering
\begin{minipage}{90mm}  %for one column use 90, 90
\centering
\includegraphics[width=90mm]{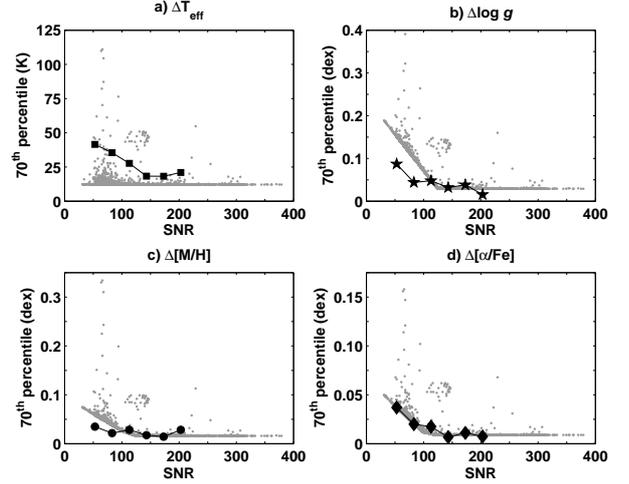}
\caption{Change in $\theta$ with SNR (in bins of 20) for the repeat observations within TGMA (black symbols) compared with the $\sigma(\theta)_{int}$ values (Section~\ref{sec:int_error}) for TGMA (grey points).  a) The $\Delta T_{\textrm{eff}}$ that 70\% of the repeat sample were less than or equal to at each SNR (black squares). The trend with SNR is shown by the black lines. b) As for a) but for variations in $\log g$ (black stars). c) As for a) but for variations in [M/H] (black circles). d) As for a) but for variations in [$\alpha$/Fe] (black diamonds).}\label{fig:snr_params_obs}
\end{minipage}
\end{figure}

\begin{figure*}[!th]
\centering
\begin{minipage}{180mm}  %for one column use 90, 90
\centering
\includegraphics[width=160mm]{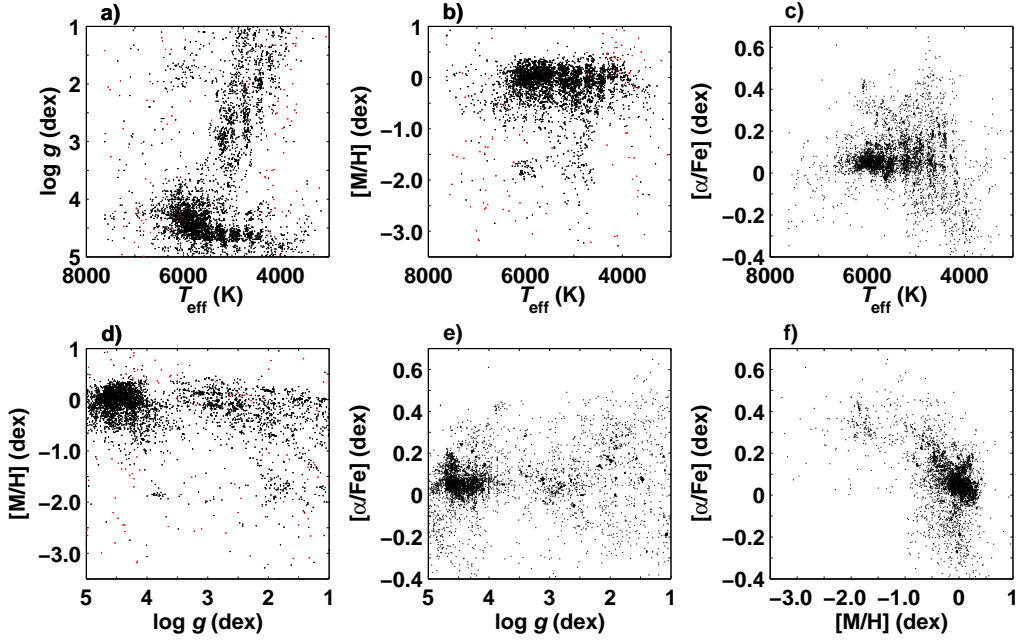}
\caption{The final FEROS stellar parameters for TGM (red) and TGMA (black) samples for a) The Hertzsrung-Russell diagram; b) [M/H] vs $T_{\textrm{eff}}$; c) [$\alpha$/Fe] vs $T_{\textrm{eff}}$; d) [M/H] vs $\log g$; e)  [$\alpha$/Fe] vs $\log g$; f) [M/H] vs [$\alpha$/Fe].}\label{fig:tlma_params}
\end{minipage}
\end{figure*}

Figure~\ref{fig:snr_params_obs} replicates the relationships in Figure~\ref{fig:snr_interr} but for the repeats sample. For each set of repeats the mean and standard deviation were calculated for the SNR and stellar parameters. The standard deviation of each of the parameters was used to represent the change in the parameters ($\Delta \theta$) for each set. The mean SNR values were binned (binsize = 25) and for each bin the 70$^{th}$ percentile of the $\Delta \theta$ was calculated. These values are shown in black. Also included in grey for each parameter are the internal errors calculated for the TGMA spectra ($\sigma(\theta)_{int}$, see Section~\ref{sec:int_error}).

In Figure~\ref{fig:snr_params_obs}a the variations in $T_{\textrm{eff}}$ with SNR for the repeats sample is greater than the calculated internal error distribution and almost traces out the upper limit on the internal error. However, in general these two independant measurements of the internal error are on the same order of magnitude confirming the internal consistency within the AMBRE:FEROS analysis.

As defined in Section~\ref{sec:int_error} the internal errors for $\log g$, [M/H] and [$\alpha$/Fe] are calculated as a piecewise function of SNR and this is reflected in the distribution of the internal errors for TGMA as shown in Figure~\ref{fig:snr_params_obs}b to d. The lower limit of the internal errors for each of these three parameters decreases with increasing SNR to the defined constant value at SNR = 125. This distribution is replicated in the variation of each parameter with SNR for the repeats sample. At low SNR there is an offset in $\log g$ and [M/H] between the two sets of values where the internal errors are of greater value. This implies that the internal errors may be overestimated at low SNR in light of the repeats sample analysis. For [$\alpha$/Fe] the variations due to the repeats sample trace the internal errors very well across the range of SNR values. But overall, as for $T_{\textrm{eff}}$, these two independant measurements of the internal error for each parameter are in very good agreement.

The distribution of the variation of the parameters with SNR for the repeats sample in Figure~\ref{fig:snr_params_obs} also replicates the analysis of the internal error due to SNR that was carried out using the synthetic spectra sample as shown in Figure~\ref{fig:snr_interr}. For the lowest SNR bin for each parameter the $\Delta \theta$ are only 2 or 3 times the equivalent value in the synthetic case which is reasonable, and as the SNR increases the $\Delta \theta$ decreases significantly also replicating the synthetic case. 

The comparison between the two measures of the internal error from the observed spectra and with the synthetic spectra analysis shows that the analysis carried out within the pipeline is consistent such that closely comparable parameters are found for different spectra of the same star, and so for different stars of similar stellar parameters, at a particular SNR value.

\section{The AMBRE:FEROS stellar parameters}\label{sec:AFstellarparams}                          
The final FEROS stellar parameters are shown in Figure~\ref{fig:tlma_params} in six combinations with which to examine the distribution of the parameters.

Figure~\ref{fig:tlma_params}a shows the Hertzsprung-Russell (HR) diagram of the TGM (red) and TGMA (black) samples. The branches of stellar evolution are distinct although broad in width with some small scatter evident throughout the figure. Figure~\ref{fig:tlma_params}b compares [M/H] with $T_{\textrm{eff}}$ and the majority of the spectra show near solar metallicities ($-1.0 <$~[M/H]~$< 0.5$~dex) consistently across the temperatures. 

Figure~\ref{fig:tlma_params}c compares [$\alpha$/Fe] with $T_{\textrm{eff}}$ and shows an interesting distribution of low ($< 0.0$~dex) and high ($> 0.2$~dex) [$\alpha$/Fe] at low temperatures but an even distribution of solar [$\alpha$/Fe] across all temperatures.

Figure~\ref{fig:tlma_params}d compares [M/H] with $\log g$ and the separation between the giants and dwarfs is evident. There is larger scatter in [M/H] at low $\log g$ (giants). 

Figure~\ref{fig:tlma_params}e compares [$\alpha$/Fe] with $\log g$ and again the scatter is larger at low $\log g$ (giants). 

Figure~\ref{fig:tlma_params}f compares [$\alpha$/Fe] with [M/H] for which there is a clear trend of enhanced [$\alpha$/Fe] ($\sim0.35$~dex) at low [M/H] ($<-1.0$~dex). For [M/H] of solar and above the [$\alpha$/Fe] decreases to solar and depleted values. 

\begin{figure}[h]
\centering
\begin{minipage}{90mm}  %for one column use 90, 90
\centering
\includegraphics[width=90mm]{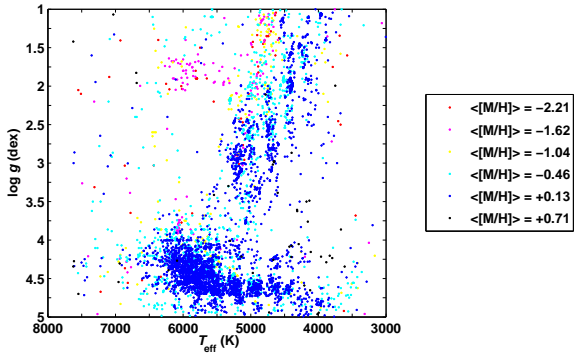}
\caption{The HR diagram of the TGM and TGMA FEROS stellar parameter samples binned by colour in [M/H]. The centre of each bin are listed in the legend.}\label{fig:tlma_final_hr}
\end{minipage}
\end{figure}

Combining three parameters in one graph provides another perspective with which to further explore this dataset. Figure~\ref{fig:tlma_final_hr} shows the HR diagram of the TGM and TGMA samples binned by colour in [M/H]. The near solar [M/H] bins contain the majority of the sample as expected from Figure~\ref{fig:tlma_params} across the Main Sequence and RGB. The samples within the more metal-poor bins lie to the left of the stellar evolution branches which agrees with stellar evolutionary tracks. Interestingly the majority of the Horizontal Branch (HB) are metal-poor ([M/H]~$=-1.62$~dex).

\section{Conclusions \& future work}\label{sec:conclusion}
The stellar parameterisation of the FEROS archive completes the first phase of the AMBRE Project. Of the 21551 FEROS object spectra that were delivered to OCA, 19181 could be analysed for their stellar parameters. The quality flags and error analysis resulted in a final total of 6508 spectra ($\sim$3087 stars) with stellar parameters to be delivered to ESO. Also delivered to ESO are the radial velocity values for 11963 spectra ($\sim$4505 stars)

Approximately 70\% of the spectra were rejected from the stellar parameterisation analysis. While 28\% were rejected due to the quality of the spectra being insufficient for analysis, 42\% were rejected due to astrophysical reasons. From this analysis it seems very likely that hot and/or fast rotating stars are the favoured observational object for FEROS. As the current synthetic grid is defined for stars cooler than 8000~K only then the high number of rejections was inevitable. In general the AMBRE synthetic spectra grid is currently configured for slow-rotating FGK stars. Non-standard stars, such as binaries and chemically peculiar stars, within the FEROS sample may still pollute the final accepted dataset as we were unable to identify such stars at this stage. 

However, stellar parameter extensions to the synthetic spectra grid are under development and identification tools are also being developed such that a larger range of spectra may be reliably analysed at a later date. 

In summary the work carried out in the analysis of the FEROS archived spectra has resulted in:

   \begin{enumerate}
      \item the development of a complex and robust analysis pipeline for the determination of stellar parameters of large spectral datasets through automated iterative spectral reduction and MATISSE analysis;
      \item the establishment of tools with which to exploit spectral atlases and libraries for the comprehensive testing and validation of the results from the AMBRE analysis pipeline;
      \item the determination of stellar parameters ($T_{\textrm{eff}}$, $\log g$, [M/H], [$\alpha$/Fe]) for 6508 of 21551 FEROS archived spectra to be made available in the ESO Archive, including quality flags.
      \item the determination of radial velocities (V$_{rad}$) for 11963 of 21551 FEROS archived spectra to be made available in the ESO Archive.
   \end{enumerate}

The AMBRE pipeline is currently being used to determine stellar parameters for the UVES and HARPS archived spectra, the next two phases of analysis in the AMBRE Project. The analysis of the FLAMES/GIRAFFE archived spectra is expected to commence in early 2012.

\begin{acknowledgements}
The AMBRE Project team members would like to thank ESO, OCA and CNES for their financial support of this project. We would also like to thank O.~Begin, Y.~Vernisse, S.~Rousset and F.~Guitton for their work on the development of the AMBRE analysis pipeline as well as the OCA Mesocentre members. We would like to thank C.~Soubiran and T.~Bensby for proving stellar parameters and spectra for testing the pipeline. We would like to thank B.~Plez for the use of Turbospectrum and the molecular linelists, and also C.~Melo for the use of the radial velocity programme. Thanks also to L.~Pasquini for initiating the project, and to M.~Romaniello and J.~Melnich for their help within ESO.

This research has made use of the SIMBAD database, operated at CDS, Strasbourg, France, as well as the NASA/IPAC Infrared Science Archive, which is operated by the Jet Propulsion Laboratory, California Institute of Technology, under contract with the National Aeronautics and Space Administration.
\end{acknowledgements}

\bibliographystyle{aa}
%Included for Gather Purpose only:
%input "\home\cworley\Projects\AMBRE\FEROS\FEROSPaper\PaperSAGA\worley_bib.bib"
%\setlinespacing{1.44}
%\bibliography{worley_bib}

\clearpage

\appendix

\section{Description of ESO Table for AMBRE:FEROS}
\begin{sidewaystable*}
\vspace{19cm}
%\begin{landscape}
%\begin{table*}[tbp]
\caption{Description of columns in the table of FEROS stellar parameters delivered to ESO.}
\label{tab:esotab_descrip}
\centering
{\tiny
\begin{tabular}{llccl}
\hline\hline
Keyword & Definition & Value range & Null value & Determination \\ 
\hline
DP\_ID & ESO data set identifier &  &  &  \\ 
OBJECT & Object designation as read in ORIGFILE &  &  &  \\ 
TARG\_NAME & Target designation as read in ORIGFILE &  &  &  \\ 
RAJ2000 & Telescope pointing (right ascension, J2000) & deg &  &  \\ 
DEJ2000  & Telescope pointing (declination, J2000) & deg &  &  \\ 
MJD\_OBS & Start of observation date & Julian Day &  &  \\ 
EXPTIME & Total integration time & sec &  &  \\ 
SNR & Signal-to-Noise Ratio as estimated by the pipeline & 0-$\infty$ & NaN &  \\ 
SNR\_FLAG & Signal-to-Noise Ratio quality flag & C,R &  & C=Crude estimate from SPA$^*$, R=Refined estimate from SPC$^\#$ \\ 
EXTREME\_EMISSION\_LINE\_FLAG & Detection of extreme emission lines. & T,F &  & T=True: detection therefore no analysis carried out,  \\ 
 &  &  &  & F=False: no detection therefore analysis carried out \\ 
EMISSION\_LINE\_FLAG & Detection of some emission lines & T,F &  & T=True: some emission lines detected but analysis carried out,  \\ 
 &  &  &  & F=False: no detection therefore analysis carried out \\ 
MEANFWHM\_LINES & Mean FWHM of absorption lines around 4500~\AA\ & 0-0.33 & NaN & FWHM measured from spectral features (m\AA) \\ 
MEANFWHM\_LINES\_FLAG & Flag on the mean FWHM & T,F &  & T=True: FWHM $>$ 0.33 or $<$ 0.11. Default FWHM values used \\ 
 &  &  &  & F=False: FWHM $<$ 0.33, $>$ 0.11 \\ 
VRAD & Stellar radial velocity & -500 to +500 & NaN & Units=kms$^{-1}$ \\ 
ERR\_VRAD & Error on the radial velocity & 0-$\infty$ & NaN & If $\sigma_{vrad} > 10$, null value used for all stellar parameters. Units=kms$^{-1}$ \\ 
VRAD\_CCF\_FWHM  & FWHM of the CCF between the spectrum and the binary mask & 0-$\infty$ & NaN & Units=kms$^{-1}$ \\ 
VRAD\_FLAG & Quality flag on the radial velocity analysis & 0,1,2,3,4,5 & -99 & 0=Excellent determination...5=Poor determination \\ 
TEFF & Stellar effective temperature ($T_{\textrm{eff}}$)  & 3000-7625 & NaN & Units=K. Null value used if $T_{\textrm{eff}}$ is outside accepted parameter \\ 
 & as estimated by the pipeline &  &  & limits or if the spectrum is rejected due to quality flags. \\ 
ERR\_INT\_TEFF & Effective temperature internal error & 0-$\infty$ & NaN & Units=K. Square root of quadrature sum of internal errors  \\ 
 &  &  &  & ($\sigma(T_{\textrm{eff}})_{int,snr}$, $\sigma(T_{\textrm{eff}})_{int,vrad}$ \& $\sigma(T_{\textrm{eff}})_{int,norm}$ \\ 
ERR\_EXT\_TEFF & Effective temperature external error & 120 & NaN & Units=K. Maximum expected error due to external sources \\ 
LOG\_G & Stellar surface gravity (log g) as estimated by the pipeline & 1-4.9 & NaN & Units=dex. Null value used if $\log g$ is outside accepted parameter \\ 
 &  &  &  & limits or if the spectrum is rejected due to quality flags. \\ 
ERR\_INT\_LOG\_G & Surface gravity internal error & 0-$\infty$ & NaN & Units=dex. Square root of quadrature sum of internal errors  \\ 
 &  &  &  & ($\sigma(\log g)_{int,snr}$, $\sigma(\log g)_{int,vrad}$ \& $\sigma(\log g)_{int,norm}$ \\ 
ERR\_EXT\_LOG\_G & Surface gravity external error & 0.2 & NaN & Units=dex. Maximum expected error due to external sources \\ 
M\_H & Mean metallicity [M/H] as estimated by the pipeline & 0-$\infty$ & NaN & Units=dex. Null value used if [M/H] is outside accepted parameter \\ 
 &  &  &  & limits or if the spectrum is rejected due to quality flags. \\ 
ERR\_INT\_M\_H & Mean metallicity internal error & 0-$\infty$ & NaN & Units=dex. Square root of quadrature sum of internal errors  \\ 
 &  &  &  & ($\sigma(\textrm{[M/H]})_{int,snr}$, $\sigma(\textrm{[M/H]})_{int,vrad}$ \& $\sigma(\textrm{[M/H]})_{int,norm}$ \\ 
ERR\_EXT\_M\_H  & Mean metallicity external error & 0.1 & NaN & Units=dex. Maximum expected error due to external sources \\ 
ALPHA & $\alpha$-elements over iron enrichment ([$\alpha$/Fe])  & -0.4 - 0.4 & NaN & Units=dex. Null value used if [$\alpha$/Fe] is outside accepted parameter  \\ 
 & as estimated by the pipeline &  &  & limits or if the spectrum is rejected due to quality flags. \\ 
ERR\_INT\_ALPHA & $\alpha$-elements over iron enrichment internal error & 0-$\infty$ & NaN & Units=dex. Square root of quadrature sum of internal errors  \\ 
 &  &  &  & ($\sigma(\textrm{[$\alpha$/Fe]})_{int,snr}$, $\sigma(\textrm{[$\alpha$/Fe]})_{int,vrad}$ \& $\sigma(\textrm{[$\alpha$/Fe]})_{int,norm}$ \\ 
ERR\_EXT\_ALPHA & $\alpha$-elements over iron enrichment external error & 0.1 & NaN & Units=dex. Maximum expected error due to external sources \\ 
CHI2 & log($\chi^2$) of the fit between the observed and the & 0-$\infty$ & NaN & Goodness of fit between final normalised  \\ 
 & reconstructed synthetic spectrum at the MATISSE parameters &  &  & and final reconstructed spectra \\ 
CHI2\_FLAG & Quality flag on the fit between the observed and the  & 0,1,2  & -99 & 0=Good fit...2=Poor fit \\ 
 & reconstructed synthetic spectrum at the MATISSE parameters &  &  &  \\ 
ORIGFILE  & ESO filename of the original spectrum being analysed &  &  &  \\ 
\hline
 &  &  &  & $^*$ = Spectral Processing B \\ 
 &  &  &  & \# = Spectral Processing C \\ 
\end{tabular}
}
%\end{table*}
%\end{landscape}
\end{sidewaystable*}

\end{document}